\documentclass[UTF8, 12pt]{article}
\usepackage[margin=1in]{geometry}
\usepackage{amssymb, amsmath}
\usepackage{graphicx,psfrag,epsf, xcolor}
\usepackage{enumerate}
\usepackage{natbib}
\usepackage{epigraph}
\usepackage{url} 
\usepackage{amsmath}
\usepackage{amssymb}
\usepackage{graphicx}
\usepackage{epstopdf}
\usepackage{inputenc}
\usepackage{caption}
\usepackage{subcaption,hyperref}
\usepackage[linesnumbered,ruled,vlined]{algorithm2e}

\newtheorem{definition}{Definition}

\newtheorem{assumption}{Assumption}
\newtheorem{theorem}{Theorem}
\newtheorem{example}{Example}
\newtheorem{proposition}{Proposition}

\title{Semiparametric Inference for Partially Identifiable Data Fusion Estimands via Double Machine Learning}
\author{Yicong Jiang, Lucas Janson}
\date{}
\usepackage{etoolbox}

\makeatletter
\patchcmd{\epigraph}{\@epitext{#1}}{\itshape\@epitext{#1}}{}{}
\makeatother

\begin{document}
	
	\maketitle

	\begin{abstract}
		Many statistical estimands of interest (e.g., in regression or causality) are functions of the joint distribution of multiple random variables. But in some applications, data is not available that measures all random variables on each subject,
		and instead the only possible approach is one of \emph{data fusion}, where multiple independent data sets, each measuring a subset of the random variables of interest, are combined for inference. 
		In general, since all random variables are never observed jointly, their joint distribution, and hence also the estimand which is a function of it, is only \emph{partially} identifiable. 
		Unfortunately, the endpoints of the partially identifiable region depend in general on entire conditional distributions, rendering them hard both operationally and statistically to estimate. To address this, we present a novel outer-bound on the region of partial identifiability (and establish conditions under which it is tight)
  that depends only on certain conditional first and second moments. This allows us to derive semiparametrically efficient estimators of our endpoint outer-bounds that only require the standard machine learning toolbox which learns conditional means. We prove asymptotic normality and semiparametric efficiency of our estimators and provide consistent estimators of their variances, enabling asymptotically valid confidence interval construction for our original partially identifiable estimand. We demonstrate the utility of our method in simulations and a data fusion problem from economics.
	\end{abstract}
	\section{Introduction} \label{sec:Intro} 
	\subsection{Motivation} \label{subsec:motiv}

	{
		\color{black}
Many questions of interest across domains relate variables that are expensive, unethical, illegal, or even physically impossible to measure simultaneously. In such cases, it may be preferable, or only possible, to perform inference via \emph{data fusion} \citep{castanedo2013review}, i.e., combining multiple data sets, each measuring a subset of the variables of interest.
For example, \cite{bostic2009housing} sought to understand the relationship between consumption ($Z$) and housing wealth ($Y$) for U.S. citizens. However, data for household consumption is collected by the U.S. Bureau of Labor Statistics’ Consumer Expenditure Survey (CEX) while housing wealth information is surveyed completely separately in the Federal Reserve Board’s Survey of Consumer Finances (SCF).
	Since these two data sources are not measured on the same individuals, $Y$ and $Z$ are never observed jointly for any individual. However, to determine, say, the correlation $\rho_{YZ}$ between $Y$ and $Z$, it is necessary to estimate $\mathbb{E}[YZ]$, which cannot be point-identified without jointly observing $(Y, Z)$.
		Luckily, both CEX and SCF contain measurements of some shared variables $X$ (though still on different individuals) including demographic variables like age and educational level. Since we observe $(Y, X)$ jointly in SCF and $(Z, X)$ jointly in CEX, the correlations $\rho_{YX}$ and $\rho_{ZX}$ are point-identified and can be used to bound $\rho_{YZ}$ via positive-semidefiniteness of $(Y,Z,X)$'s correlation matrix: $\rho_{YZ} \in \Bigl[\rho_{YX}\rho_{ZX} - \sqrt{(1 - \rho_{YX}^2)(1 - \rho_{ZX}^2)}, \rho_{YX}\rho_{ZX} + \sqrt{(1 - \rho_{YX}^2)(1 - \rho_{ZX}^2)}\Bigr]$.
		Thus $\rho_{YZ}$ is \emph{partially} identifiable, in that it is impossible to precisely determine its value regardless of the size of either data set, yet it \emph{is} possible to bound $\rho_{YZ}$ within a range of values, whose width approaches $2\sqrt{(1 - \rho_{YX}^2)(1 - \rho_{ZX}^2)}$ as both data set sizes approach infinity.
		If either $\rho_{YX}$ or $\rho_{ZX}$ is close to 1, this range will be quite narrow and thus highly informative about the value of $\rho_{YZ}$ despite never observing $(Y,Z)$ jointly.
		
		
		Data fusion problems are common across many different fields, but they share a similar structure. Typically, these problems involve three (possibly multivariate) random variables $Y$, $Z$, and $X$, and an estimand that is a functional of their joint distribution. The estimand often involves terms of the form $\mathbb{E}[h(Y, Z, X)]$, where $h$ is a known function. However, only separate observations of $(Y, X)$ and $(Z, X)$ are available instead of joint observations of $(Y, Z, X)$, resulting in partial identifiability of the estimand. Here are a few more illustrative examples to motivate our work.
		
		
		Data fusion is often needed for private, sensitive, or confidential information (see e.g., \cite{ruggles2019differential, santos2020differential, muellera672proposed, kenny2021impact, kallus2022assessing}). For instance, consider the study of discrimination in loan applications. Here, $Y$ might represent whether an individual receives loan approval, $Z$ a sensitive attribute like the applicant's race, and $X$ a non-sensitive proxy for $Z$, such as the individual's name. A goal might be to measure the relationship between race $Z$ and loan approval $Y$. However, certain fairness regulations, like the U.S. Equal Credit Opportunity Act, prohibit the collection of racial information, preventing $(Y, Z)$ from being observed jointly.
		Recognizing that loan application data contains $(Y, X)$, while $(Z, X)$ can be obtained through public Census data which do not include individual credit information like $Y$, data fusion provides a way forward.

		Causal inference is another area where data fusion can be necessary. In the potential outcome framework \citep{rubin1974estimating}, the counterfactual outcomes for both treatment and control cannot be observed simultaneously (so in our notation, let $Y$ be the treatment outcome and $Z$ the control outcome for an individual). The covariates $X$ in causal inference, on the other hand, are observed for all the individuals. Hence, one can only observe $(Y, X)$ (in the treatment group) or $(Z, X)$ (in the control group) simultaneously, but not $(Y, Z, X)$ jointly. The most common estimand in causal inference, the average treatment effect $\mathbb{E}[Y-Z] = \mathbb{E}[Y] - \mathbb{E}[Z]$, only depends on the marginal distributions of $Y$ and $Z$, so this particular estimand has no issue of partial identifiability and can be inferred without data fusion. 
		However, researchers sometimes want to estimate more complex estimands that depend on the joint distribution of potential outcomes $Y$ and $Z$ \citep{FAN201742}, and since these outcomes are never observed together, such estimands require data fusion. For example, it is sometimes natural to consider certain outcomes on a multiplicative scale, such as income or survival time, and in such cases researchers may want to estimate the average \emph{relative} treatment effect $\mathbb{E}[Y/Z]$.
        \footnote{Although researchers often address $\mathbb{E}[Y/Z]$'s partial identifiability by log-transforming (or other related transformation, see, e.g., \cite{chen2023logs}) their outcomes and then performing inference on the identifiable average treatment effect of the transformed potential outcomes, such an approach changes the estimand, including possibly even its direction, impacting the interpretability of its inferences.}

		Yet another example of data fusion arises when validation studies are used in epidemiology (see, e.g., \cite{fox2020common, marshall1990validation, wacholder1993validation}), but in the interest of space, we defer the details of this example to Appendix~\ref{appendix:val}.
	}

	\subsection{Problem Statement}
	All the aforementioned motivating examples are data fusion problems involving partial identification and can be mathematically formulated as follows. Suppose that there are two datasets $\mathcal{D}_Y$ and $\mathcal{D}_Z$ to be fused, with $n_Y$ and $n_Z$ observations respectively, so in total $n = n_Y + n_Z$ independent observations are available. 
	Both the datasets contain common random variables $X \in \mathbb{R}^{p_X}$, while only in the first dataset $\mathcal{D}_Y$ are the random variables $Y \in \mathbb{R}^{p_Y}$ observed and only in the second dataset $\mathcal{D}_Z$ are the random variables $Z \in \mathbb{R}^{p_Z}$ observed. 
	To unify the notation across datasets, define $R$ as the inclusion indicator of dataset $\mathcal{D}_Y$, i.e., $R = 1$ if the observation is from $\mathcal{D}_Y$ and $R = 0$ otherwise. Then, the entire set of observations from the two datasets can be written as $(X_i, R_i Y_i, (1 - R_i)Z_i, R_i)_{i=1}^{n}$. Suppose that the uncensored full data $(X_i, Y_i, Z_i, R_i)_{i=1}^{n}$ are independently and identically distributed, and let $\mathbb{P}_0$ be the true joint distribution of a single such observation $(X, Y, Z, R)$.
	Our target is to estimate and provide inference on the estimand $\theta = \mathbb{E}_{\mathbb{P}_0}[h(Y, Z, X)]$ for a certain known scalar-valued function $h$.\footnote{Although some estimands of interest (e.g., $\mathrm{Corr}(Y, Z)$) may be functions of multiple such expectations, we focus on targets that can be written as just one expectation for expositional simplicity and because it captures the main technical and methodological challenges for our data fusion setting.
	}
	$\theta$ is generally not fully identifiable because $(Y, Z, X)$ are not observed jointly, but it is partially identifiable via the identifiability of the pairwise distributions of $(Y, X)$ and $(Z, X)$ from $\mathcal{D}_Y$ and $\mathcal{D}_Z$, respectively.
	
	In this paper, we focus on the missing-at-random case, where $R \perp \!\!\!\perp (Y, Z)\mid X$. In causal inference or survey sampling, this corresponds to the standard unconfoundedness assumption.
	This is our main structural assumption about the data-generating process, and without it, there is very little information that can be shared between the two data sets, as the unobserved $Y$'s corresponding to dataset $\mathcal{D}_Z$ can have arbitrary distribution (and vice versa).
	Notably, the missing-at-random assumption only implies the identifiability of the joint distributions of $(Y,X)$ and $(Z,X)$ but not the full joint distribution of $(Y,Z,X)$, so it does not obviate the partial identifiability of $\theta = \mathbb{E}_{\mathbb{P}_0}[h(Y, Z, X)]$.

	\subsection{Related Work} \label{subsec:rel_work}
	

	The literature on \emph{statistical matching} studies the same problem as data fusion and contains a rich body of work (see, e.g., \cite{d2006statistical, rassler2012statistical} for reviews of the area). Common nonparametric approaches in the statistical matching literature make additional identification assumptions, such as conditional independence between $Y$ and $Z$ given $X$, or the presence of an instrumental variable \citep{stock2003introduction,baiocchi2014instrumental}. Another example is \cite{li2023efficient}'s assumption of alignment conditions across different datasets, which generalizes such conditional independence assumptions. 
	All such approaches' assumptions are sufficient to ensure full identifiability of $\theta$, while in this paper we avoid such assumptions, resulting in $\theta$ being only partially identifiable.
	Another class of approaches leverages parametric modeling.
	For instance, \cite{bickel1993efficient,robins1995semiparametric,hasminskii2006asymptotic,evans2018doubly} assume parametric models which make the estimand identifiable, while other works such as \cite{pacini2019two} also adopt parametric models in a partially identifiable setting.
	In contrast to these works, our paper adopts a semiparametric approach that always results in partial identifiability and does not rely on any parametric models.

	Data fusion is also studied in probability theory and econometrics \citep{makarov1982estimates, frank1987best, manski2003partial, molinari2008partial,beresteanu2012partial, fan2017partial, firpo2019partial, russell2021sharp} and it is closely related to ecological inference \citep{goodman1953ecological, wakefield2001ecological, king2004ecological,  imai2008bayesian, 10.1093/oxfordhb/9780199286546.003.0024,
		cho2008cross,james2009r, king2013solution, manski2018credible,  https://doi.org/10.3982/QE778}. In those areas, the main focus is on characterizing the partly identifiable region for the estimand (a population quantity), instead of statistical inference (based on a finite sample), which is the primary focus of our paper.

	Two data fusion papers conduct statistical inference for model-free partial identification bounds and hence are particularly closely related to our work. 
	First, \cite{fan2016estimation} investigates the data fusion problem for causal inference. They provide asymptotic inference in the case where the counterfactuals $Y$ and $Z$ are \emph{binary}, which simplifies the partially identifiable bounds and facilitates inference. 
	In contrast, our paper addresses the more challenging setting where $Y$ and $Z$ can be non-binary and continuous, and in particular addresses challenges that are unique to this setting and do not arise in the binary setting considered in \cite{fan2016estimation}; see Section~\ref{sec:bound} for details.
	Second, \cite{ji2023model} considers data fusion formulated very similarly to our paper but takes quite a different methodological approach. These methodological differences lead to both advantages and disadvantages, and we defer a detailed methodological comparison to Section~\ref{sec:ji} after we have introduced our method, and we also compare the methods numerically throughout Section~\ref{sec:empirical}.

	\subsection{Our Contribution}
	

We propose a novel semiparametric inference method for data fusion that addresses statistical, computational, and operational challenges inherent to this challenging yet commonplace problem. Our method takes a double machine learning approach that allows the user to leverage state-of-the-art modeling (including machine learning) under only assumptions on its rate of estimation consistency. In order to enable this, we introduce novel bounds on the partially identifiable parameter set that only depend on a small set of conditional moments, making them more tractable for semiparametric inference than the exact partially identifiable bounds which depend on entire conditional distributions, yet our bounds often approximate the exact bounds well. Our inference is proved to be asymptotically valid and semiparametrically efficient for our tractable bounds, leading to narrow and robust inference for challenging data fusion estimands which we demonstrate and compare to alternatives in various simulations and an economics application relating consumption to wealth.

	\subsection{Notation}
We will use $\mathbb{E}[\cdot]$ and $\mathrm{Var}(\cdot)$ to respectively denote the expectation and variance (covariance matrix when the input is multidimensional) under $\mathbb{P}_0$. For two random variables $V$ and $W$, 
we denote the marginal distribution of $V$ and the conditional distribution of $V\mid W$, both under $\mathbb{P}_0$, by $P_0(V)$ and $P_0(V\mid W)$, respectively. 
For a positive semi-definite matrix $A$, $\sqrt{A}$ denotes any positive semi-definite matrix satisfying $\sqrt{A}\sqrt{A} = A$. 

	\section{Partially Identifiable Bounds} \label{sec:bound}

\subsection{Tight Identifiable Bounds}\label{sec:tightbounds}
We start by stating the tightest identifiable bounds on $\theta$, which by our problem setup can only depend on the distribution $\mathbb{P}_0=P_0(Y,X,Z)$ via the conditional distributions $P_0(Y\mid X, R=1)$ and $P_0(Z\mid X, R=0)$.
Without loss of generality, we first focus on the upper bound:
	\begin{align} \label{equ:thetau}
		\theta = \mathbb{E}[h(Y, Z, X)] &= \mathbb{E}[\mathbb{E}[h(Y, Z, X)\mid X]] \notag \\
		&\le  \mathbb{E}\left[\sup_{\mathbb{Q}_X \in \mathcal{C}(P_0(Y\mid X), P_0(Z\mid X))} \mathbb{E}_{\mathbb{Q}_X}[h(Y, Z, X)\mid X]\right] \notag \\
		&= \mathbb{E}\left[\sup_{\mathbb{Q}_X \in \mathcal{C}(P_0(Y\mid X, R = 1), P_0(Z\mid X, R=0))} \mathbb{E}_{\mathbb{Q}_X}[h(Y, Z, X)\mid X]\right]\overset{\mathrm{def}}{=} \theta_\mathrm{U},
	\end{align}
	where $\mathcal{C}(\mu_Y, \mu_Z)$ denotes the copula of the two marginal distributions $\mu_Y$, $\mu_Z$, i.e., the collection of all joint distributions (in this case for $(Y,Z)$) whose marginal distributions match $\mu_Y$ and $\mu_Z$, respectively. And the first equality in the last line follows from our missing-at-random assumption that $R \perp \!\!\!\perp (Y, Z)\mid X$.
	Since the inequality in \eqref{equ:thetau} is tight, $\theta_{\mathrm{U}}$ is the tightest identifiable upper bound of $\theta$.
	The tight lower bound is defined and derived analogously:
	\begin{align} \label{equ:thetal}
		\theta \ge \theta_\mathrm{L} \overset{\mathrm{def}}{=} \mathbb{E}\left[\inf_{\mathbb{Q}_X \in \mathcal{C}(P_0(Y\mid X, R = 1), P_0(Z\mid X, R = 0))} \mathbb{E}_{\mathbb{Q}_X}[h(Y, Z, X)\mid X]\right].
	\end{align}
	%
	As desired, both $\theta_\mathrm{L}$ and $\theta_\mathrm{U}$ are functions only of $P_0(Y\mid X, R = 1)$ and $P_0(Z\mid X, R = 0)$, which are identifiable. 
$\theta_\mathrm{L}$ and $\theta_\mathrm{U}$ are the endpoints of the exact partially identifiable region for $\theta$ and thus provide the most natural targets for statistical inference, since the best confidence interval that is achievable for $\theta$ must consist of a lower confidence bound for $\theta_\mathrm{L}$ and an upper confidence bound for $\theta_{\mathrm{U}}$. However, the infimum/supremum in the expressions for $\theta_\mathrm{L}$ and $\theta_\mathrm{U}$ render them challenging to estimate. In particular, $\theta_\mathrm{L}$ and $\theta_\mathrm{U}$ depend on the \emph{entire} conditional distributions $P_0(Y\mid X, R = 1)$ and $P_0(Z\mid X, R = 0)$, unlike more standard estimands such as in M-estimation or quantile regression, where typically the estimand depends on only a finite number of (conditional) moments or quantiles. From an estimation and inference standpoint, this detailed dependence on $P_0(Y\mid X, R = 1)$ and $P_0(Z\mid X, R = 0)$ could make the problem far harder: setting aside the conditioning on $X$, estimating a finite number of moments or quantiles can generally be done at parametric rates under quite mild nonparametric assumptions, while estimating an entire distribution function can typically only be done at a parametric rate under parametric assumptions, and under nonparametric assumptions the rate depends on the smoothness of the density, degrading rapidly as the smoothness decreases (see, e.g., \cite{diaz2017efficient}). This smoothness-dependent rate particularly impacts the ability to quantify uncertainty of an estimand (necessary for inference) in the absence of smoothness assumptions. Operationally, estimating conditional moments, and to some extent also conditional quantiles, is a standard and ubiquitous task in supervised learning, and thus there exists an extremely rich array of tools for such estimation. On the other hand, it is much less standard to estimate an entire conditional distribution, and since such estimation would be necessary for even consistent estimation of $\theta_\mathrm{L}$ and $\theta_\mathrm{U}$, nevermind inference, there is an additional roadblock to inference for $\theta$ via $\theta_\mathrm{L}$ and $\theta_\mathrm{U}$ that would make it operationally hard for an analyst to leverage the many standard tools in machine learning for this task.

	\subsection{Cauchy--Schwarz Bounds}
    \label{subsec:cs_bounds}

As suggested at the end of the previous subsection, we would statistically and operationally prefer to work with estimands that do not depend on all of $P_0(Y\mid X, R = 1)$ and $P_0(Z\mid X, R = 0)$. To this end, this subsection presents outer bounds for $\theta_\mathrm{L}$ and $\theta_\mathrm{U}$ that depend only on the first two conditional moments of $P_0(Y\mid X, R = 1)$ and $P_0(Z\mid X, R = 0)$, making these outer bounds considerably more tractable to perform inference on. We will also discuss when and how tight these outer bounds are. 

For a value $x\in\mathbb{R}^{p_X}$,
functions $f: \mathbb{R}^{p_Y}\times \mathbb{R}^{p_X} \rightarrow \mathbb{R}^{p_f}$ and $g: \mathbb{R}^{p_Z}\times \mathbb{R}^{p_X} \rightarrow \mathbb{R}^{p_g}$, and 
distributions $\mu_{Y}$ and $\mu_{Z}$ defined on $\mathbb{R}^{p_Y}$ and $\mathbb{R}^{p_Z}$, respectively, 
let $\mathcal{C}^{(2)}_{f,g,x}(\mu_{Y},\mu_{Z})$ denote the collection of all joint distributions $\mathbb{Q}$ for $(Y,Z)$ such that
\begin{align*} 
\mathbb{E}_{\mathbb{Q}}[f(Y,x)] = \mathbb{E}_{\mu_Y}[f(Y,x)], \quad & \quad \mathbb{E}_{\mathbb{Q}}[f(Y,x)f(Y,x)^T] = \mathbb{E}_{\mu_Y}[f(Y,x)f(Y,x)^T], \\
\mathbb{E}_{\mathbb{Q}}[g(Z,x)] = \mathbb{E}_{\mu_Z}[g(Z,x)], \quad & \quad \mathbb{E}_{\mathbb{Q}}[g(Z,x)g(Z,x)^T] = \mathbb{E}_{\mu_Z}[g(Z,x)g(Z,x)^T].
\end{align*}
So $\mathcal{C}^{(2)}_{f,g,x}(\mu_{Y},\mu_{Z})$ can be thought of as a relaxation of a copula, in that instead of requiring $\mathbb{Q}\in\mathcal{C}^{(2)}_{f,g,x}(\mu_{Y},\mu_{Z})$ to match the entire marginal distributions $\mu_Y,\mu_Z$, it only requires $\mathbb{Q}$ to match the first two moments of some functions $f$ and $g$. We can now define second-order approximations of $\theta_{\mathrm{L}}$ and $\theta_{\mathrm{U}}$ analogously to the tight bounds in Equation~\eqref{equ:thetau}--\eqref{equ:thetal}:
\begin{equation}\label{equ:csl}
\theta^{(2)}_{f,g,\mathrm{L}} \overset{\mathrm{def}}{=} \mathbb{E}\left[\inf_{\mathbb{Q}_X \in \mathcal{C}_{f,g,X}^{(2)}(P_0(Y\mid X, R = 1), P_0(Z\mid X, R = 0))} \mathbb{E}_{\mathbb{Q}_X}[h(Y, Z, X)\mid X]\right],
\end{equation}
\begin{equation}\label{equ:csu}
\theta^{(2)}_{f,g,\mathrm{U}} \overset{\mathrm{def}}{=} \mathbb{E}\left[\sup_{\mathbb{Q}_X \in \mathcal{C}_{f,g,X}^{(2)}(P_0(Y\mid X, R = 1), P_0(Z\mid X, R = 0))} \mathbb{E}_{\mathbb{Q}_X}[h(Y, Z, X)\mid X]\right].
\end{equation}
Note ${\mathcal{C}_{f,g,X}^{(2)}(P_0(Y\mid X, R = 1), P_0(Z\mid X, R = 0))} \supseteq {\mathcal{C}(P_0(Y\mid X, R = 1), P_0(Z\mid X, R = 0))}$ for any $f$ and $g$, which guarantees that $\theta^{(2)}_{f,g,\mathrm{L}} \le \theta_{\mathrm{L}}$ and $\theta^{(2)}_{f,g,\mathrm{U}} \ge \theta_{\mathrm{U}}$, and thus $\theta^{(2)}_{f,g,\mathrm{L}} \le \theta \le \theta^{(2)}_{f,g,\mathrm{U}}$. Unfortunately, Equations~\eqref{equ:csl}--\eqref{equ:csu} are still not in particularly appealing form for inference, and hence our main result of this section is to derive simple closed form expressions for $\theta^{(2)}_{f,g,\mathrm{L}}$ and $\theta^{(2)}_{f,g,\mathrm{U}}$ in terms of only the conditional means and variances of $f(Y,X)$ and $g(Z,X)$, for $f$ and $g$ chosen carefully based on the following notion of \emph{decomposability}.

\begin{definition}\label{def:decomposability} A scalar-valued function $h(y,z,x)$ is \emph{decomposable} if it can be written as $h(y,z,x)=f(y,x)^Tg(z,x)$ for some functions $f$ and $g$ with finite-dimensional outputs. The definition of the term `$(f,g)$-decomposable' is identical except that it also specifies $f$ and $g$.
\end{definition}
Decomposability is often met for estimands of interest, including $\mathbb{E}[YZ]$ (by letting $f(y,x)=y$ and $g(z,x)=z$), the average relative treatment effect $\mathbb{E}[Y/Z]$ presented in Section~\ref{subsec:motiv} (by letting $f(y,x)=y$ and $g(z,x)=1/z$), and many more; see Appendix~\ref{app:decomposable_examples} for more examples. 
Furthermore, even more generally, \emph{any} non-pathological $h(y,z,x)$ can be arbitrarily well-approximated by $f(y,x)^Tg(z,x)$ via basis decomposition. For instance, if for any fixed $z$ and $x$, $h(y, z, x) \in L^2(\mathbb{R}^{p_Y})$, then $h(y, z, x)$ can be expressed as the basis expansion $h(y, z, x) = \sum_{i=1}^{\infty} \phi_i(y) \cdot \left(\int_y h(y, z, x) \phi_i(y) dy\right)$, where $\{\phi_i(y): i \in \mathbb{Z}^+\}$ is any complete orthonormal basis of $\mathbb{R}^{p_Y}$, e.g., the Fourier basis. Define $f(y, x) = \left(\phi_i(y)\right)_{i = 1}^{p_f}$ and $g(z, x) = \left(\int_y h(y, z, x) \phi_i(y) dy\right)_{i = 1}^{p_f}$, then $f(y,x)^Tg(z,x)$ can approximate $h(y, z, x)$ arbitrarily well as long as $p_f$ is large enough.
 Now considering decomposable $h$, we can state our main theoretical result.

\begin{theorem} \label{lemma:cs}
    If $h$ is $(f,g)$-decomposable, then
    \begin{align}
	 		\theta_{f,g,\mathrm{L}}^{(2)} &= \mathbb{E}[\mathbb{E}[f(Y, X)\mid X]^T \mathbb{E}[g(Z, X)\mid X]] - \notag \\ &\qquad \quad \mathbb{E}\left[\mathrm{tr}\left(\sqrt{\sqrt{\mathrm{Var}[g(Z, X)\mid X]}\mathrm{Var}[f(Y, X)\mid X]\sqrt{\mathrm{Var}[g(Z, X)\mid X]}}\right)\right], \\
	 			\theta_{f,g,\mathrm{U}}^{(2)} &= \mathbb{E}[\mathbb{E}[f(Y, X)\mid X]^T \mathbb{E}[g(Z, X)\mid X]] + \notag \\ &\qquad\quad \mathbb{E}\left[\mathrm{tr}\left(\sqrt{\sqrt{\mathrm{Var}[g(Z, X)\mid X]}\mathrm{Var}[f(Y, X)\mid X]\sqrt{\mathrm{Var}[g(Z, X)\mid X]}}\right)\right].
	 	\end{align}
\end{theorem}
We defer the proof of Theorem~\ref{lemma:cs} to Appendix~\ref{appendix:cs}. For the remainder of the paper we will leave $f$ and $g$ fixed and assume $h$ is $(f,g)$-decomposable, and hence simplify the notation by defining $\theta_\mathrm{L}^{(\mathrm{CS})}\overset{\mathrm{def}}{=}\theta_{f,g,\mathrm{L}}^{(2)}$ and $\theta_\mathrm{U}^{(\mathrm{CS})}\overset{\mathrm{def}}{=}\theta_{f,g,\mathrm{U}}^{(2)}$, where `CS' stands for \emph{Cauchy--Schwarz}, inspired by the central role of the Cauchy--Schwarz inequality in the proof of Theorem~\ref{lemma:cs}. We now turn to the tightness of $\theta_\mathrm{L}^{(\mathrm{CS})}$ and $\theta_\mathrm{U}^{(\mathrm{CS})}$.

\begin{proposition}\label{prop:tight}
    If there exist measurable functions $U(x)$ and $V(x)$ such that for any $x$, the conditional distribution, given $X=x$, of $f(Y,X)$ is the same as that of $U(X) g(Z, X) + V(X)$,
    then the Cauchy--Schwarz bounds are tight, i.e.,
    $\theta_\mathrm{L}^{(\mathrm{CS})} = \theta_\mathrm{L}$ and $\theta_\mathrm{U}^{(\mathrm{CS})} = \theta_\mathrm{U}$.
\end{proposition}
 The proof of Proposition~\ref{prop:tight} is deferred to Appendix~\ref{appendix:prop:tight}. Proposition~\ref{prop:tight} indicates that the Cauchy--Schwarz bound coincides with the tight bound when $P_0(f(Y, X)\mid X)$ and $P_0(g(Z, X)\mid X)$ are have the same distribution up to $X$-dependent location and scale parameters. This is quite a rich nonparametric class of model families that includes many common parametric families, such as if both $P_0(f(Y, X)\mid X)$ and $P_0(g(Z, X)\mid X)$ are multivariate Gaussian (with means and covariance matrices allowed to depend arbitrarily on $X$), or if they are both Exponentially distributed (with scales allowed to depend arbitrarily on $X$). But it also includes many nonparametric model families that would be harder to describe in words but may describe real data more accurately, and more importantly, may provide a reasonable second-order approximation to an even larger class of model families that do not strictly satisfy the conditions of Proposition~\ref{prop:tight}. In these cases, Proposition~\ref{prop:tight} tells us that there is little or no loss to using the Cauchy--Schwarz bounds in place of the tight bounds. However, the Cauchy--Schwarz bounds could be less tight when $f$ and $g$'s conditional distributions are far from being separated by only an $X$-dependent location and scale function.

The main benefit of Theorem~\ref{lemma:cs} is that $\theta_\mathrm{L}^{(\mathrm{CS})}$ and $\theta_\mathrm{U}^{(\mathrm{CS})}$ have relatively simple forms depending only on conditional first and second moments; both such moments can be fitted via a straightforward adaptation of conditional mean fitting, which nearly all machine learning algorithms are explicitly designed for. This makes $\theta_\mathrm{L}^{(\mathrm{CS})}$ and $\theta_\mathrm{U}^{(\mathrm{CS})}$ amenable to a semiparametric double machine learning approach \citep{chernozhukov2018double} that allows the statistical and operational benefits of leveraging machine learning for inference, as we detail in the following section.
	


	\section{Inference via Double Machine Learning} \label{sec:theory}
	
	
	This section will provide doubly-robust and efficient estimation methods for the bounds defined in Section~\ref{sec:bound} through double machine learning. We focus on the case that $p_f = p_g = 1$ to avoid tedium. The result can be easily generalized to higher $p_f=p_g$. 
	
	
	\subsection{Doubly-robust Estimation} \label{subsec:dr}
    
	\sloppy
	We first focus on the Cauchy--Schwarz upper bound $\theta_\mathrm{U}^{(\mathrm{CS})}$. The lower bound can be handled similarly. The key idea for obtaining a doubly-robust estimator is to alleviate the influence of the nuisance parameter.  Denote ${m_Y(x) = \mathbb{E}_{\mathbb{P}_0}[f(Y, X)\mid X = x]}$, ${m_Z(x) = \mathbb{E}_{\mathbb{P}_0}[g(Z, X)\mid X = x]}$.  
 ${e(x) = \mathbb{E}_{\mathbb{P}_0}[R\mid X = x]}$, ${v_Y(x) = \mathrm{Var}_{\mathbb{P}_0}[f(Y, X)\mid X = x]}$, and ${v_Z(x) = \mathrm{Var}_{\mathbb{P}_0}[g(Z, X)\mid X = x]}$.
We state the \emph{efficient influence function} for $\theta_\mathrm{U}^{(\mathrm{CS})}$ as follows.
	\begin{align}
		\varphi^{(\mathrm{CS})}_\mathrm{U}(y, z , x, r) &= 
		\frac{r}{e(x)}\varphi^{(\mathrm{CS})}_{Y, X, \mathrm{U}}(y, x)
		+ \frac{1 - r}{1 - e(x)}\varphi^{(\mathrm{CS})}_{Z, X, \mathrm{U}}(z, x) 
		 + M^{\left(\mathrm{CS}\right)}_\mathrm{U} \left(x \right) -\theta_\mathrm{U}^{(\mathrm{CS})}, 
	\end{align}
	where the first term corresponds to the contribution of $Y$, with
	\begin{align}
		&\varphi^{(\mathrm{CS})}_{Y, X, \mathrm{U}}(y, x) = \left(f\left(y, x\right) - m_Y(x)\right)m_Z(x)   + \frac{1}{2}\left[ \left(f\left(y, x\right) - m_Y(x)\right)^2 - v_Y(x) \right]\notag   \sqrt{\frac{v_Z(x)}{v_Y(x)}},  
	\end{align}
	the second term corresponds to the contribution of $Z$, with
	\begin{align}
		&\varphi_{Z, X, \mathrm{U}}^{\left(\mathrm{CS}\right)}\left(z, x\right) = \left(g\left(z, x\right) - m_Z(x)\right)m_Y(x) +\frac{1}{2}\left[ \left(g\left(z, x\right) - m_Z(x)\right)^2 - v_Z(x) \right] \notag \sqrt{\frac{v_Y(x)}{v_Z(x)}},  
	\end{align}
	and the third term corresponds to the contribution of $X$, with
	\begin{align*}
		M^{\left(\mathrm{CS}\right)}_\mathrm{U}\left(x\right) &=  m_Y(x)m_Z(x) + \sqrt{v_Y(x)v_Z(x)}.
	\end{align*}

	The efficient influence functions capture the first-order impact of the distribution of $(Y, Z, X)$ on the estimand. Also, they represent the hardest direction for estimation. Therefore, including them in the estimator can help reduce the bias and lead to a better rate of convergence. See Appendix~\ref{apendix:eif} for a brief introduction to the efficient influence function. Note that, except for the last term $\theta_\mathrm{U}^{(\mathrm{CS})}$, $\varphi^{(\mathrm{CS})}_\mathrm{U}(y, z, x, r)$ depends only on the first two conditional moment functions of $Y, Z, R \mid X$  and it consists of three main components: the contribution of $Y$, $Z$, and $X$. Within each component, there are two parts: the impact of the conditional mean in the Cauchy--Schwarz bound, and the impact of the standard deviation. Note that because the standard deviation is the square root of a functional of a probability distribution, its contribution to the influence function involves the derivative of the square root, resulting in a conditional standard deviation in the denominator. Similar to standard positivity assumptions on inverse propensity scores in causal inference, we will require these denominators to not be too concentrated near 0; see Assumption~\ref{as:pos} in Subsection~\ref{subsec:asy} for the formal statement. Also see Appendix~\ref{appendix:ex4} for an example of the singular behavior of an estimand containing such a square root, demonstrating the necessity of such positivity assumptions.
	
With the efficient influence functions in hand, we now apply standard semiparametric inference techniques (see, e.g., \citet{kennedy2022semiparametric} for a recent review) to derive a Neyman orthogonal estimator via cross-fitting, as detailed in Algorithm~\ref{alg:est}, which assumes for expositional simplicity that $K$ evenly divides $n$. 
	
\begin{algorithm}
\caption{Semiparametric inference for $\theta$ via Cauchy--Schwarz bounds}\label{alg:est}

	Split the data  $\{(X_i, Y_i, Z_i, R_i): i \in [n]\}$ evenly into $K$ folds  $\{(X_i, Y_i, Z_i, R_i): i \in I_k\}$
 
  \For{$k= 1$ \KwTo $K$}{
	Estimate  $m_Y(x), m_Z(x), v_Y(x), v_z(x), e(x)$ (e.g., via machine learning) using all the data except the $k$th fold $I_k$. 
  Denote the corresponding estimates by $\hat{m}_Y^{(-k)}(x) , \hat{m}_Z^{(-k)}(x), \hat{v}_Y^{(-k)}(x),\hat{v}_Z^{(-k)}(x), \hat{e}^{(-k)}(x)$, respectively.\label{line:ml}
  
  Calculate the plug-in estimators  on 
		$I_k$:
  $$\hat{\theta}_\mathrm{L}^{(\mathrm{CS}, k)} = \frac{K}{n}\sum_{i \in I_k}\left[\hat{m}_Y^{(-k)}(X_i)\hat{m}_Z^{(-k)}(X_i) - \sqrt{\hat{v}_Y^{(-k)}(X_i)\hat{v}_Z^{(-k)}(X_i)} \right],$$
  $$\hat{\theta}_\mathrm{U}^{(\mathrm{CS}, k)} = \frac{K}{n}\sum_{i \in I_k}\left[\hat{m}_Y^{(-k)}(X_i)\hat{m}_Z^{(-k)}(X_i) + \sqrt{\hat{v}_Y^{(-k)}(X_i)\hat{v}_Z^{(-k)}(X_i)} \right],$$
  
	 Calculate the debiased estimator on 
		$I_k$,
        $$ \hat{\theta}_\mathrm{L}^{(*, \mathrm{CS}, k)} = \hat{\theta}_\mathrm{L}^{(\mathrm{CS}, k)} + \frac{K}{n}\sum_{i \in I_k}  \hat{\varphi}^{(\mathrm{CS},k)}_\mathrm{L}(Y_i, Z_i , X_i, R_i) ,\  \hat{\theta}_\mathrm{U}^{(*, \mathrm{CS}, k)} = \hat{\theta}_\mathrm{U}^{(\mathrm{CS}, k)} + \frac{K}{n}\sum_{i \in I_k}  \hat{\varphi}^{(\mathrm{CS},k)}_\mathrm{U}(Y_i, Z_i , X_i, R_i) ,$$
		where $\hat{\varphi}^{(\mathrm{CS},k)}_\mathrm{L}, \hat{\varphi}^{(\mathrm{CS},k)}_\mathrm{U}$ are the plug-in estimates of $\varphi^{(\mathrm{CS})}_\mathrm{L}, \varphi^{(\mathrm{CS})}_\mathrm{U}$ that replace 
  all conditional moment functions by their $\hat{\cdot}^{(-k)}$ estimates
  and $\theta_\mathrm{L}^{(\mathrm{CS})}, \theta_\mathrm{U}^{(\mathrm{CS})}$ with $\hat{\theta}_\mathrm{L}^{(\mathrm{CS}, k)},\hat{\theta}_\mathrm{U}^{(\mathrm{CS}, k)}$. 
       }
		Aggregate the debiased estimators:
		\begin{align*}
    \hat{\theta}_{\mathrm{L}}^{(\mathrm{CS})} =\frac{1}{K} \sum_{k=1}^{K}  \hat{\theta}_\mathrm{L}^{(*, \mathrm{CS}, k)}, \qquad
     \hat{\theta}_{\mathrm{U}}^{(\mathrm{CS})} =\frac{1}{K} \sum_{k=1}^{K}  \hat{\theta}_\mathrm{U}^{(*, \mathrm{CS}, k)}.
		\end{align*}
    	
		Estimate the variance of the debiased estimators: 
  $$\widehat{V}_\mathrm{L} = \frac{1}{n}\sum_{k=1}^{K}\sum_{i \in I_k}\left[ \hat{\varphi}^{(\mathrm{CS},k)}_\mathrm{L}(Y_i, Z_i , X_i, R_i) \right]^2, \qquad \widehat{V}_\mathrm{U} = \frac{1}{n}\sum_{k=1}^{K}\sum_{i \in I_k}\left[ \hat{\varphi}^{(\mathrm{CS},k)}_\mathrm{U}(Y_i, Z_i , X_i, R_i) \right]^2.$$

      
      Calculate the $1 - \alpha$ lower confidence bound (LCB), $\hat{\theta}_{\mathrm{LCB}}^{(\mathrm{CS})}$, and the $1 - \alpha$ upper confidence bound (UCB), $\hat{\theta}_{\mathrm{UCB}}^{(\mathrm{CS})}$  of $\theta$:
      $$\hat{\theta}_{\mathrm{LCB}}^{(\mathrm{CS})} = \hat{\theta}_{\mathrm{L}}^{(\mathrm{CS})} - q_{1 -  \alpha / 2}\sqrt{\widehat{V}_\mathrm{L}}, \qquad  \hat{\theta}_{\mathrm{UCB}}^{(\mathrm{CS})} = \hat{\theta}_{\mathrm{U}}^{(\mathrm{CS})} + q_{1 -  \alpha / 2}\sqrt{\widehat{V}_\mathrm{U}},$$
      where $q_{1 - \alpha / 2}$ is the $1 -  \alpha / 2$ quantile of the standard Gaussian distribution $\mathcal{N}(0, 1)$.
	
  \Return $\left[\hat{\theta}_{\mathrm{LCB}}^{(\mathrm{CS})},\hat{\theta}_{\mathrm{UCB}}^{(\mathrm{CS})}\right]$, the $1 - \alpha$ confidence interval for $\theta$.

\end{algorithm}

Note that the estimation in Line~\ref{line:ml} in Algorithm~\ref{alg:est} can be performed using only the most standard supervised/machine learning paradigm of conditional mean estimation: First, the propensity score function $\mathbb{E}[R\mid X]$ can be estimated by fitting a machine learning algorithm to a data set with the $R_i$'s as the response and the $X_i$'s as the covariates. Second, the conditional mean of $f(Y,X)\mid X,R=1$ can be estimated by fitting a machine learning algorithm to a data set comprised of the data points for which $R_i=1$, with the $f(Y_i,X_i)$'s as the response and the $X_i$'s as the covariates. Letting this fitted conditional mean function of $f(Y,X)\mid X,R=1$ be denoted by $\hat{m}_Y(x)$, the conditional variance of $f(Y,X)\mid X,R=1$ can then be estimated by fitting a machine learning algorithm to a data set comprised of the data points for which $R_i=1$, with the $(f(Y_i,X_i)-\hat{m}_Y(X_i))^2$ as the response and the $X_i$'s as the covariates. An analogous strategy works for the conditional mean of $g(Z,X)\mid X, R=0$. Hence in deploying Algorithm~\ref{alg:est}, the user can leverage the vast majority of the field of supervised/machine learning, thus providing a high degree of modeling flexibility and operational convenience. 

In the next section, we lay out the conditions under which the confidence interval $\left[\hat{\theta}_{\mathrm{LCB}}^{(\mathrm{CS})},\hat{\theta}_{\mathrm{UCB}}^{(\mathrm{CS})}\right]$ returned by Algirhtm~\ref{alg:est} is valid. 

\subsection{Asymptotic Theory} \label{subsec:asy}

To prove the validity of our inference, we need several assumptions.


\begin{assumption}[Missing at random] \label{as:mar}
	$R \perp\!\!\!\perp (Y, Z) \mid X$.
\end{assumption}
Missing at random, or unconfoundedness, is often assumed in missing data or causal inference literature. In our problem, this assumption guarantees that it is possible to fuse the two datasets. Otherwise, if the missing indicator $R$ depends on $Y$ or $Z$, the distribution of $Y$ or $Z$ might be systematically different in the two datasets, and even partial identification of the parameter could be impossible.

\begin{assumption}[Finite moments] \label{as:finite-m}
	The following expectations are finite, for all $k$: \hspace{.2cm}${\mathbb{E}[f(Y, X)^{16}]}$, ${\mathbb{E}[g(Z, X)^{16}]}$, ${\mathbb{E}[[\hat{m}_Y^{(-k)}(X) - m_Y(X)]^{16}] }$, ${\mathbb{E}[[\hat{m}_Z^{(-k)}(X) - m_Z(X)]^{16}] }$, ${\mathbb{E}[[\hat{v}_Y^{(-k)}(X)^{0.5} - v_Y(X)^{0.5}]]^{16}}$, ${\mathbb{E}[[\hat{v}_Z^{(-k)}(X)^{0.5} - v_Z(X)^{0.5}]^{16}] }$, ${\mathbb{E}[[\hat{v}_Y^{(-k)}(X)^{-0.5} - v_Y(X)^{-0.5}]^{16}] }$, and ${\mathbb{E}[[\hat{v}_Z^{(-k)}(X)^{-0.5} - v_Z(X)^{-0.5}]^{16}] }$.
\end{assumption}
Assumptions similar to Assumption~\ref{as:finite-m} are prevalent in the semiparametric inference literature, although typically they only require finite fourth moments. The reason we require finite 16th moments is that the Cauchy--Schwarz bounds require estimation of the product of variances, a fourth-order function of the original random variables, so we are essentially requiring finite fourth moments of fourth-order functions (and $4\times 4 = 16$).

\begin{assumption}[Positivity] \label{as:pos}
	The following expectations are finite: ${\mathbb{E}[\frac{1}{\mathbb{E}[R\mid X]}^4]}$, ${\mathbb{E}[\frac{1}{\mathrm{Var}(f(Y, X)\mid X)}^8]}$, and ${\mathbb{E}[\frac{1}{\mathrm{Var}(g(Z, X)\mid X)}^8]}$.
\end{assumption}
Positivity, or a non-overlapping condition, is a common assumption in missing data and causal inference literature. Specifically, in our problem, the assumption on the propensity score, $\mathbb{E}[\frac{1}{\mathbb{E}[R\mid X]}^4] < \infty$, ensures that the sizes of the two datasets are of the same scale so that their information contents are of the same order. Otherwise, the smaller dataset will dominate the error in estimation, and one could essentially treat the larger dataset as infinite and only study the error coming from the smaller dataset, which would not really be in the spirit of data fusion. Less standard are our conditional variance positivity assumptions, which arise because of the square root in the Cauchy--Schwarz bound, $\sqrt{\mathrm{Var}(f(Y, X)\mid X)}$ and $\sqrt{\mathrm{Var}(g(Z, X)\mid X)}$, causing there to be $\mathrm{Var}(f(Y, X)\mid X)^{-0.5}$ and $\mathrm{Var}(g(Z, X)\mid X)^{-0.5}$ terms in the influence functions, which have a singularity at 0 we must avoid.

\begin{assumption}[Consistency of non-parametric estimation] \label{as:cons}
	The following expectations are $o(1)$, for all $k$: ${\mathbb{E}[[\hat{v}_Y^{(-k)}(X)^{0.5} - v_Y(X)^{0.5}]^4]}$, ${\mathbb{E}[[\hat{v}^{(-k)}_Z(X)^{0.5} - v_Z(X)^{0.5}]^4]}$, ${\mathbb{E}[[\hat{v}^{(-k)}_Y(X)^{-0.5} - v_Y(X)^{-0.5}]^4]}$, ${\mathbb{E}[[\hat{v}^{(-k)}_Z(X)^{-0.5} - v_Z(X)^{-0.5}]^4]}$, and ${\mathbb{E}[[\frac{1}{\hat{e}^{(-k)}(x)} - \frac{1}{e(x)}]^4]}$.
\end{assumption}
\begin{assumption}[Efficiency of non-parametric estimation]\label{as:effi}
	The following quantities are $o(n^{-1/4})$, for all $k$: 
	${\left[\mathbb{E}[[\hat{m}_Y^{(-k)}(X) - m_Y(X)]^8]\right]^{\frac{1}{8}}}$,
	${\left[\mathbb{E}[[\hat{m}_Z^{(-k)}(X) - m_Z(X)]^8]\right]^{\frac{1}{8}}}$,
	${\left[\mathbb{E}[[\hat{v}_Y^{(-k)}(X) - v_Y(X)]^4]\right]^{\frac{1}{4}}}$,
	${\left[\mathbb{E}[[\hat{v}_Z^{(-k)}(X) - v_Z(X)]^4]\right]^{\frac{1}{4}}}$, 
	${\left[\mathbb{E}[[\frac{1}{\hat{e}^{(-k)}(x)} - \frac{1}{e(x)}]^2]\right]^{\frac{1}{2}}}$.
\end{assumption}
Assumption \ref{as:cons} and \ref{as:effi} require the estimators to be consistent, with error of order $o(n^{-1/4})$, which is considerably weaker than the parametric rate of $O(n^{-1/2})$. Often in semiparametric inference, the required rates of consistency are stated as products of the errors of a pair of estimators achieving a rate of $o(n^{-1/2})$, relaxing the requirement that both estimators achieve a rate of $o(n^{-1/4})$ and instead allowing one to achieve a worse rate as long as the other achieves a correspondingly better rate so the product of their errors achieves a rate of $o(n^{-1/2})$. In fact, the same is true in our setting, but the error products do not have a simple form (see Appendix~\ref{appendix:est}), so we preferred to present a simplification that separates them out in Assumption~\ref{as:effi}. We give empirical evidence of robustness when one conditional moment is estimated better than the other in Section~\ref{subsec:simu:imbal}.

With these assumptions, we can now establish the inferential properties of our procedure through the following two theorems.

\begin{theorem} \label{thm:est}
	Assume that Assumptions \ref{as:mar}--\ref{as:effi} hold. Then, estimators $\hat{\theta}_{\mathrm{U}}^{(\mathrm{CS})}$ and $\hat{\theta}_{\mathrm{L}}^{(\mathrm{CS})}$ are asymptotically normal and semiparametric efficient for their estimands $\theta_{\mathrm{U}}^{(\mathrm{CS})}$ and $\theta_{\mathrm{L}}^{(\mathrm{CS})}$:
	$$ \sqrt{n}(\hat{\theta}_{\mathrm{U}}^{(\mathrm{CS})} - \theta_{\mathrm{U}}^{(\mathrm{CS})}) \overset{d}{\to} \mathcal{N}(0, \mathbb{E}[\varphi^{(\mathrm{CS})}_\mathrm{U}(Y, Z , X, R )^2]),$$
	
	$$ \sqrt{n}(\hat{\theta}_{\mathrm{L}}^{(\mathrm{CS})} - \theta_{\mathrm{L}}^{(\mathrm{CS})}) \overset{d}{\to} \mathcal{N}(0, \mathbb{E}[\varphi^{(\mathrm{CS})}_\mathrm{L}(Y, Z , X, R )^2]).$$
\end{theorem}


\begin{theorem}\label{thm:var}
	Assume that Assumptions \ref{as:mar}-\ref{as:effi} hold. Then,
	$$\widehat{V}_\mathrm{U} \overset{p}{\to} \mathbb{E}[\varphi^{(\mathrm{CS})}_\mathrm{U}(Y, Z , X, R)^2], \qquad\widehat{V}_\mathrm{L} \overset{p}{\to} \mathbb{E}[\varphi^{(\mathrm{CS})}_\mathrm{L}(Y, Z , X, R)^2].$$
\end{theorem}

The proofs of Theorems~\ref{thm:est} and \ref{thm:var} are given in Appendices~\ref{appendix:est} and \ref{appendix:var}, respectively. An immediately corollary of Theorems \ref{thm:est}--\ref{thm:var} is the asymptotic coverage of the confidence interval from Algorithm~\ref{alg:est}:
$$\liminf_{n \to \infty}\mathbb{P}\left(\theta \in [\hat{\theta}_{\mathrm{LCB}}^{(\mathrm{CS})}, \hat{\theta}_{\mathrm{UCB}}^{(\mathrm{CS})}]\right) \ge 1 - \alpha.$$

%
%
%


\subsection{Comparison to \cite{ji2023model}} \label{sec:ji}

\citet{ji2023model} considers nearly the same statistical problem as we do, but their proposed optimization-based method, Dualbounds, is quite different from ours.  Dualbounds constructs an optimization problem that depends on the estimated conditional distributions $P_0(Y\mid X)$ and $P_0(Z\mid X)$ in such a way that as the estimation error decreases, the inference procedure derived from their optimization becomes tighter and tighter around the tight partially identifiable bounds. 

Perhaps the primary advantage \citet{ji2023model}'s approach has over our method is that, at least in principle, its inferential bounds can always approach the tight bounds, whereas our method will only do so under the conditions laid out in Proposition~\ref{prop:tight}. In practice, however, such tightness requires consistent estimation of $P_0(Y\mid X)$ and $P_0(Z\mid X)$, a task we argue in this paper is both statistically and operationally challenging. This challenge also seems to be recognized by \citet{ji2023model}, who propose as a solution to, like us, only model first and second conditional moments; indeed the empirical results in their paper exclusively consider estimators characterized entirely by first and second conditional moments. If an analyst is already restricting themselves to only estimate conditional first and second moments, our method has a number of distinct advantages: it is semiparametrically efficient (with particular benefits over Dualbounds when the estimation errors are imbalanced as may often be the case in data fusion; see Section~\ref{subsec:simu:imbal} and Appendix~\ref{appendix:lin_qua}), operationally simple (aside from moment estimation, Algorithm~\ref{alg:est} performs only simple arithmetic operations with no tuning parameters except $K$), and computationally fast (about $600$ times faster in a $n=1000$, $p_X=20$ example, due to the many internal optimizations Dualbounds runs; see Appendix~\ref{appendix:simu}).

Another noteworthy difference between our methods is the allowable forms of estimand. In our notation, an advantage of Dualbounds is that it does not require $h$ to be decomposable, but a disadvantage is that it requires $Y$ and $Z$ to be one-dimensional, while we do not restrict the dimension of $Y$, $Z$, $f(Y,X)$, or $g(Z,X)$. We argue after Definition~\ref{def:decomposability} that decomposability is a mild restriction, though arguably so is Dualbounds' restriction to one-dimensional $Y$ and $Z$. With that said, one of \citet{ji2023model}'s primary motivating estimands (and the estimand in all of its empirical results) is the average treatment effect under selection bias, which in fact cannot be expressed in terms of one-dimensional $Y$ and $Z$ without an extra assumption; in Example~\ref{example:lee} in Appendix~\ref{app:decomposable_examples} we show how it can be expressed without this assumption in terms of expectations of decomposable estimands with \emph{two}-dimensional $Y$ and $Z$, thus fitting into the framework of our paper.

For more detail on the comparison between our work and \citet{ji2023model}, see Appendix~\ref{app:jietal}. The next section also provides empirical comparisons with \citet{ji2023model} in a range of settings.

\section{Numerical Experiments} \label{sec:empirical}

In this section, we perform several numerical experiments to demonstrate the good performance of our method. All code to implement our method and replicate our numerical results can be found at \url{https://github.com/yicong-jiang/DataFusion.git}.

\subsection{Simulations}\label{subsec:simu}

\subsubsection{Linear Model} \label{subsec:simu:imbal}
Our first simulation is a proof of concept of our method in a low-dimensional simulation based on an epidemiological validation study modeled via a Gaussian linear model. The results are as we would hope: across noise levels and correlation strengths, our method has excellent coverage and width close to that of the partially identifiable lower bound $\theta_\mathrm{U}-\theta_\mathrm{L}$. These results are descriptive and validating but tell us little new beyond the theory from Section~\ref{subsec:asy}, so we defer them (and the details of the experiment) to Appendix~\ref{appendix:val}, while focusing on more informative settings in the main text. 

Next, we study our method and compare it to Dualbounds in a heavy-tailed linear model simulation. We focus in particular on the case with disparate noise levels between the two data sets, as a model for the common data fusion setting when one data set is of higher quality or sample size than the other. Formally, suppose $Y, Z \in \mathbb{R}$ and $X \in \mathbb{R}^{20}$, with 
$$ Y = \beta_Y^T X + \sigma_Y \epsilon_Y, \qquad Z = \beta_Z^T X + \sigma_Z \epsilon_Z, $$
where $\epsilon_Y^{1/3} \mid X$ and $\epsilon_Z^{1/3} \mid X$ both follow a $\mathcal{N}(0, 15^{-1/3})$ distribution (which makes $\mathrm{Var}(\epsilon_Y)=\mathrm{Var}(\epsilon_Z)=1$). 
It follows from Proposition~\ref{prop:tight} that this is a setting where the tight bounds and the Cauchy--Schwarz bounds are the same, i.e., $\theta_\mathrm{L}^{(\mathrm{CS})} = \theta_\mathrm{L}$ and $\theta_\mathrm{U}^{(\mathrm{CS})} = \theta_\mathrm{U}$. We generate $R \mid  X \sim \mathrm{Bern}(0.5)$ and set the total sample size to $n = 1000$.
We generate $\beta_Y = \beta_Z = \beta$  from the uniform distribution on the unit sphere  and $X \sim \mathcal{N}(0, I)$ so that $\mathrm{Var}(\beta_Y^T X) = \mathrm{Var}(\beta_Z^T X) = 1$. We set $Z$'s noise level $\sigma_Z = 0.2$, and vary $Y$'s noise level $\sigma_Y$. 

To make comparison with Dualbounds as fair as possible, we implement our method to match the defaults in the Dualbounds software (and use those defaults for our Dualbounds implementation as well). In particular, we provide our method with the known propensity score (i.e., $\hat{e}^{(-k)}(x)=e(x)$), use cross-validated ridge regression as our base machine learner for the conditional mean, and estimate the conditional variance as homoskedastic (i.e., we use the scalar empirical variance of the residuals of the fitted conditional mean). We also set $K$ in Algorithm~\ref{alg:est} to $2$.

\begin{figure}[t]
	\centering
	\includegraphics[width=\textwidth]{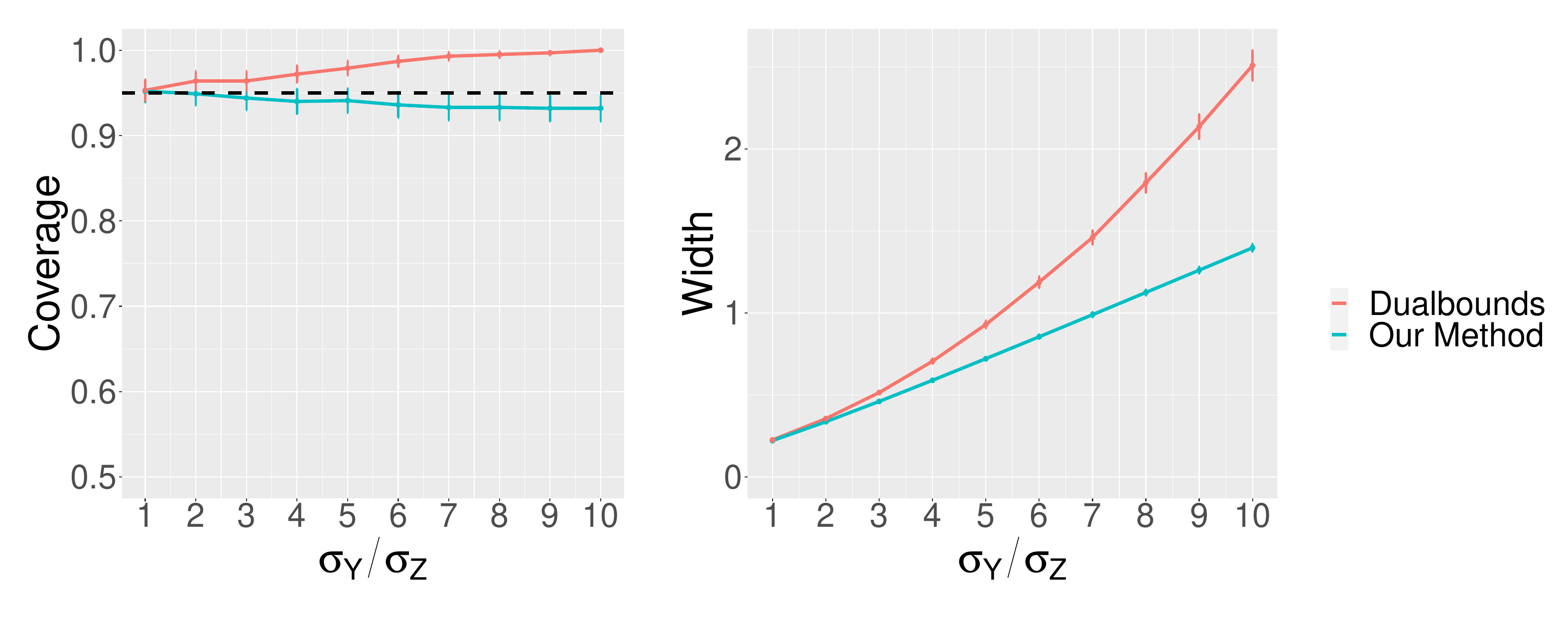}
	\caption{
    Influence of the noise level $\sigma_Y$ on our method's and Dualbounds' coverage (left) and width (right) for 95\% confidence intervals in the simulation of Section~\ref{subsec:simu:imbal}. Error bars represent $\pm 1.96$ Monte Carlo standard errors. 
    }
	\label{fig:ols_imbal}
\end{figure}

Figure~\ref{fig:ols_imbal} shows the coverage and width of our method's and Dualbounds's 95\% confidence regions. 
Throughout Section~\ref{sec:empirical}, ``Coverage" refers to the (empirical) probability that the confidence interval includes the entire partially identifiable region $[\theta_\mathrm{L}, \theta_\mathrm{U}]$ defined in Equations~\eqref{equ:thetau}--\eqref{equ:thetal}. 
The left panel of Figure~\ref{fig:ols_imbal} shows our method's coverage close to the nominal level for all values of the noise ratio $\sigma_Y/\sigma_Z$, only dipping below 95\% by a couple percentage points for high noise ratios. Dualbounds's coverage matches the nominal level when $\sigma_Y/\sigma_Z=1$ but climbs to very close to 100\% as the noise ratio increases to 10. The right panel of Figure~\ref{fig:ols_imbal} shows that Dualbounds's conservativeness for large $\sigma_Y/\sigma_Z$ comes at a significant cost in terms of confidence interval width: although the two methods have the same width when $\sigma_Y/\sigma_Z=1$, Dualbounds's width grows to nearly double that of our method by $\sigma_Y/\sigma_Z=10$. 
The apparent linearity of the width of our method can be explained by its double-robustness leading to widths scaling like the product of estimation errors, in this case, $\sigma_Y\sigma_Z$, which is linear in $\sigma_Y/\sigma_Z$ when $\sigma_Z$ is held constant. The superlinearity of the width of Dualbounds seems to align with bias bounds proved in \cite{ji2023model} which are suggestive of a quadratic relationship with $\sigma_Y^2$. See Appendix~\ref{appendix:lin_qua} for more detailed theoretical investigation of the scaling of both methods' width curves in this simulation.
Appendix~\ref{appendix:simu} repeats this experiment with the heavy-tailed residuals replaced by Gaussian, so that Dualbounds' conditional models are well-specified; the results are qualitatively similar but the differences between the widths of the two methods are less pronounced.

\subsubsection{Average Relative Treatment Effect} \label{subsec:simu_ape}
Having just considered in Section~\ref{subsec:simu:imbal} a relatively favorable setting for our method where $(\theta_\mathrm{L}^{(\mathrm{CS})},\theta_\mathrm{U}^{(\mathrm{CS})}) = (\theta_\mathrm{L},\theta_\mathrm{U})$ and the data had moderate tails, we now turn to a much less favorable setting where the Cauchy--Schwarz bounds are not tight and the data has \emph{very} heavy-tails. In particular, 
in this subsection we consider the estimation of the average relative treatment effect in causal inference, discussed at the end of Section~\ref{subsec:motiv}: $\theta = \mathbb{E}[Y/Z]$ where $Y$ represents the potential outcome under treatment and $Z$ the potential outcome under control. 

We let $X\in\mathbb{R}^{20}$ and
$$ \begin{pmatrix}
	\log(Y) \\ \log(Z) 
\end{pmatrix} 
\sim
\mathcal{N}\left(
\begin{pmatrix}
	\beta_1^T X\\ \beta_0^T X
\end{pmatrix}, \sigma^2
\begin{pmatrix}
	1 & \rho \\ \rho & 1
\end{pmatrix}
\right).
$$ 
We generate $R\mid X \sim \mbox{Bern}((1 + \exp(- \beta_3^TX))^{-1})$ and set the total sample size to $n=1000$. We generate $\beta_1, \beta_2, \beta_3 \sim \mathcal{N}\left(0, \frac{0.5^2}{20}I_{20}\right)$ and  $X\sim \mathcal{N}(0, \Sigma)$ with $\Sigma_{ij} = 0.3^{|i - j|}$.
In this case, $f(Y, X) = Y$ and $g(Z, X) = 1 / Z$ and our (unidentifiable) estimand\footnote{An alternative standard approach is to perform inference on the identifiable average treatment effect after log transformation, $\tilde{\theta}:=\mathbb{E}[\log(Y(1)) - \log(Y(0))]$, but in this case $\tilde{\theta}=0$ regardless of the parameters, rendering such inference uninformative.} is
\begin{equation}\label{eq:multmodel}
\theta  = \mathbb{E}\left[Y/Z\right] = \exp\{\sigma^2 (1 - \rho) + 0.5(\beta_1 - \beta_0)^T\Sigma(\beta_1 - \beta_0) \}.
\end{equation}

For this simulation, both our method and Dualbounds are not given the true propensity score $\mathbb{E}[R\mid X]$ and instead estimate it via cross-validated logistic regression with combined ridge and lasso penalties (via python's \texttt{sklearn.LogisticRegressionCV} function). Both methods are given the correctly specified parametric conditional model \eqref{eq:multmodel} and estimate $\beta_0$ and $\beta_1$ via cross-validated ridge regression on the log-transformed outcomes and estimate $\sigma^2$ via the mean squared error of those estimates. 

\begin{figure}[t]
	\centering
	\includegraphics[width=\textwidth]{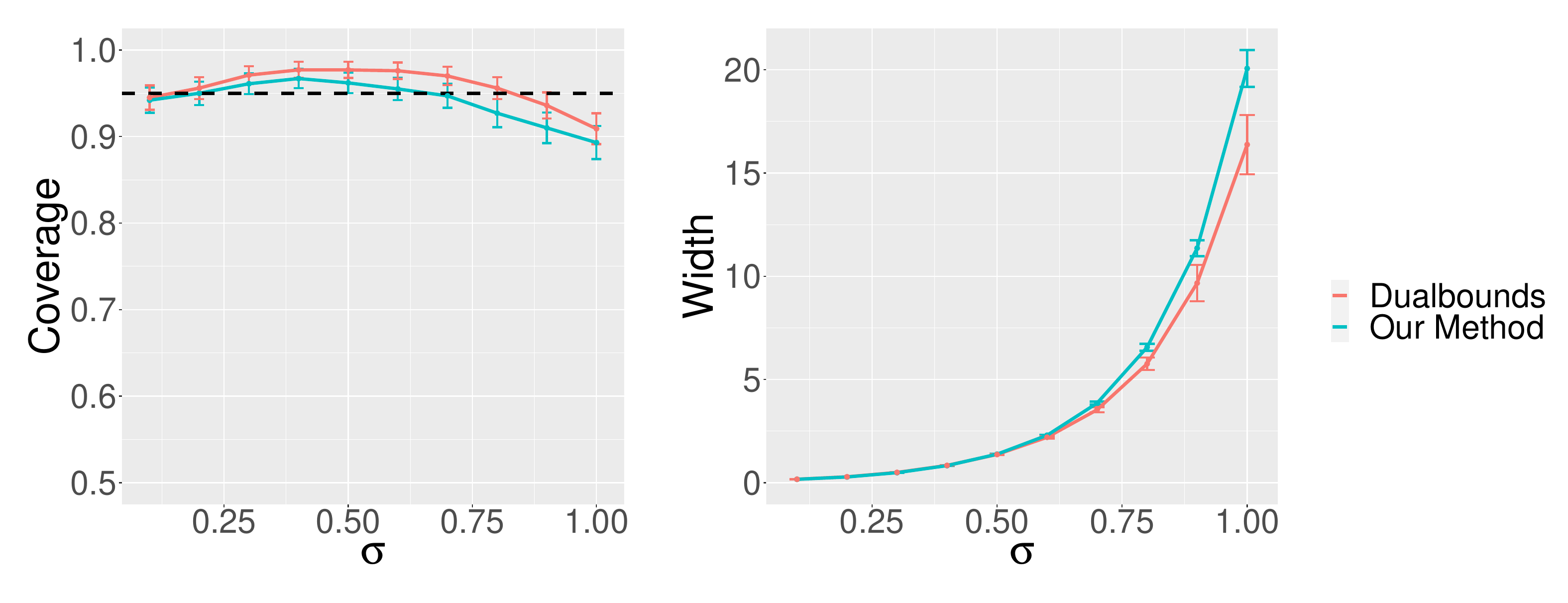}
	\caption{Influence of the noise level $\sigma$ on our method's and Dualbounds' coverage (left) and width (right) for 95\% confidence intervals in the simulation of Section~\ref{subsec:simu_ape}. Error bars represent $\pm 1.96$ Monte Carlo standard errors.
 }
	\label{fig:lognorm}
\end{figure}

Figure~\ref{fig:lognorm} shows the coverage and width of the two methods' 95\% confidence intervals as we vary $\sigma$, which controls both the noise level and the heaviness of the tails of the data. 
It can be seen that when $\sigma $ is moderate (i.e. $\sigma \le 0.7$), the two methods perform quite similarly, both maintaining 95\% coverage and almost identical width.
As $\sigma$ grows beyond 0.7, both methods' widths rapdily increase, and we also see both methods' coverage start to dip.
Since the $m$th moment of $Y$ and $Z$ is proportional to $\exp(0.5m^2\sigma^2)$, as $\sigma$ increases, the moments of $Y$ and $Z$ escalate rapidly. For instance, when $\sigma  = 1$, the $4$th moment of $Y$ is $\exp(6) \approx 403$ times as large as when $\sigma  = 0.5$, and the $16$th moment of $Y$ is $\exp(96) \approx 5\times 10^{41}$ times as large as when $\sigma  = 0.5$. Despite these ballooning moments, our method's coverage degrades only slightly, maintaining coverage above 90\%.
Additionally, although Dualbounds is in principle able to approach the tight bounds while in this setting our method is not, the two methods' widths remain very close, suggesting the Cauchy--Schwarz bounds our method relies on do not hamper its performance by much even when Proposition~\ref{prop:tight} does not hold.

\subsection{Relating Consumption to Wealth} \label{subsec:real}
We consider the data fusion problem introduced in \cite{bostic2009housing} as described in Section~\ref{sec:Intro}, and which was also analyzed in \cite{evans2018doubly}. Recall that to study the relationship between consumption ($Z$) and housing (net) wealth ($Y$), we need to fuse two datasets:  the U.S. Bureau of Labor Statistics’ Consumer Expenditure Survey (CEX), which contains $Z$, and the Federal Reserve Board’s Survey of Consumer Finances (SCF), which contains $Y$.  
The two datasets share numerous common demographic variables ($X$), including attributes strongly correlated with wealth and consumption, such as wage. Consequently, $X$ is able to explain $Y$ and $Z$ reasonably well, providing hope for informative inference on their partially identifiable relationship.
The target parameter is the ordinary least squares coefficient $\beta_Z$ for $Z$ in the regression of $Y$ on $X$ and $Z$.
Following \cite{evans2018doubly}, we use the data for the third quarter of 1979.

We consider both linear regression (fitted via cross-validated ridge regression as in Section~\ref{subsec:simu}) and random forests to model both the first and second conditional moments, and logistic regression (fitted as in Section~\ref{subsec:simu_ape}) for estimating the propensity score. 
Since the target parameter is a function of several identifiable or partially identifiable estimands, we use the delta method during the inference process; see Appendix~\ref{appendix:delta} for more details.
Our 95\% confidence interval is $[0.31, 1.10]$ for linear regression and $[0.35, 1.05]$ for random forests, respectively. Both confidence intervals largely agree with one another and are bounded above 0, illustrating a moderately positive impact of consumption on wealth which is in line with economic theory \citep{mankiwMacroeconomics2013}. 
We also ran the same analysis with truncation applied to $\mathbb{E}[R\mid X]$ to keep it away from 0 and 1 (a standard technique applied to propensity scores in causal inference \citep{glynn2010introduction}) and/or with quadratic terms added (since \cite{evans2018doubly} added quadratic terms to their model). The results were all fairly consistent, with all confidence intervals overlapping and bounded above 0; see Appendix~\ref{appendix:real} for a full table of results. In contrast to our nonparametric approach that handles the partial identifiability, \cite{evans2018doubly}'s analogous analysis assumes a parametric model with particular quadratic terms that ensure $\beta_Z$ is identifiable. They reach a qualitatively similar conclusion, though their confidence interval is narrower thanks to their stronger assumptions; see Appendix~\ref{appendix:real} for further details.




Dualbounds can also be applied to this data, though it is not immediate how to apply it to $\beta_Z$ which is a a function of a mix of identifiable and partially identifiable estimands. So to compare with Dualbounds, we consider the related (but purely partially identifiable) estimand $\mathbb{E}[YZ]$.
When using linear regression to estimate the conditional moments, our 95\% confidence interval is $[-2.13, 1.03]$ while Dualbounds' is $[-2.29,1.15]$ (9\% wider) and when using random forests, our 95\% confidence interval is $[-1.97, 1.10]$ while Dualbounds' is $[-2.18,1.38]$ (16\% wider). 

\section*{Acknowledgments}
YJ and LJ were partially supported by DMS-2045981 and DMS-2134157.

\bibliographystyle{chicago}
\bibliography{partial.bib}

\newpage

\appendix

\allowdisplaybreaks

\section{Estimands with Decomposable $h$}\label{app:decomposable_examples}

In this section, we provide several examples of decomposable functions. Among them, Example~\ref{example:CDF}, Example~\ref{example:varITE}, Example~\ref{example:CTE}, and Example~\ref{example:lee}  are closely related to or come from the examples in Section 2.5 of \cite{ji2023model}.

\begin{example}[Regression MSE]
		Consider the case where $h(y, z, x) = (f(y) - g(z, x))^2$. Denote $h_1(y, z, x) = f(y)^2 + g(z, x)^2$, $h_2(y, z, x) = f(y)g(z, x)$. Then $h = h_1 - 2h_2$, $\mathbb{E}[h_1(Y, Z, X)]$ is identifiable, and $h_2$ is decomposable. As a result, $\theta = \mathbb{E}[h(Y, Z, X)]$ can be partially identified by Cauchy--Schwarz bound. 
	\end{example}

  \begin{example}[Joint CDF at fix point] \label{example:CDF}
     Consider the joint CDF of $Y$ and $Z$ at a fix point $(y_0, z_0)$, $\theta = \mathbb{P}(Y \le y_0, Z \le z_0)$. Since $\theta = \mathbb{E}[I_{Y \le y_0} I_{Z \le z_0}]$, we can define $h(y, z, x) = I_{y \le y_0} I_{z \le z_0}$, $f(y, x) = I_{y \le y_0}$, $g(z, x) = I_{y \le y_0}$. Then $h(y, z, x) = f(y, x) g(z, x)$ is decomposable.
 \end{example}
 
\begin{example}[Ecological Inference]
     In ecological inference, one major goal is to evaluate $\theta = \mathbb{P}[Y = y \mid Z = z, X = x]$ with the knowledge of $\mathbb{P}[Y = y\mid X = x]$ and $\mathbb{P}[Z = z\mid X = x]$ \citep{cross2002regressions}. Note that $\theta = \frac{\mathbb{E}[I_{Y = y}I_{Z = z}I_{X = x}]}{\mathbb{P}[Z = z, X = x]}$. Denote $h(y, z, x) = I_{Y = y}I_{Z = z}I_{X = x}$, $f(y, x) = I_{Y = y}$, $g(z, x) = I_{Z = z}I_{X = x}$. Then $\theta = \frac{\mathbb{E}[h(y, z, x)]}{\mathbb{P}[Z = z, X = x]} = \frac{\mathbb{E}[f(y, x)g(z, x)]}{\mathbb{P}[Z = z, X = x]}$, with the denominator being identifiable and numerator being the expectation of the decomposable function.
\end{example}

 \begin{example}[Average Percentage Effect]
     In causal inference, one may be interested in the treatment effect under the multiplicative scale. Mathematically, suppose that $Y(1), Y(0)$ are the potential outcomes. One may hope to estimate the average percentage effect $\theta = \mathbb{E}[Y(1) / Y(0)]$. Denote $Y = Y(1), Z = Y(0)$, and let $X$ be the observed covariates. Define $h(y, z, x) = y / z$, $f(y, x) = y, g(z, x) = 1 / z$, then $\theta = \mathbb{E}[h(Y, Z, X)] = \mathbb{E}[f(Y, X)g(Z, X)]$, with $h$ being decomposable. 
 \end{example}

 \begin{example}[Variance of ITE] \label{example:varITE}
     In causal inference, we are often interested in the diversity of the individual treatment effect (ITE), which is mathematically characterized by the variance of ITE, $\theta = \mathrm{Var}(Y(1) - Y(0)) = \mathbb{E}[Y(1)^2] + \mathbb{E}[Y(0)^2] - \mathbb{E}[(Y(1) - Y(0))]^2 - 2\mathbb{E}[Y(1)Y(0)]$. The first part of $\theta$,  $\mathbb{E}[Y(1)^2] + \mathbb{E}[Y(0)^2] - \mathbb{E}[(Y(1) - Y(0))]^2$, is identifiable. The second part of $\theta$, $-2\mathbb{E}[Y(1)Y(0)]$, is the expectation of a decomposable function if we denote $Y = Y(1), Z = Z(0), f(Y, X) = Y, g(Z, X) = Z$, and $h(Y, Z, X) = YZ = f(Y, X)g(Z, X)$.
 \end{example}
    \begin{example}[Subgroup treatment effect] \label{example:CTE}
      In causal inference, researchers may be interested in the average treatment effect on a certain sub-population. For instance, \cite{kaji2023assessing} introduce  the subgroup treatment effect, $\theta = \mathbb{E}[Y(1) - Y(0) \mid Y(0) \le c]$, where $Y(1), Y(0)$ are potential outcomes and $c$ is a constant. Denote $Y = Y(1), Z = Y(0)$, then $\theta = \frac{\mathbb{E}[YI_{Z \le c}]}{\mathbb{E}[I_{Z \le c}]} - \mathbb{E}[Z \mid Z \le c]$. Both $\mathbb{E}[I_{Z \le c}]$ and $\mathbb{E}[Z \mid Z \le c]$ are identifiable. The numerator $\mathbb{E}[YI_{Z \le c}]$ is partially identifiable, and can be expressed as the expectation of decomposable function $h$, where $h(Y, Z, X) = YI_{Z \le c}$ and can be decomposed as $f(Y, X)g(Z, X)$, with $f(Y, X) = Y$, and $g(Z, X) = I_{Z \le c}$.
 \end{example}
 \begin{example}[Lee bounds] \label{example:lee}
     In causal inference, selection bias is very common. For instance, \cite{lee2005training} considers a randomized experiment studying the causal effect of a training program on wages. However, when the wages are measured, some participants have not yet found a job, so their wages are missing. Therefore, the average treatment effect (ATE) of the whole population is not identifiable, and the researchers focus on the employed sub-population. Mathematically, suppose $Y(1), Y(0)$ are the potential outcomes of the wages, with $Y(1)$ corresponding to joining the training and $Y(0)$ corresponding to not joining; suppose $S(1)$ and $S(0)$ are the potential outcomes of the employment status, being 1 if the individual is employed and 0 otherwise. Then the ATE of the employed population can be defined as $\theta = \mathbb{E}[Y(1) - Y(0)\mid S(1) = S(0) = 1]$.  Note that $\mathbb{E}[Y(1) - Y(0)\mid S(1) = S(0) = 1] = \mathbb{E}[Y(1)S(1) \cdot S(0) - S(1) \cdot Y(0)S(0)] / \mathbb{E}[S(1)S(0)]$. Define $Y = (Y(1), S(1))$, $Z = (Y(0), S(0))$, $f_1(Y, X) = (Y(1)S(1), S(1))$, $g_1(Z, X) = (S(0), Y(0)S(0))$, $f_2(Y, X) = S(1)$, $g_2(Z, X) = S(0)$. Then $\theta$ can be expressed as $\mathbb{E}[f_1(Y, X)^Tg_1(Z, X)] / \mathbb{E}[f_2(Y, X)g_2(Z, X)]$, with both the numerator and denominator being the expectation of decomposable functions.
 \end{example}

\section{Introduction to Efficient Influence Function} \label{apendix:eif}

We introduce the \emph{Efficient Influence Function} in Subsection~\ref{subsec:dr}.
The efficient influence function is a key concept in  semiparametric statistics theory. It has two key properties:
\begin{enumerate}
	\item (zero impact) For any distribution $\mathbb{P}$, 
	$$ \mathbb{E}_{\mathbb{P}}[\varphi^{(\mathrm{CS})}_\mathrm{U}(Y, Z , X, R; \mathbb{P})] = 0$$
	\item (first-order influence) For any one-dimensional distribution class $\mathbb{P}_\epsilon$ satisfying $d\mathbb{P}_\epsilon(y, z, x, r) = (1 + \epsilon S(y, z, x, r)) d\mathbb{P}_0(y, z, x, r)$, we have
	$$\frac{\partial}{\partial \epsilon} \theta_\mathrm{U}^{(\mathrm{CS})}(\mathbb{P}_\epsilon) = \mathbb{E}_{\mathbb{P}_0}[S(Y, Z, X, R)\varphi^{(\mathrm{CS})}_\mathrm{U}(Y, Z , X, R; \mathbb{P}_0)]$$
	where $\theta_\mathrm{U}^{(\mathrm{CS})}(\mathbb{P}_\epsilon)$ is the value of $\theta_\mathrm{U}^{(\mathrm{CS})}$ when the expectation is taking with respect to $\mathbb{P}_\epsilon$.
\end{enumerate}

The first property of the influence function enables it to lay no impact on the mean of estimators when adding it to the estimator so that one can flexibly leverage it in the estimation process. The second property of the influence function assimilates it to the derivative of the target parameter $\theta$ with respect to the joint distribution of $(Y, Z, X, R)$. Furthermore, for any one-dimensional distribution class $\mathbb{P}_\epsilon$ satisfying $d\mathbb{P}_\epsilon(y, z, x, r) = (1 + \epsilon S(y, z, x, r)) d\mathbb{P}_0(y, z, x, r)$, the Cram\'er-Rao lower bound for $\theta_\mathrm{U}^{(\mathrm{CS})}$ is as follows:
\begin{align}
	\frac{(\frac{\partial}{\partial \epsilon} \theta_\mathrm{U}^{(\mathrm{CS})})^2}{\mathbb{E}_{\mathbb{P}_0}[S(Y, Z, X, R)^2]} = \frac{(\mathbb{E}_{\mathbb{P}_0}[S(Y, Z, X, R)\varphi^{(\mathrm{CS})}_\mathrm{U}(Y, Z , X, R; \mathbb{P}_0)])^2}{\mathbb{E}_{\mathbb{P}_0}[S(Y, Z, X, R)^2]} \le \mathbb{E}_{\mathbb{P}_0}[\varphi^{(\mathrm{CS})}_\mathrm{U}(Y, Z , X, R; \mathbb{P}_0)^2] \notag
\end{align}

The equality holds when $S = \varphi^{(\mathrm{CS})}_\mathrm{U}(Y, Z , X, R; \mathbb{P}_0)$, which indicates that $\varphi^{(\mathrm{CS})}_\mathrm{U}$ represents the "hardest" direction for estimation. Hence, if one can annihilate this direction's influence, the error will reduce dramatically.

\section{More Detailed Comparison with \citet{ji2023model}}\label{app:jietal}


  {
  \color{black}
  In Section~\ref{sec:ji} of the main paper, we describe connections and differences between our proposed method and the Dualbounds method proposed in \cite{ji2023model}. Here we elaborate further on these points.
  
  \subsection{Preliminaries on Dualbounds}\label{appendix:ji:intro}
  In order to better compare our method with \cite{ji2023model}, we restate their method Dualbounds in our framework. Similar to our algorithm, Dualbounds aims at partially identifying $\theta = \mathbb{E}[h(Y, Z, X)]$ via identifiable upper and lower bounds. Here we focus on the upper bound and the lower bound can be derived similarly. To obtain the upper bound, they consider choosing functions $\nu_1(y, x)$ and $\nu_0(z, x)$ such that $\nu_1(y, x) + \nu_0(z, x) \ge h(y, z, x)$ for any $y, z, x$ on the support of the joint distribution of $(Y, Z, X)$. Then, $\mathbb{E}[\nu_1(Y, X)] + \mathbb{E}[\nu_0(Z, X)]$ is naturally an identifiable upper bound for $\theta$. To make the bound tight, the authors of \cite{ji2023model} propose to optimize $\nu_1$, $\nu_0$ on certain function class $\mathcal{C}$ as follows,
  \begin{equation}\label{equ:ji:opt}
      \min_{\nu_1, \nu_0 \in \mathcal{C}}[\mathbb{E}[\nu_1(Y, X)] + \mathbb{E}[\nu_0(Z, X)]] \qquad \text{s.t.} \quad \nu_1(y, x) + \nu_0(z, x) \ge h(y, z, x)
  \end{equation}
or equivalently, for each $x$ on the support of $X$ and function class $\tilde{\mathcal{C}}$, solve
  \begin{equation}\label{equ:ji:opt:equ}
      \min_{\nu_{1x}, \nu_{0x} \in \tilde{\mathcal{C}}}[\mathbb{E}[\nu_{1x}(Y) \mid X = x] + \mathbb{E}[\nu_{0x}(Z) \mid X = x]] \qquad \text{s.t.} \quad \nu_{1x}(y) + \nu_{0x}(z) \ge h(y, z, x)
  \end{equation}

  To estimate the upper bound, \cite{ji2023model} needs to (1) estimate the nuisance parameter $\nu_1, \nu_0$ with estimator $\hat{\nu}_1, \hat{\nu}_0$, and (2) estimate the upper bound $\mathbb{E}[\nu_1(Y, X) + \nu_0(Z, X)]$ via the sample mean of $\hat{\nu}_1(Y, X) + \hat{\nu}_0(Z, X)$. As illustrated in \citet[Theorem 3.2]{ji2023model}, in Step (1), the error in estimating the conditional distribution $Y, Z \mid X$ will propagate to the optimization problem \eqref{equ:ji:opt:equ}, rendering error in $\hat{\nu}_1, \hat{\nu}_0$ and bias in the final upper bound estimator. \citet[Theorem 3.2]{ji2023model} upper-bounds the bias as the product of the error in estimating $Y, Z \mid X$ and $\nu_1, \nu_0$. Because the error in estimating $Y, Z \mid X$ directly propagates to the error in estimating $\nu_1, \nu_0$, the bias bound seems to be approximately of the same order as the \emph{squared} estimation error of $Y, Z\mid X$; \citet[Lemma 3.3]{ji2023model} proves this under certain conditions. 

  Step (2) takes sum of the sample means of the estimates $\hat{\nu}_1$ and $\hat{\nu}_0$ over the observed data points, 
  resulting in an asymptotic variance (in the consistent estimation setting) of $O\left(\frac{1}{n}\left(\mathrm{Var}(\nu_1(Y, X)) + \mathrm{Var}(\nu_0(Z, X))\right)\right)$ for the Dualbounds upper bound estimator. This variance depends on the choice of function class $\mathcal{C}$ over which $\nu_1, \nu_0$ are optimized. In general, the relationship between $\mathcal{C}, \nu_1, \nu_0$ and the variance of the upper bound estimator is complicated and as far as we know is not known to be semi-parametric efficient.

  \subsection{
  The Impact of Estimation Error Confidence on Bound Width} \label{appendix:lin_qua}
    By the definition of confidence bounds, both our and \cite{ji2023model}'s confidence bounds' widths
    can be decomposed into three parts:
    \begin{enumerate}
        \item The width of the tightest-possible bounds: $\theta_{\mathrm{U}}^{(\mathrm{CS})}- \theta_{\mathrm{L}}^{(\mathrm{CS})}$ for our method and  $\theta_{\mathrm{U}} - \theta_{\mathrm{L}}$ for \cite{ji2023model}.
        
        \item The bias of the estimation, $\mathbb{E}[\hat{\theta}_{\mathrm{U}}^{(\mathrm{CS})} - \theta_{\mathrm{U}}^{(\mathrm{CS})}]$ for our upper bound and $\mathbb{E}[\hat{\nu}_1(Y, X) - \nu^*_1(Y, X)] + \mathbb{E}[\hat{\nu}_0(Z, X) - \nu^*_0(Z, X)]$ for \citet{ji2023model}'s upper bound (the bias for the lower bound is defined similarly), where $\nu^*_1,\nu^*_0$ are the tightest choices for $\nu_1,\nu_0$, defined by Equation~\eqref{equ:ji:opt} when the function class $\mathcal{C}$ is taken to be all measurable functions. 
        
        \item The width caused by the standard deviation of the estimation, $q_{1 - \alpha / 2} \times \sqrt{\mathrm{Var}[\hat{\theta}_{\mathrm{U}}]}$  and $q_{1 - \alpha / 2} \times \sqrt{\mathrm{Var}[\hat{\theta}_{\mathrm{L}}]}$, assuming that the asymptotic normality holds for both methods, as proved in Theorem~\ref{thm:est} in our paper and \citet[Theorem 3.4]{ji2023model}.
    \end{enumerate}

  The first part, the theoretical bounds, are close for both methods in a number of cases, as argued in Section~\ref{subsec:cs_bounds} and Proposition~\ref{prop:tight}. In the simulations in Section~\ref{subsec:simu:imbal}, because $Y \mid X$ and $Z \mid X$ are the same up to location and scale parameters, as shown in 
  Proposition~\ref{prop:tight}, our Cauchy--Schwarz are the same as the tight bounds, which match the bounds in \cite{ji2023model}. Note that by definition, the width of the Cauchy--Schwarz bound is $2\mathbb{E}[\sqrt{\mathrm{Var}[Y \mid X]}\sqrt{\mathrm{Var}[Z \mid X]}]$. In the simulations in Section~\ref{subsec:simu:imbal},$\sqrt{\mathrm{Var}[Y \mid X]} = \sigma_Y, \sqrt{\mathrm{Var}[Z \mid X]} = \sigma_Z$, hence the theoretical bounds for the two methods are exactly the same and are of order $O(\sigma_Y\sigma_Z)$.

  The second part, the bias of the estimation, is determined by the doubly robust nature of the methods.  In our method, due to double machine learning, the bias primarily consists of terms that are products of the estimation errors for \( Y \mid X \) and \( Z \mid X \). 
   Mathematically, the bias of our Cauchy--Schwarz bound can be calculated as follows,

  \begin{align*}
  &\mathbb{E}[\hat{\theta}_{\mathrm{U}}^{(\mathrm{CS})} - \theta_{\mathrm{U}}^{(\mathrm{CS})}] \\
  &= \mathbb{E}\left[\frac{1}{K} \sum_{k=1}^{K}  \hat{\theta}_\mathrm{U}^{(*, \mathrm{CS}, k)}- \theta_{\mathrm{U}}^{(\mathrm{CS})}\right] \\
  &= \frac{1}{K}\sum_{k=1}^{K}\mathbb{E}\left[ \hat{\theta}_\mathrm{U}^{(*, \mathrm{CS}, k)}- \theta_{\mathrm{U}}^{(\mathrm{CS})}\right] \\
  &= \frac{1}{K}\sum_{k=1}^{K}\mathbb{E}\left[ \hat{\theta}_\mathrm{U}^{(\mathrm{CS}, k)} + \frac{K}{n}\sum_{i \in I_k}  \hat{\varphi}^{(\mathrm{CS},k)}_\mathrm{U}(Y_i, Z_i , X_i, R_i)- \theta_{\mathrm{U}}^{(\mathrm{CS})}\right] \\
  &= \frac{1}{n}\sum_{k=1}^{K}\sum_{i \in I_k}\mathbb{E}\left[\frac{R_i}{\hat{e}^{(-k)}(X_i)}\hat\varphi^{(\mathrm{CS},k)}_{Y, X, \mathrm{U}}(Y_i, X_i)
		+ \frac{1 - R_i}{1 - \hat{e}^{(-k)}(X_i)}\hat{\varphi}^{(\mathrm{CS}, k)}_{Z, X, \mathrm{U}}(Z_i, X_i) 
		 + \hat{M}^{\left(\mathrm{CS}, k\right)}_\mathrm{U} \left(X_i \right) - \theta_{\mathrm{U}}^{(\mathrm{CS})}\right]\\
  &(R\perp\!\!\!\perp Y | X) \\
  & = \frac{1}{n}\sum_{k=1}^{K}\sum_{i \in I_k}\mathbb{E}\Big[\frac{\mathbb{E}[R_i \mid X_i]}{\hat{e}^{(-k)}(X_i)}\mathbb{E}[\hat\varphi^{(\mathrm{CS},k)}_{Y, X, \mathrm{U}}(Y_i, X_i) \mid X_i]
		+ \frac{\mathbb{E}[1 - R_i \mid X_i]}{1 - \hat{e}^{(-k)}(X_i)} \mathbb{E}[\hat{\varphi}^{(\mathrm{CS}, k)}_{Z, X, \mathrm{U}}(Z_i, X_i) \mid X_i]  \\
		 & \qquad + \hat{M}^{\left(\mathrm{CS}, k\right)}_\mathrm{U} \left(X_i \right) - \theta_{\mathrm{U}}^{(\mathrm{CS})}\Big] \\
   & = \frac{1}{n}\sum_{k=1}^{K}\sum_{i \in I_k}\mathbb{E}\Big[\frac{e(X_i)}{\hat{e}^{(-k)}(X_i)}\mathbb{E}[\hat\varphi^{(\mathrm{CS},k)}_{Y, X, \mathrm{U}}(Y_i, X_i) \mid X_i]
		+ \frac{1 - e(X_i)}{1 - \hat{e}^{(-k)}(X_i)} \mathbb{E}[\hat{\varphi}^{(\mathrm{CS}, k)}_{Z, X, \mathrm{U}}(Z_i, X_i) \mid X_i]  \\
		 & \qquad + \hat{m}_Y^{(-k)}(X_i)\hat{m}_Z^{(-k)}(X_i) + \sqrt{\hat{v}_Y^{(-k)}(X_i)}\sqrt{\hat{v}_Z^{(-k)}(X_i)} \\
         & \qquad - m_Y(X_i)m_Z(X_i) - \sqrt{v_Y(X_i)}\sqrt{v_Z(X_i)}\Big],   
  \end{align*} 
where in the last line, we can replace $\theta_U^{(\mathrm{CS})}$ with $m_Y(X_i)m_Z(X_i) - \sqrt{v_Y(X_i)}\sqrt{v_Z(X_i)}$ because by definition, $\mathbb{E}[m_Y(X_i)m_Z(X_i) - \sqrt{v_Y(X_i)}\sqrt{v_Z(X_i)}] = \theta_U^{(\mathrm{CS})}$.

To simplify the expression, consider the known propensity score setting assumed in most of \cite{ji2023model} and used in the simulation in Section~\ref{subsec:simu:imbal}, i.e., assume for all $k$ that $\hat{e}^{(-k)}(x)=e(x)$. Consequently

\begin{align} \label{equ:D2:1}
  \mathbb{E}[\hat{\theta}_{\mathrm{U}}^{(\mathrm{CS})} - \theta_{\mathrm{U}}^{(\mathrm{CS})}] &= \frac{1}{n}\sum_{k=1}^{K}\sum_{i \in I_k}\mathbb{E}\Big[\mathbb{E}[\hat\varphi^{(\mathrm{CS},k)}_{Y, X, \mathrm{U}}(Y_i, X_i) \mid X_i]
		+  \mathbb{E}[\hat{\varphi}^{(\mathrm{CS}, k)}_{Z, X, \mathrm{U}}(Z_i, X_i) \mid X_i] \notag \\
		 & \qquad + \hat{m}_Y^{(-k)}(X_i)\hat{m}_Z^{(-k)}(X_i) + \sqrt{\hat{v}_Y^{(-k)}(X_i)}\sqrt{\hat{v}_Z^{(-k)}(X_i)} \notag \\
		 & \qquad - m_Y(X_i)m_Z(X_i) - \sqrt{v_Y(X_i)}\sqrt{v_Z(X_i)}\Big].  
  \end{align} 

Note that
\begin{align} \label{equ:D2:2}
    \mathbb{E}[\hat\varphi^{(\mathrm{CS},k)}_{Y, X, \mathrm{U}}(Y_i, X_i) \mid X_i] &= \mathbb{E}\Big[(f(Y_i, X_i) - \hat{m}^{(-k)}_Y(X_i))\hat{m}^{(-k)}_Z(X_i) \notag \\
    &\qquad + \frac{1}{2} ((f(Y_i, X_i) - \hat{m}^{(-k)}_Y(X_i))^2 - \hat{v}^{(-k)}_Y(X_i))\sqrt{\frac{\hat{v}^{(-k)}_Z(X_i)}{\hat{v}^{(-k)}_Y(X_i)}} \mid X_i \Big] \notag\\
    &=(m_Y(X_i) - \hat{m}^{(-k)}_Y(X_i))\hat{m}^{(-k)}_Z(X_i) \notag \\
    &\qquad + \frac{1}{2} \bigl(v_Y(X_i) - \hat{v}^{(-k)}_Y + (m_Y(X_i) - \hat{m}^{(-k)}_Y(X_i))^2\bigr)\sqrt{\frac{\hat{v}^{(-k)}_Z(X_i)}{\hat{v}^{(-k)}_Y(X_i)}},
\end{align}
and an analogous result holds for $\mathbb{E}[\hat\varphi^{(\mathrm{CS},k)}_{Z, X, \mathrm{U}}(Y_i, X_i) \mid X_i]$.
Plugging 
these expressions into \eqref{equ:D2:1}, we have
\begin{align} 
  &\mathbb{E}[\hat{\theta}^{(\mathrm{CS})}_{\mathrm{U}} - \theta_{\mathrm{U}}^{(\mathrm{CS})}] = \frac{1}{n}\sum_{k=1}^{K}\sum_{i \in I_k}\mathbb{E}\left[-(m_Y(X_i) - \hat{m}^{(-k)}_Y(X_i))(m_Z(X_i) - \hat{m}^{(-k)}_Z(X_i))\right] \label{equ:D2:4:1}\\
   &\qquad + \frac{1}{n}\sum_{k=1}^{K}\sum_{i \in I_k}\frac{1}{2}\mathbb{E}\left[\frac{v_Y(X_i)- \hat{v}^{(-k)}_Y(X_i)}{\sqrt{\hat{v}^{(-k)}_Y(X_i)}} \sqrt{\hat{v}^{(-k)}_Z(X_i)} + \frac{v_Z(X_i)- \hat{v}^{(-k)}_Z(X_i)}{\sqrt{\hat{v}^{(-k)}_Z(X_i)}} \sqrt{\hat{v}^{(-k)}_Y(X_i)}\right] \label{equ:D2:4:2}\\
   &\qquad -  \frac{1}{n}\sum_{k=1}^{K}\sum_{i \in I_k}\frac{1}{2}\mathbb{E}\left[\frac{(m_Y(X_i) - \hat{m}^{(-k)}_Y(X_i))^2}{\sqrt{\hat{v}^{(-k)}_Y(X_i)}} \sqrt{\hat{v}^{(-k)}_Z(X_i)} + \frac{(m_Z(X_i) - \hat{m}^{(-k)}_Z(X_i))^2}{\sqrt{\hat{v}^{(-k)}_Z(X_i)}} \sqrt{\hat{v}^{(-k)}_Y(X_i)}\right] \label{equ:D2:4:3}
  \end{align} 

 
In the simulation in Section~\ref{subsec:simu:imbal}, we adopt the cross-validated ridge regression. According to the corresponding asymptotic theory of it (see e.g., \cite{liu2019ridge}), when $\sigma_Y, \sigma_Z \le O(\sqrt{\frac{n}{p}})$, we have $m_Y(X_i) - \hat{m}^{(-k)}_Y(X_i) = O_P(\sqrt{\frac{p}{n}}\sigma_Y(1 + \sqrt{\frac{p}{n}}\sigma_Y)) = O_P(\sqrt{\frac{p}{n}}\sigma_Y)$, and similarly $m_Z(X_i) - \hat{m}^{(-k)}_Z(X_i) = O_P(\sqrt{\frac{p}{n}}\sigma_Z)$; we have $\mathbb{E}[v_Y(X_i)]- \hat{v}^{(-k)}_Y(X_i) = O_P(\sqrt{\frac{p}{n}}\sigma_Y^2(1 + O(1)) = O_P(\sqrt{\frac{p}{n}}\sigma_Y^2)$, and similarly $\mathbb{E}[v_Z(X_i)]- \hat{v}^{(-k)}_Z(X_i)  = O_P(\sqrt{\frac{p}{n}}\sigma_Z^2)$. Therefore, \eqref{equ:D2:4:1}, \eqref{equ:D2:4:2}, and \eqref{equ:D2:4:3} are all of order $O(\frac{p}{n}\sigma_Y\sigma_Z)$. In all, the bias of our Cauchy--Schwarz upper bound is $O(\frac{p}{n}\sigma_Y\sigma_Z)$, which is bilinear in $\sigma_Y,\sigma_Z$.

  In contrast, as explained in Appendix~\ref{appendix:ji:intro}, under certain conditions (e.g., the space of $Y, Z$ is finite), \cite{ji2023model} bounds the bias of Dualbounds by a quantity of the same order as the \emph{squared} estimation error of $Y, Z\mid X$. These conditions do not hold in our simulations and we were unable to apply the theoretical bias bounds in \cite{ji2023model} to our simulations. Nevertheless, if we naively extrapolate such results to our simulations to bound the bias of Dualbounds by the squared estimation error, we would get a bound of $O(\frac{p}{n}(\sigma_Y ^ 2+ \sigma_Z^2))$, which could explain the nonlinearity of the width curves in Section~\ref{subsec:simu:imbal} and Appendix~\ref{appendix:simu}. 

  The third part, the standard deviation of the estimation, can be evaluated via the asymptotic distribution of the estimators. For our method, according to Theorem~\ref{thm:var}, the asymptotic variance of the estimators is of order $O\left(\frac{1}{n}\left[\mathbb{E}\left[\left[\varphi^{(\mathrm{CS})}_{Y, X, \mathrm{U}}(Y, X)\right]^2\right] +  \mathbb{E}\left[\left[\varphi^{(\mathrm{CS})}_{Z, X, \mathrm{U}}(Z, X)\right]^2\right]+  \mathbb{E}\left[\left[\varphi^{(\mathrm{CS})}_{Y, X, \mathrm{L}}(Y, X)\right]^2\right]+  \mathbb{E}\left[\left[ \varphi^{(\mathrm{CS})}_{Z, X, \mathrm{L}}(Z, X)\right]^2\right]\right]\right)$ $+O\left(\frac{1}{n}\mathbb{E}\left[[M_{\mathrm{U}}(X) - \theta_{\mathrm{U}}]^2\right]\right)$. Without the loss of generality, we can focus the second moment of $\varphi^{(\mathrm{CS})}_{Y, X, \mathrm{U}}(Y, X)$ and $M_{\mathrm{U}}(X) - \theta_{\mathrm{U}}$. The second moment of other influence functions can be evaluated similarly.

Firstly, for $\varphi^{(\mathrm{CS})}_{Y, X, \mathrm{U}}(Y, X)$, we have
  \begin{align} \label{equ:compare:var}
  \mathbb{E}[\varphi^{(\mathrm{CS})}_{Y, X, \mathrm{U}}(Y, X)]^2 
  &\le 2  \mathbb{E}\left[\left[ \left(f\left(Y, X\right) - m_Y(X)\right)m_Z(X)\right]^2\right] 
  \notag \\ &\quad 
  +2\mathbb{E}\left[\left[ \frac{1}{2}\left[ \left(f\left(Y, X\right) - m_Y(X)\right)^2 - v_Y(X) \right]\notag   \sqrt{\frac{v_Z(X)}{v_Y(X)}}\right]^2\right]  \notag \\
  &= 2 \mathbb{E}\left[ v_Y(X)m_Z(X)^2\right] + \frac{1}{2}\mathbb{E}\left[v_Z(X)\frac{\mathbb{E}\left[ \left(f\left(Y, X\right) - m_Y(X)\right) \mid X\right]^4 - v_Y(X)^2 }{v_Y(X)}\right] 
  \end{align}

  In Section~\ref{subsec:simu:imbal}, $v_Y(x) = \sigma_Y^2$, $v_Z(x) = \sigma_Z^2$, $\mathbb{E}[m_Z(X)^2] = \mathbb{E}[(\beta_Z^TX)^2] = 1$ by design, and $\mathbb{E}\left[ \left(f\left(Y, X\right) - m_Y(X)\right) \mid X\right]^4 = \mathbb{E}[\epsilon_Y^4] = O(\sigma_Y^4)$. Plugging these into  \eqref{equ:compare:var}, we can derive that the second moment of $\varphi^{(\mathrm{CS})}_{Y, X, \mathrm{U}}(Y, X) = O(\sigma_Y^2 + \sigma_Y\sigma_Z)=O(\sigma_Y^2+\sigma_Z^2)$. The second moment of other influence functions can also be derived similarly, and are also $O(\sigma_Y^2+\sigma_Z^2)$.

  Secondly, for $M_{\mathrm{U}}(X) - \theta_{\mathrm{U}}$, we have
  \begin{align*}
      &\mathbb{E}\left[[M_{\mathrm{U}}(X) - \theta_{\mathrm{U}}]^2\right] \\
      &= \mathbb{E}\left[\left[m_Y(X)m_Z(X) + \sqrt{v_Y(X)}\sqrt{v_Z(X)} - \mathbb{E}\left[m_Y(X)m_Z(X) + \sqrt{v_Y(X)}\sqrt{v_Z(X)}\right]\right]^2\right] \\
      &\le 2\mathbb{E}\left[\left[m_Y(X)m_Z(X)   - \mathbb{E}\left[m_Y(X)m_Z(X) \right]\right]^2\right] + 2\mathbb{E}\left[\left[ \sqrt{v_Y(X)}\sqrt{v_Z(X)} - \mathbb{E}\left[\sqrt{v_Y(X)}\sqrt{v_Z(X)}\right]\right]^2\right]
  \end{align*}
  In Section~\ref{subsec:simu:imbal}, $$\mathbb{E}\left[\left[m_Y(X)m_Z(X)   - \mathbb{E}\left[m_Y(X)m_Z(X) \right]\right]^2\right] = \mathbb{E}\left[\left[(\beta_Y^TX)(\beta_Z^TX) - \mathbb{E}\left[(\beta_Y^TX)(\beta_Z^TX) \right]\right]^2\right] = 1, $$
  and $\mathbb{E}\left[\left[ \sqrt{v_Y(X)}\sqrt{v_Z(X)} - \mathbb{E}\left[\sqrt{v_Y(X)}\sqrt{v_Z(X)}\right]\right]^2\right] = 0$ since $v_Y(X), v_Z(X)$ are constant functions. Therefore, the second moment of $M_{\mathrm{U}}(X) - \theta_{\mathrm{U}} = 1$.

  In all, in Section~\ref{subsec:simu:imbal}, the width of our confidence bound is linear with respect to $\sigma_Y$, while that of Dualbounds may be quadratic due to (at least) the bias, which matches the results  in Section~\ref{subsec:simu:imbal} illustrated by Figure~\ref{fig:ols_imbal}:  as $\sigma_Y / \sigma_Z$ increases, the width of our confidence bounds grows linearly, while the width of Dualbounds appears to grow superlinearly. 

  \subsection{Other comparisons} 
  \label{appendix:ji_other}

  Besides the comparison on the impact of disparate noise of the datasets on the performance of the algorithms, below are several comparisons of the methods in some other aspects.

\begin{enumerate}
    \item The primary focus of \cite{ji2023model} is on randomized trials within causal inference, whereas our framework addresses the data fusion problem, which inherently involves observational data. Although \cite{ji2023model} touches upon observational studies in Section~{3.4}, they do not provide results regarding tightness and cross-fitting. This is understandable as randomized settings are common in causal inference, but in the context of data fusion, where individuals in the datasets are typically not randomly assigned, it is crucial to fully address the challenges posed by observational data. Additionally, although when addressing randomized experiments, \cite{ji2023model} only requires mild conditions for validity, in the case of observational studies, \cite{ji2023model} need similar assumptions as our work (see, e.g., \citet[Theorem 3.5]{ji2023model}).

    \item As mentioned in Section~\ref{sec:ji}, \cite{ji2023model} considers a general estimand \( \theta = \mathbb{E}[h(Y, Z, X)] \) where \( Y \) and \( Z \) are one-dimensional. 
    We provide here a bit more detail about why \cite{ji2023model} may face challenges in generalizing to multi-dimensional cases. In their Section~{4.2}, they propose discretizing the space in which \( Y \) and \( Z \) lie, followed by solving a linear programming problem whose parameters increase with the number of fragments in the discretized space. Consequently, as the dimensionality of \( Y \) and \( Z \) increases, the number of fragments in the discretized space rises exponentially, and the computational cost grows rapidly. Although they mention using basis functions in Section~{4.1} to address the curse of dimensionality, this approach is theoretically and computationally challenging, and it is not adopted in their empirical studies. 
\end{enumerate}

\section{The Singular Behavior of Estimand with Square Root: An Example} \label{appendix:ex4}

Suppose that 
$X, X_1, X_2, ..., X_n \overset{iid}{\sim}\mathcal{N}(\theta^2, 1)$, and the estimand of interest is $\theta \ge 0$. The MLE estimator of $\theta$ is 
$$\hat{\theta} = \sqrt{\max\{\bar{X}, 0\}},$$ 
where the sample mean $\bar{X} = \sum_{i=1}^{n}X_i / n$. 

Notice that $\bar{X} \sim \theta^2 + X / \sqrt{n}$, hence $$\hat{\theta} \sim n^{-0.25} \sqrt{\max\{\sqrt{n}\theta^2 + X, 0\}}.$$ 
If $\theta = o(n^{-0.25})$, $\hat{\theta}$ will have a $n^{-0.25}$ rate, which is significantly smaller than the parametric rate. This toy example indicates why positivity assumption can be necessary when the estimand contains the square root of a functional of the probability distribution.

\section{Supplementary Results for Empirical Study} 

\subsection{Linear Model with Gaussian Tail} \label{appendix:simu}
In this subsection, we consider a similar simulation setting as in Section~\ref{subsec:simu:imbal}, where the only difference is that $\epsilon_Y \mid X, \epsilon_Z \mid X \sim \mathcal{N}(0, 1)$, so that the tail of $f(Y, X)\mid X$ and $g(Z, X)\mid X$ becomes lighter. The corresponding results, coverage of the two algorithms, and the width of the corresponding 95\% confidence bounds are exhibited in Figure~\ref{fig:ols_apendix}. Because \cite{ji2023model} recommends using Gaussian models for estimating the conditional distribution of $f(Y, X)\mid X$ and $g(Z, X)\mid X$, adopting a Gaussian linear model makes their recommended model perfectly specified. However, even in this case, we can observe from Figure~\ref{fig:ols_apendix} that our algorithm still outperforms theirs. This demonstrates that our method's outperformance of Dualbounds' is not caused by specific noise distribution or restricted to the heavy-tailed scenario but rendered by the intrinsic merit that our method can remain efficient when the sample size or the noise level of the two datasets is imbalanced, while Dualbounds seems to suffer in this situation. It is also notable that our method's runtime is approximately 0.00613 seconds, compared to 3.72 seconds for Dualbounds using the same model, making our algorithm nearly 600 times faster.
\begin{figure}[t]
	\centering
	\includegraphics[width=\textwidth]{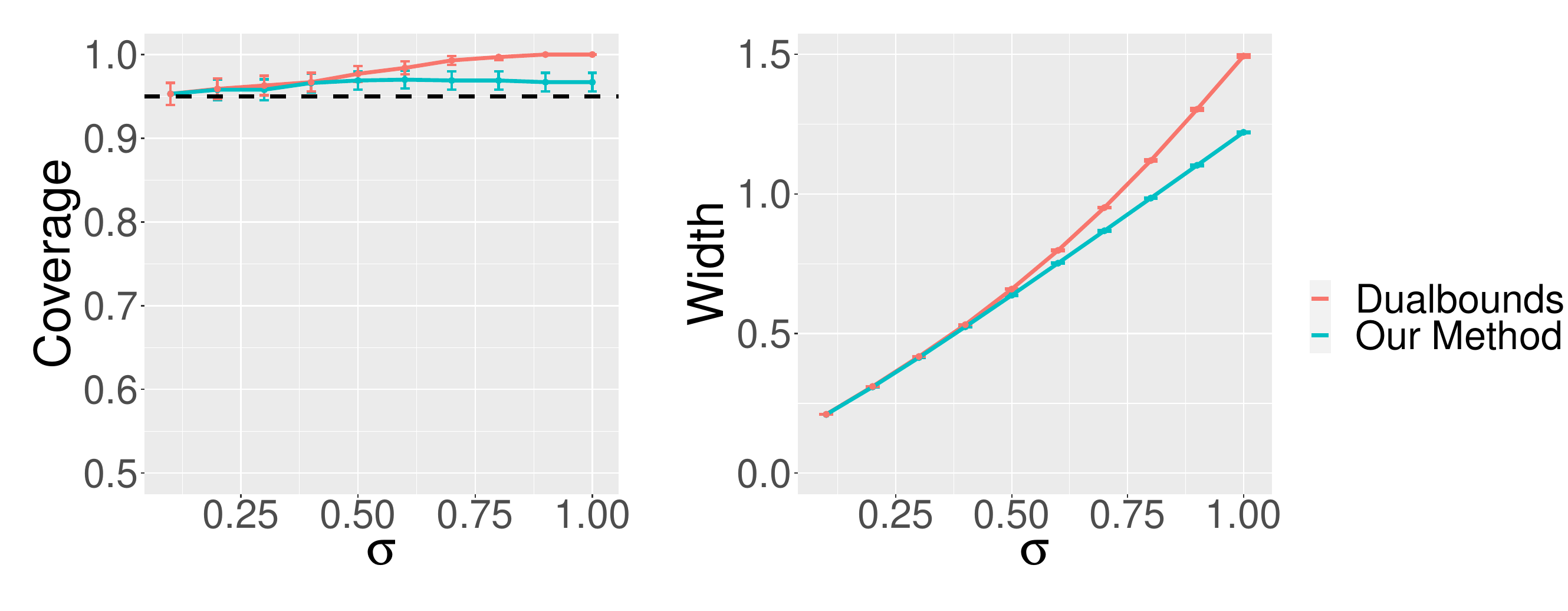}
	\caption{
    Influence of the noise level $\sigma_Y$ on our method's and Dualbounds' coverage (left) and width (right) for 95\% confidence intervals in the simulation of Appendix~\ref{appendix:ex4}. Error bars represent $\pm 1.96$ Monte Carlo standard errors. 
    }
	\label{fig:ols_apendix}
\end{figure}

\subsection{Data fusion and Validation Study} \label{appendix:val}
In this subsection, we consider the application of data fusion in epidemiology studies
where the estimand of interest is usually determined by several expensive medical attributes $Y$ and $Z$ that are too costly or not possible to measure jointly.
To address this problem, researchers often conduct a large-scale study called the main study, with only part of the expensive attributes, $Y$, being accurately measured through a state-of-the-art approach (i.e. \emph{gold standard}), and the rest, $Z$, being replaced with a less expensive but possibly lower quality alternative $X$. 
For example, researchers may be interested in the relationship between cancer and the infection history of certain viruses. Ideally, one hopes to directly study the correlation between the gold standard, the cancer precursor conditions $Y$, and the precise medical record $Z$. However, both $Y$ and $Z$ consume enormous medical resources, so in large-scale studies, one will only reserve $Y$ and substitute $Z$ by the self-reported health record $X$, which is often inaccurate.

 We illustrate our method in a simulated validation study through the following experiment. 
 Suppose that the gold standard $Z = (Z_1, Z_2) \in \mathbb{R}^2$ and its alternative $X = (X_1, X_2) \in \mathbb{R}^2$.
 Let $R$ be the indicator that the data comes from the main study. We consider a linear model as follows: 

$$(Z_1, Z_2) \sim \mathcal{N}\left(0, \begin{pmatrix} 1 & \rho \\ \rho & 1 \end{pmatrix} \right);$$

$${(X_1, X_2)}\mid {(Z_1, Z_2)} \sim \mathcal{N} \left({(Z_1, Z_2)}, \sigma^2 \begin{pmatrix} 1 & \tau \\ \tau & 1 \end{pmatrix} \right);$$

$${Y}\mid ({Z_1}, {Z_2}, {X_1}, {X_2}) \sim \mathcal{N} \left(\beta_1 {Z_1} + \beta_2 {Z_2}, \sigma^2_\epsilon \right);$$

$$R \mid X \sim \mathrm{Bern}(0.5).$$

Because $Z_1, Z_2$ are the gold standard, in the true model, all the impact of $X_1, X_2$ on $Y$ is blocked by $Z_1, Z_2$, but we do not assume that this information is known in the inference process. The target estimator is the regression parameter $\beta_1$. Notice that $\beta_1 = \mathbb{E}\left[{Y}{\frac{(Z_1 - \rho Z_2)}{1 - \rho^2}}\right] $. Since $Y$, $Z_1$, and $Z_2$ are not jointly observable, $\beta_1$ can only be partially identified.  Due to the linearity of Gaussian distribution, the tight bounds and Cauchy--Schwarz bounds coincide. 

We choose the sample size $n = 2000$, correlation parameter $\tau = 0.3$, error scale $\sigma_\epsilon = 0.5$, and signal strength $\beta_1 = \beta_2 = 1$. We perform 500 replications for our experiment, and adopt standard linear regression 
as our machine learning algorithm. 

As the noise parameter $\sigma$ and correlation parameter $\rho$ varies, the signal strength in model $Z\mid X$ changes, and correspondingly, the coverages of the  LCB and UCB (the probabilities that LCB/UCB is less than the theoretical value of the Cauchy--Schwarz lower/upper bound, respsctively) and the widths of the confidence intervals are illustrated in Figures~\ref{fig:ols} and \ref{fig:ols_width}. Figure~\ref{fig:violin} exhibits the violin plots of the LCB and UCB. The figures show that the Cauchy--Schwarz bounds provide a meaningful and informative identification region that is bounded away from zero when  $\sigma$ is moderate. The coverage of our inference matches the nominal level. 
Also, Figures~\ref{fig:ols_width} and \ref{fig:violin} show that the estimation error of our bounds is small compared to the width of the theoretical identification region, indicating the efficiency of our algorithm.

\begin{figure}[h]
	\centering
	\includegraphics[width=\textwidth]{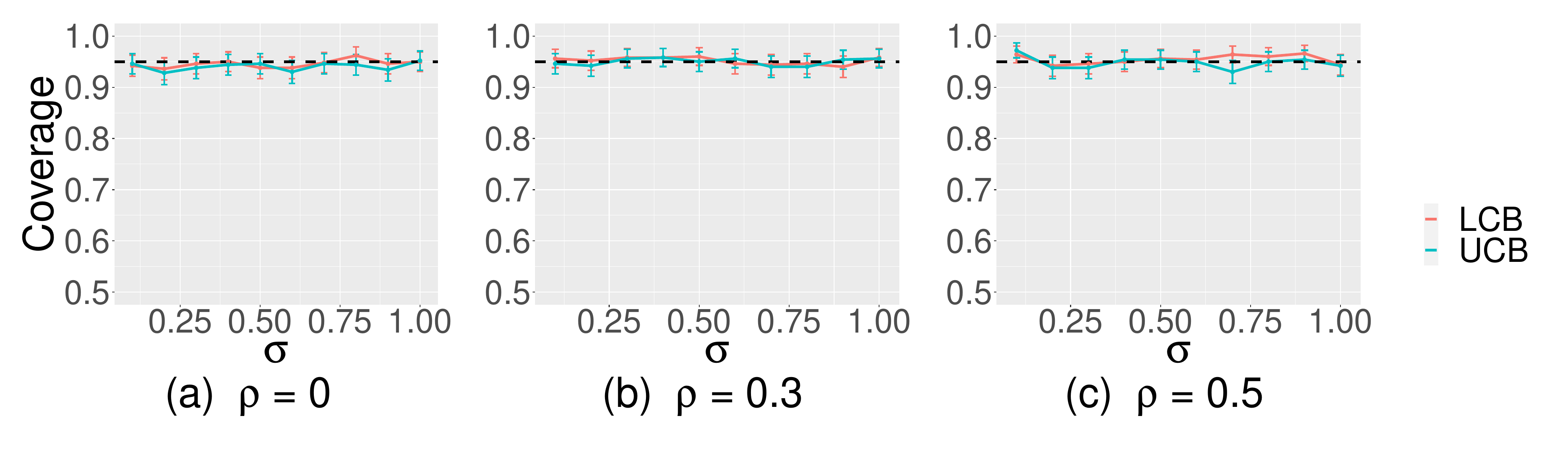}
	\caption{Influence of the noise level $\sigma$ and the correlation parameter $\rho$ on coverage in the simulation of Appendix~\ref{appendix:val}. Error bars represent $\pm 1.96$ Monte Carlo standard errors. 
    }
	\label{fig:ols}
\end{figure}

\begin{figure}[h]
	\centering
	\includegraphics[width=\textwidth]{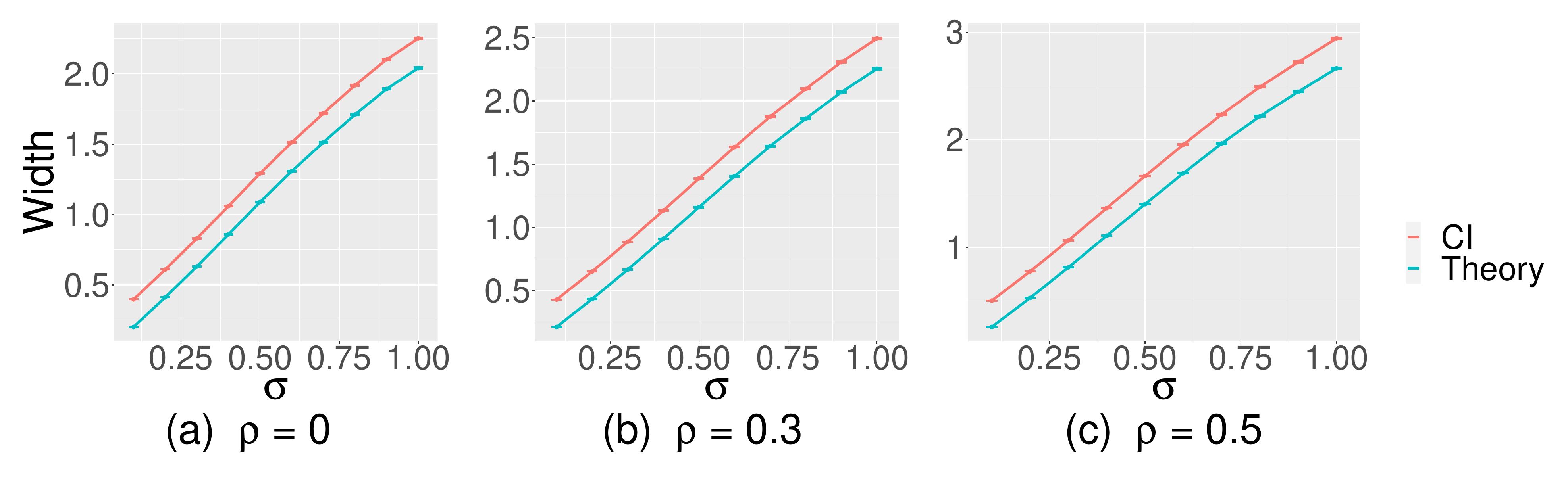}
	\caption{Influence of the noise level $\sigma$ and the correlation parameter $\rho$ on width.``Theory" stands for the difference of the upper and lower bounds.``CI" stands for the difference of the upper confidence bound for the upper bound and the lower confidence bound for the lower bound. 
    Error bars represent $\pm 1.96$ Monte Carlo standard errors.
    }
	\label{fig:ols_width}
\end{figure}

\begin{figure}[h]
	\centering
	\includegraphics[width=0.9\textwidth]{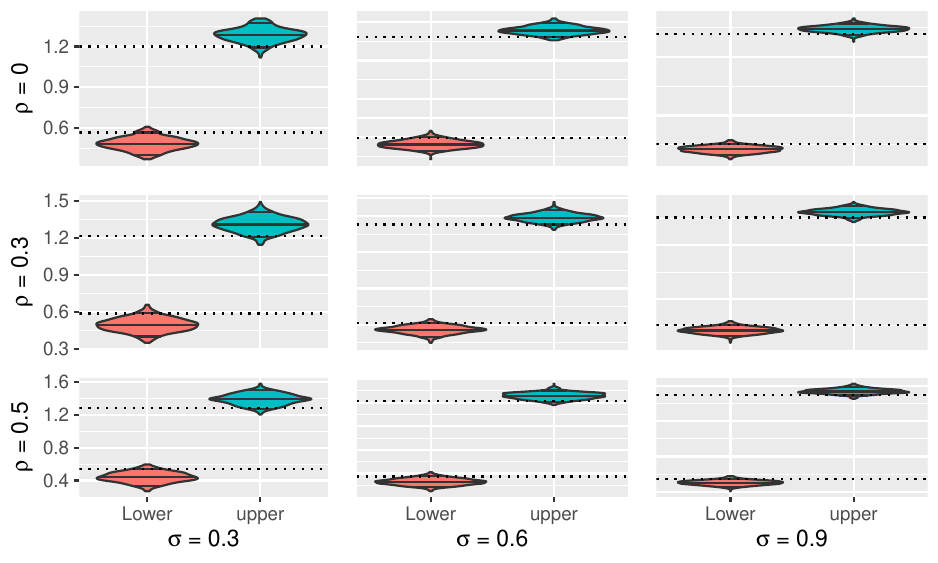}
	\caption{The violin plots of the LCB and UCB with varying $\rho$ and $\sigma$, where the $0.05, 0.50, 0.95$ quantiles of the LCB and UCB are illustrated as solid lines in the plot, and the dashed line stands for the theoretical value of the Cauchy--Schwarz lower and upper bounds.}
	\label{fig:violin}
\end{figure}

\subsection{Sensitivity Analysis of our Real Data Analysis}\label{appendix:real}

In Section~\ref{subsec:real}, when estimating the conditional mean and variances, we do not include the high-order terms in our machine learning algorithms. In \cite{evans2018doubly}, the authors include quadratic terms in their model for identifiability and also better model-fitting. In order to compare with it,  we show our results on 95\% confidence identification region when including the quadratic terms as follows:
\begin{table}[h]
	\centering
	\begin{tabular}{|c|c|}
		\hline
		Linear  & Random Forest  \\
		\hline
		[0.29, 1.44]       &  [0.54, 1.59] \\
		\hline
	\end{tabular}
\end{table}

As mentioned in \cite{evans2018doubly}, in their model, the corresponding inverse probability weighting (IPW) estimator and doubly robust (DR) estimator derive very different results, with IPW suggesting ``a negative association between household net worth and total expenditure", and DR implying a positive one. On the other hand, the results of the DR estimator are close to that of imputation-based methods. This implies that inverse probability weighting may be less valid in this study. In fact, we observe that when using simple logistic regression (as adopted in \cite{evans2018doubly}), the propensity score estimation of missingness in the study can be extremely close to 0 or 1, resulting in the violation of the positivity assumption and an estimator with extremely large variance. To address this issue, besides the study in the main paper, we also apply another version of our method with truncation on the propensity score. Precisely, we clip the estimated propensity score to $[0.05, 0.95]$. With this modification,  our results on 95\% confidence identification region are as follows:  

\begin{table}[h]
	\centering
	\begin{tabular}{|c|c|c|c|}
		\hline
		Linear  &  Linear (with quadratic) & Random Forest  & Random Forest (with quadratic) \\
		\hline
		[0.31, 1.09] & [0.34, 1.30]       &  [0.54, 1.05] & [0.55, 1.51] \\ 
		\hline
	\end{tabular}
\end{table}

The above results show that the existence of quadratic terms or truncation modification does not significantly influence our partial identification result or alter the conclusion about the association between household net worth and total expenditure. This demonstrates our method's robustness to various machine learning algorithms and approaches. 

\section{Function of Partially Identifiable Estimands} \label{appendix:delta}
In the main paper, we assume that the target parameter is of the form $\theta = \mathbb{E}[h(Y, Z, X)]$. In practice, this may not be satisfied. For instance, the OLS parameter in Section~\ref{subsec:real} can not be directly expressed in this form. 
Consider the OLS parameter $\theta = e_{p_X + p_Z}^T \mathbb{E}[\tilde{X}\tilde{X}^T]^{-1} \mathbb{E}[\tilde{X}Y]$, where $e_{p_X + p_Z}$ stands for the $p_X + p_Z$ dimensional column vector whose entries are all 0 except the last entry being 1, and $\tilde{X} = (X^T Z^T)^T$. This estimand fails to fit into our theory directly but can be addressed by slightly generalizing our method. Denote $\theta^{(XZ)} = \mathbb{E}[\tilde{X}\tilde{X}^T]$, $\theta^{(XY)} = \mathbb{E}[XY]$, $\theta^{(YZ)} = \mathbb{E}[YZ]$, function $v(A) = e_{p_X + p_Z}^T A^{-1}$ for matrix $A$, then we can rewrite $\theta$ as
$$ \theta = v(\theta^{(XZ)}) \cdot
\begin{pmatrix}
    \theta^{(XY)} \\ \theta^{(YZ)}
\end{pmatrix} $$

Note that $\theta^{(XZ)}$ and $\theta^{(XY)}$ are identifiable; furthermore, $\theta^{(YZ)} = \mathbb{E}[YZ]$ is an expectation of a decomposable function and can be partially identified between the Cauchy--Schwarz  bounds $\theta^{(YZ)}_{\mathrm{L}}$ and $\theta^{(YZ)}_{\mathrm{U}}$. Consequently, we have the Cauchy--Schwarz  bounds for $\theta$, $\theta^{(\mathrm{CS})}_{\mathrm{L}}$ and $\theta^{(\mathrm{CS})}_{\mathrm{U}}$, defined as follows,

\begin{align*}
    \theta^{(\mathrm{CS})}_{\mathrm{L}} &= s_{\mathrm{L}}(\theta^{(XZ)}, \theta^{(XY)}, \theta^{(YZ)}_{\mathrm{L}}, \theta^{(YZ)}_{\mathrm{U}}) &\overset{\mathrm{def}}{=} v(\theta^{(XZ)}) \cdot
\begin{pmatrix}
    \theta^{(XY)} \\ \theta^{(YZ)}_{\mathrm{L}}
\end{pmatrix} - \mathrm{ReLU}(- v(\theta^{(XZ)})) \cdot
\begin{pmatrix}
   0 \\ \theta^{(YZ)}_{\mathrm{U}} - \theta^{(YZ)}_{\mathrm{L}}
\end{pmatrix}, \\
\theta^{(\mathrm{CS})}_{\mathrm{U}}  &= s_{\mathrm{U}}(\theta^{(XZ)}, \theta^{(XY)}, \theta^{(YZ)}_{\mathrm{L}}, \theta^{(YZ)}_{\mathrm{U}}) &\overset{\mathrm{def}}{=} v(\theta^{(XZ)}) \cdot
\begin{pmatrix}
    \theta^{(XY)} \\ \theta^{(YZ)}_{\mathrm{L}}
\end{pmatrix} + \mathrm{ReLU}(v(\theta^{(XZ)})) \cdot
\begin{pmatrix}
   0 \\ \theta^{(YZ)}_{\mathrm{U}} - \theta^{(YZ)}_{\mathrm{L}}
\end{pmatrix},
\end{align*}
where $\mathrm{ReLU}(\alpha) = (\alpha_1I_{\alpha_1 > 0}, \ldots, \alpha_{p_X + p_Z}I_{\alpha_{p_X + p_Z} > 0})$ is the entrywise rectified linear unit.

With the above closed-form expression, one can adopt the plug-in estimator with the following steps. 

\begin{enumerate}
    \item Estimate $ \theta^{(YZ)}_{\mathrm{L}}$ and $ \theta^{(YZ)}_{\mathrm{U}}$ with Algorithm~\ref{alg:est} to obtain estimators $ \hat{\theta}^{(YZ)}_{\mathrm{L}}$ and $ \hat{\theta}^{(YZ)}_{\mathrm{U}}$. 
    \item Note that $\theta^{(XZ)}$ and $ \theta^{(XY)}$ fit within our framework with the lower and upper Cauchy--Schwarz bounds the same and both equal to the estimand (i.e. $\theta = \theta_{\mathrm{L}}^{(\mathrm{CS})} = \theta_{\mathrm{U}}^{(\mathrm{CS})}$), so they can also be estimated via Algorithm~\ref{alg:est} with all the conditional variance terms ($\hat{v}^{(-k)}_Y$ and $\hat{v}^{(-k)}_Z$) removed. Denote the corresponding estimators to be $\hat{\theta}^{(XZ)}$ and $ \hat{\theta}^{(XY)}$.
    \item Calculate the plug-in estimator, $$\hat{\theta}^{(\mathrm{CS})}_{\mathrm{L}} = s_{\mathrm{L}}(\hat{\theta}^{(XZ)}, \hat{\theta}^{(XY)}, \hat{\theta}^{(YZ)}_{\mathrm{L}}, \hat{\theta}^{(YZ)}_{\mathrm{U}}), \qquad \hat{\theta}^{(\mathrm{CS})}_{\mathrm{U}} = s_{\mathrm{U}}(\hat{\theta}^{(XZ)}, \hat{\theta}^{(XY)}, \hat{\theta}^{(YZ)}_{\mathrm{L}}, \hat{\theta}^{(YZ)}_{\mathrm{U}}).$$
\end{enumerate}

The asymptotic distributions of $\hat{\theta}^{(\mathrm{CS})}_{\mathrm{L}}$ and $\hat{\theta}^{(\mathrm{CS})}_{\mathrm{U}}$ can be found via the delta method. The mathematical formulation is as follows.

Similar to the asymptotic normality result in Theorem~\ref{thm:est}, we have

$$ \sqrt{n} \begin{pmatrix}
    \hat{\theta}^{(XZ)} - \theta^{(XZ)} \\
    \hat{\theta}^{(XY)} - \theta^{(XY)} \\
    \hat{\theta}^{(YZ)}_{\mathrm{L}} - \theta^{(YZ)}_{\mathrm{L}} \\
    \hat{\theta}^{(YZ)}_{\mathrm{U}} - \theta^{(YZ)}_{\mathrm{U}} 
\end{pmatrix} \overset{\mathrm{d}}{\to} \mathcal{N}(0, V),$$
where $V$ is the covariance matrix of the influence functions of $\theta^{(XZ)}, \theta^{(XY)}, \theta^{(YZ)}_{\mathrm{L}}, \theta^{(YZ)}_{\mathrm{U}}$ and can be estimated similarly as $\hat{V}_{\mathrm{L}}$ and $\hat{V}_{\mathrm{U}}$ in Algorithm~\ref{alg:est}. Denote the estimator of $V$ by $\widehat{V}$.

Therefore, by the delta method\footnote{Rigorously, in order to use delta method, which relies on the intermediate point theorem, we need the entries of $(\theta^{(XZ)})^{-1}$ are non-zero to guarantee that $s_{\mathrm{L}}$ and $s_{\mathrm{U}}$ are differentiable at $(\theta^{(XZ)}, \theta^{(XY)}, \theta^{(YZ)}_{\mathrm{L}}, \theta^{(YZ)}_{\mathrm{U}})$. This is always satisfied when $Z$ is 1-dimensional, which is the case in  in Section~\ref{subsec:real}.}, we have,
 $$ \sqrt{n} (\hat{\theta}^{(\mathrm{CS})}_{\mathrm{L}} - \theta^{(\mathrm{CS})}_{\mathrm{L}}) \overset{\mathrm{d}}{\to} \mathcal{N}\left(0, \nabla^T s_{\mathrm{L}} \cdot V \cdot \nabla s_{\mathrm{L}}\right),$$
$$ \sqrt{n} (\hat{\theta}^{(\mathrm{CS})}_{\mathrm{U}} - \theta^{(\mathrm{CS})}_{\mathrm{U}}) \overset{\mathrm{d}}{\to} \mathcal{N}\left(0, \nabla^T s_{\mathrm{U}} \cdot V \cdot \nabla s_{\mathrm{U}}\right).$$

To conclude, we can calculate the $1 - \alpha$ lower confidence bound (LCB), $\hat{\theta}_{\mathrm{LCB}}^{(\mathrm{CS})}$, and the $1 - \alpha$ upper confidence bound (UCB), $\hat{\theta}_{\mathrm{UCB}}^{(\mathrm{CS})}$  of the estimated $\theta$:
      $$\hat{\theta}_{\mathrm{LCB}}^{(\mathrm{CS})} \overset{\mathrm{def}}{=} \hat{\theta}_{\mathrm{L}}^{(\mathrm{CS})} - q_{1 -  \alpha / 2}\sqrt{\nabla^T s_{\mathrm{L}} \cdot \widehat{V} \cdot \nabla s_{\mathrm{L}}}, \qquad  \hat{\theta}_{\mathrm{UCB}}^{(\mathrm{CS})} \overset{\mathrm{def}}{=} \hat{\theta}_{\mathrm{U}}^{(\mathrm{CS})} + q_{1 -  \alpha / 2}\sqrt{\nabla^T s_{\mathrm{U}} \cdot \widehat{V} \cdot \nabla s_{\mathrm{U}}},$$
      where $q_{1 - \alpha / 2}$ is the $1 -  \alpha / 2$ quantile of the standard Gaussian distribution $\mathcal{N}(0, 1)$. Similar to Theorem~\ref{thm:var}, we have 
$$\liminf_{n \to \infty}\mathbb{P}\left(\theta \in [\hat{\theta}_{\mathrm{LCB}}^{(\mathrm{CS})}, \hat{\theta}_{\mathrm{UCB}}^{(\mathrm{CS})}]\right) \ge 1 - \alpha.$$



\section{Proof of Lemma~\ref{lemma:cs}} \label{appendix:cs}
We focus on the upper bound $\theta_\mathrm{U}^{(2)}$. The statement for the lower bound $\theta_\mathrm{L}^{(2)}$ can be proved similarly (i.e., by considering the upper bound of $-\theta$).

Denote 
$$\epsilon_Y = \Bigl(\mathrm{Var}[f(Y, X)\mid X]\Bigr)^{-0.5}\Bigl(f(Y, X) - \mathbb{E}[f(Y, X)\mid X]\Bigr),$$
$$\epsilon_Z = \Bigl(\mathrm{Var}[g(Z, X)\mid X]\Bigr)^{-0.5}\Bigl(g(Z, X) - \mathbb{E}[g(Z, X)\mid X]\Bigr).$$

Then we have $\mathbb{E}[\epsilon_Y\mid X] = \mathbb{E}[\epsilon_Z\mid X] = 0$, $\mathrm{Var}[\epsilon_Y\mid X] = \mathrm{Var}[\epsilon_Z\mid X] = I_{p_f}$.

Notice that 
\begin{align}
    \theta &= \mathbb{E}[h(Y, Z, X)] \notag \\
    &= \mathbb{E}[f(Y, X)^Tg(Z, X)] \notag \\
    &= \mathbb{E}\left[\left( \mathbb{E}[f(Y, X)\mid X] +  \sqrt{\mathrm{Var}[f(Y, X)\mid X]}\epsilon_Y\right)^T \left(\mathbb{E}[g(Z, X)\mid X] +  \sqrt{\mathrm{Var}[g(Z, X)\mid X]}\epsilon_Z\right)\right] \notag \\
    &= \mathbb{E}[\mathbb{E}[f(Y, X)\mid X]^T \mathbb{E}[g(Z, X)\mid X]] + \label{equ:19}\\
    &\qquad \mathbb{E}[\epsilon_Y^T\sqrt{\mathrm{Var}[f(Y, X)\mid X]}\sqrt{\mathrm{Var}[g(Z, X)\mid X]} \epsilon_Z] + \label{equ:20}
      \\
     &\qquad \mathbb{E}\left[\mathbb{E}[f(Y, X)\mid X]^T \sqrt{\mathrm{Var}[g(Z, X)\mid X]}\epsilon_Z\right] + \mathbb{E}\left[\mathbb{E}[g(Z, X)\mid X]^T \sqrt{\mathrm{Var}[f(Y, X)\mid X]}\epsilon_Y\right] \label{equ:21}
    \end{align}

Since $\mathbb{E}[\epsilon_Y\mid X] = \mathbb{E}[\epsilon_Z\mid X] = 0$, by the definition of conditional expectation, the terms in \eqref{equ:21} equal to 0.
\eqref{equ:19} is a functional of the first conditional moments of $f(Y, X)$ and $g(Z, X)$. Therefore, it suffices to study \eqref{equ:20}.

Suppose that the singular decomposition of $\sqrt{\mathrm{Var}[f(Y, X)\mid X]}\sqrt{\mathrm{Var}[g(Z, X)\mid X]}$ is
$U(X) D(X) V(X)^T$, where $U(X)^TU(X) = V(X)^TV(X) = I_{p_f}$, and $D(X) = \mathrm{diag}(d_1(X), ..., d_{p_f}(X))$ is a diagonal matrix.

Define $\Tilde{\epsilon}_Y = U(X)^T\epsilon_Y$, $\Tilde{\epsilon}_Z = V(X)^T\epsilon_Z$, then $\mathbb{E}[\Tilde{\epsilon}_Y\mid X] = \mathbb{E}[\Tilde{\epsilon}_Z\mid X] = 0$, $\mathrm{Var}[\Tilde{\epsilon}_Y\mid X] = \mathrm{Var}[\Tilde{\epsilon}_Z\mid X] = I_{p_f}$, and we can rewrite \eqref{equ:20} as follows,

\begin{align*}
    \mathbb{E}[\epsilon_Y^T\sqrt{\mathrm{Var}[f(Y, X)\mid X]}\sqrt{\mathrm{Var}[g(Z, X)\mid X]} \epsilon_Z] &= \mathbb{E}[\Tilde{\epsilon}_Y^T D(X)\Tilde{\epsilon}_Z] \\
    &= \mathbb{E}[\mathbb{E}[\Tilde{\epsilon}_Y^T D(X)\Tilde{\epsilon}_Z\mid X]] \\
    &= \mathbb{E}\left[\mathbb{E}\left[\left(D(X)^{0.5}\Tilde{\epsilon}_Y\right)^T\left(D(X)^{0.5}\Tilde{\epsilon}_Z\right)\mid X\right]\right] \\
    & = \mathbb{E}\left[\sum_{k=1}^{p_f}\mathbb{E}\left[\left(d_k(X)^{0.5}(\Tilde{\epsilon}_Y)_k\right)\left(d_k(X)^{0.5}(\Tilde{\epsilon}_Z)_k\right)\mid X\right]\right] \\
    \text{(Cauchy--Schwarz)}& \le \mathbb{E}\left[\sum_{k=1}^{p_f}\sqrt{\mathbb{E}\left[d_k(X)(\Tilde{\epsilon}_Y)_k^2 \mid X\right]}\sqrt{\mathbb{E}\left[d_k(X)(\Tilde{\epsilon}_Z)_k^2 \mid X\right]}\right] \\
    &= \mathbb{E}\left[\sum_{k=1}^{p_f}\sqrt{d_k(X)}\sqrt{d_k(X)}\right] \\
    &= \mathbb{E}\left[\mathrm{tr}(D(X))\right] 
\end{align*}

Notice that 
\begin{align*}
    &\sqrt{\mathrm{Var}[g(Z, X)\mid X]}\mathrm{Var}[f(Y, X)\mid X]\sqrt{\mathrm{Var}[g(Z, X)\mid X]} \\ &= \left(\sqrt{\mathrm{Var}[f(Y, X)\mid X]}\sqrt{\mathrm{Var}[g(Z, X)\mid X]}\right)^T\left(\sqrt{\mathrm{Var}[f(Y, X)\mid X]}\sqrt{\mathrm{Var}[g(Z, X)\mid X]}\right) \\
    &= V(X)D(X)^2V(X)^T
\end{align*}

Hence 

$$ \mathrm{tr}(D(X)) = \mathrm{tr}\left(\sqrt{V(X)D(X)^2V(X)^T}\right) = \mathrm{tr}\left(\sqrt{\sqrt{\mathrm{Var}[g(Z, X)\mid X]}\mathrm{Var}[f(Y, X)\mid X]\sqrt{\mathrm{Var}[g(Z, X)\mid X]}}\right)$$

In all, we have

$$\eqref{equ:20} \le \mathbb{E}\left[\mathrm{tr}\left(\sqrt{\sqrt{\mathrm{Var}[g(Z, X)\mid X]}\mathrm{Var}[f(Y, X)\mid X]\sqrt{\mathrm{Var}[g(Z, X)\mid X]}}\right)\right]$$

Therefore, combining \eqref{equ:19}, \eqref{equ:20}, and \eqref{equ:21}, we have
\begin{align}
    \theta &\le \mathbb{E}[\mathbb{E}[f(Y, X)\mid X]^T \mathbb{E}[g(Z, X)\mid X]] + \notag \\ &\quad \mathbb{E}\left[\mathrm{tr}\left(\sqrt{\sqrt{\mathrm{Var}[g(Z, X)\mid X]}\mathrm{Var}[f(Y, X)\mid X]\sqrt{\mathrm{Var}[g(Z, X)\mid X]}}\right)\right] \label{equ:l<r}
\end{align}

Since the RHS of \eqref{equ:l<r} only relies on the first second moments of $f(Y, X)\mid X$ and $g(Y, X)\mid X$, which are fixed in the optimization problem for defining $\theta_\mathrm{U}^{(2)}$, we can conclude that

\begin{align}
    \theta_\mathrm{U}^{(2)} &\le \mathbb{E}[\mathbb{E}[f(Y, X)\mid X]^T \mathbb{E}[g(Z, X)\mid X]] + \notag \\ &\quad \mathbb{E}\left[\mathrm{tr}\left(\sqrt{\sqrt{\mathrm{Var}[g(Z, X)\mid X]}\mathrm{Var}[f(Y, X)\mid X]\sqrt{\mathrm{Var}[g(Z, X)\mid X]}}\right)\right]
\end{align}

On the other hand, if we choose $\Tilde{\epsilon}_Y = \Tilde{\epsilon}_Z = \Tilde{\epsilon}$, where $\Tilde{\epsilon}$ is an arbitrary random vector satisfying $\mathbb{E}[\Tilde{\epsilon} \mid X] =  0$, $\mathrm{Var}[\Tilde{\epsilon} \mid X] = I_{p_f}$, then all the inequalities in our previous proof degenerate to equalities, and the corresponding
$$ f(Y, X) =  \mathbb{E}[f(Y, X)\mid X] +  \sqrt{\mathrm{Var}[f(Y, X)\mid X]}U(X)\Tilde{\epsilon},$$
$$ g(Z, X) =  \mathbb{E}[f(Y, X)\mid X] +  \sqrt{\mathrm{Var}[f(Y, X)\mid X]}U(X)\Tilde{\epsilon},$$
satisfies the constraints in the optimization problem for defining $\theta_\mathrm{U}^{(2)}$ (in other words, the first two conditional moments of $f(Y, X)\mid X$ and $g(Y, X)\mid X$, are matched). Therefore, \eqref{equ:l<r} is tight, and we can conclude that

\begin{align}
    \theta_\mathrm{U}^{(2)} &= \mathbb{E}[\mathbb{E}[f(Y, X)\mid X]^T \mathbb{E}[g(Z, X)\mid X]] + \notag \\ &\quad \mathbb{E}\left[\mathrm{tr}\left(\sqrt{\sqrt{\mathrm{Var}[g(Z, X)\mid X]}\mathrm{Var}[f(Y, X)\mid X]\sqrt{\mathrm{Var}[g(Z, X)\mid X]}}\right)\right]
\end{align}

This concludes the proof.

\section{Proof of Proposition~\ref{prop:tight}}
\label{appendix:prop:tight}

Note that because $f(Y,X) \mid X \sim [U(X) \cdot g(Z, X) + V(X)]\mid X$,  in Section~\ref{appendix:cs}, $\tilde{\epsilon}_Y \sim \tilde{\epsilon}_Z$, hence the proof still holds if we replace $\theta_\mathrm{U}^{(2)}$ with the tight bound $\theta_\mathrm{U}$. Therefore,
$$ \theta_\mathrm{U} = \theta_\mathrm{U}^{(2)} = \theta_\mathrm{U}^{(\mathrm{CS})}$$.

Similarly, we have $ \theta_\mathrm{L} = \theta_\mathrm{L}^{(\mathrm{CS})}$

\section{Proof of Theorem~\ref{thm:est}} \label{appendix:est}

We focus on the upper bound $\theta_\mathrm{U}^{(\mathrm{CS})}$. The statement for the lower bound $\theta_\mathrm{L}^{(\mathrm{CS})}$ can be proved similarly (i.e., by considering the upper bound of $-\theta$).

Recall that the debiased estimator of $ \theta_\mathrm{U}^{(\mathrm{CS})}(\mathbb{P}_{Y, X}, \mathbb{P}_{Z, X})$, 
\begin{align}
	\hat{\theta}_{\mathrm{U}}^{(\mathrm{CS})} 
	&=   \sum_{i \in I_k} \frac{w_k}{n_k}\left[
	\frac{R_i}{\hat{e}^{(-k)}(X_i)}\varphi^{(\mathrm{\mathrm{CS}})}_{Y, X}(Y_i, X_i; \hat{\mathbb{P}}^{(-k)})
	+ \frac{1 - R_i}{1 - \hat{e}^{(-k)}(X_i)}\varphi^{(\mathrm{CS})}_{Z, X}(Z_i, X_i; \hat{\mathbb{P}}^{(-k)}) + M^{\left(\mathrm{CS}\right)}\left(x;\mathbb{\mathbb{P}}^{(-k)} \right)\right]
\end{align}

where 

$$ M^{\left(\mathrm{CS}\right)}\left(x;\mathbb{\mathbb{P}} \right) =  \mathbb{E}_{\mathbb{\mathbb{P}}}\left[f\left(Y\right)\mid X = x\right]\mathbb{E}_{\mathbb{\mathbb{P}}}[g\left(Z\right)\mid X = x]  + \sqrt{\mathrm{var}_{\mathbb{\mathbb{P}}}\left(f\left(Y\right)\mid X = x\right)}\sqrt{\mathrm{var}_{\mathbb{\mathbb{P}}}\left(g(Z)\mid X = x\right)}, $$
\begin{align}
	\varphi^{(\mathrm{CS})}_{Y, X}(y, x; \mathbb{P}) =& \left(f\left(y\right) - \mathbb{E}_{\mathbb{\mathbb{P}}}\left[f\left(Y\right)\mid X = x\right]\right)\mathbb{E}_{\mathbb{\mathbb{P}}}\left[g\left(Z\right)\mid X = x\right] \notag \\ &  + \frac{1}{2}\left[ \left(f\left(y\right) - \mathbb{E}_{\mathbb{\mathbb{P}}}\left[f\left(Y\right)\mid X = x\right]\right)^2 - \mathrm{var}_{\mathbb{\mathbb{P}}}\left[f\left(Y\right)\mid X = x\right] \right]\notag   \cdot \sqrt{\frac{\mathrm{var}_{\mathbb{\mathbb{P}}}\left[g\left(Z\right)\mid X = x\right]}{\mathrm{var}_{\mathbb{\mathbb{P}}}\left[f\left(Y\right)\mid X = x\right]}},   \notag
\end{align}

\begin{align}
	\varphi_{Z, X}^{\left(\mathrm{CS}\right)}\left(z, x;\mathbb{\mathbb{P}}\right) =& \left( g\left(z\right) - \mathbb{E}_{\mathbb{\mathbb{P}}}\left[g\left(Z\right)\mid X = x\right]\right)\mathbb{E}_{\mathbb{\mathbb{P}}}\left[f\left(Y\right)\mid X = x\right] \notag \\ &  +\frac{1}{2}\left[ \left(g\left(z\right) - \mathbb{E}_{\mathbb{\mathbb{P}}}\left[g\left(Z\right)\mid X = x\right]\right)^2 - \mathrm{var}_{\mathbb{\mathbb{P}}}\left[g\left(Z\right)\mid X = x\right] \right] \notag  \cdot \sqrt{\frac{\mathrm{var}_{\mathbb{\mathbb{P}}}\left[f\left(Y\right)\mid X = x\right]}{\mathrm{var}_{\mathbb{\mathbb{P}}}\left[g\left(Z\right)\mid X = x\right]}},  \notag
\end{align}
 notation $\hat{\mathbb{P}}^{(-k)}$ meaning that the first two moments of $f(Y) \mid X$ and $g(Z) \mid X$ is estimated using the $k$th data fold $I_k$ (which can also be regarded as the estimated conditional distribution $f(Y) \mid X$ and $g(Z) \mid X$ using $I_k$; note that we do not need the whole conditional distribution for our algorithm, but with it, the proof is more comprehensible),  and $w_k = I_k$ being the size of the $k$th data fold. Here to make the proof's insights clearer, we use notations such as $\varphi^{(\mathrm{CS})}_{Y, X}(y, x; \mathbb{P})$ in place of $\varphi^{(\mathrm{CS}, k)}_{Y, X}(y, x)$.

For random function $h(y, z, x, r)$, define $\mathbb{E}_k h(x, y, z, r) = \frac{1}{n_k}\sum_{i \in I_k} h(Y_i, Z_i, X_i, R_i)$.
For random function $h_Y(y, x)$, define $\mathbb{E}_{Y, X}h_Y(y, x) = \int h_Y(y, x) d\mathbb{P}_0(dy, dx)$;
similarly,  for random function $h_Z(z, x)$, define $\mathbb{E}_{Z, X}h_Z(z, x) = \int h_Z(z, x) d\mathbb{P}_{Z, X}(dz, dx)$;
for random function $h_X(x)$, define $\mathbb{E}_{X}h_X(x) = \int h_X(x) d\mathbb{P}_X(dx)$, where $\mathbb{P}_X$ is the true  marginal distribution of $X$.
In other words, $\mathbb{E}_{Y, X}$,  $\mathbb{E}_{Z, X}$, $\mathbb{E}_{X}$ is taking expectation only with respect to only the variable of random function $h_Y, h_Z, h_X$ and ignore the randomness of themselves. 

One can rewrite $\hat{\theta}_{\mathrm{U}}^{(\mathrm{CS})}$ as follows:
\begin{align}
	\hat{\theta}_{\mathrm{U}}^{(\mathrm{CS})} = \theta_{\mathrm{U}}^{(\mathrm{CS})} + T_0  + \sum_{k=1}^{K} w_k (T_{1k} + T_{2k}) \notag
\end{align}

where
\begin{align}
	T_0 &=  \sum_{k=1}^{K} \frac{w_k}{n_k} \sum_{i \in I_k}  \frac{R_i}{e(X_i)} \varphi_{Y, X}^{(\mathrm{CS})}(Y_i, X_i;\mathbb{P}_0)  
	+ \sum_{k=1}^{K} \frac{w_k}{n_k} \sum_{i \in I_k} \frac{1 - R_i}{1 - e(X_i)} \varphi_{Z, X}^{(\mathrm{CS})}(Z_i, X_i;\mathbb{P}_0)  \notag \\
	&\qquad + \sum_{k=1}^{K} \frac{w_k}{n_k} \sum_{i \in I_k}   \Bigl[M^{\left(\mathrm{CS}\right)}\left(x;\mathbb{\mathbb{P}}_0 \right) - \theta_{\mathrm{U}}^{(\mathrm{CS})}\Bigr]
\end{align}

\begin{align}
	T_{1k} &= (\mathbb{E}_k - \mathbb{E}_{Y, X}) \left(\frac{R_i}{\hat{e}^{(-k)}(X_i)}\varphi_{Y, X}^{(\mathrm{CS})}(\cdot;\hat{\mathbb{P}}^{(-k)}) - \frac{R_i}{e(X_i)}\varphi_{Y, X}^{(\mathrm{CS})}(\cdot;\mathbb{P}_0)\right)  \notag \\
	&\quad + (\mathbb{E}_k - \mathbb{E}_{Z, X}) \left(\frac{1 - R_i}{1 - \hat{e}^{(-k)}(X_i)}\varphi_{Z, X}^{(\mathrm{CS})}(\cdot;\hat{\mathbb{P}}^{(-k)}) - \frac{1 - R_i}{1 - e(X_i)} \varphi_{Z, X}^{(\mathrm{CS})}(\cdot;\mathbb{P}_0)\right) \notag \\
	&\quad + (\mathbb{E}_k - \mathbb{E}_{X}) \left(M^{\left(\mathrm{CS}\right)}\left(x;\hat{\mathbb{P}}^{(-k)}\right)  - M^{\left(\mathrm{CS}\right)}\left(x;\mathbb{\mathbb{P}}_0 \right)\right)
\end{align}

\begin{align}
	T_{2k} &= \mathbb{E}_{Y, X} \Bigl[\frac{R_i}{\hat{e}^{(-k)}(X_i)} \varphi_{Y, X}^{(\mathrm{CS})}(\cdot;\hat{\mathbb{P}}^{(-k)})\Bigr] +  \mathbb{E}_{Z, X} \Bigl[\frac{1 - R_i}{1 - \hat{e}^{(-k)}(X_i)} \varphi_{Z, X}^{(\mathrm{CS})}(\cdot;\hat{\mathbb{P}}^{(-k)})\Bigr]  +   \mathbb{E}_{X} \Bigl[M^{\left(\mathrm{CS}\right)}\left(x;\hat{\mathbb{P}}^{(-k)} \right) \Bigr]
\end{align}

\subsection{Analysis of $T_0$}
Since $|n_k - w_kn| \le 1$, $\Bigl|\frac{w_k}{n_k} - \frac{1}{n}\Bigr| \le \frac{1}{n n_k}$. Hence,
\begin{align} \label{equ:T_0}
	\sqrt{n}T_0 &= \frac{1}{\sqrt{n}} \sum_{k=1}^{K} \sum_{i \in I_k}  \frac{R_i}{e(X_i)} \varphi_{Y, X}^{(\mathrm{CS})}(Y_i, X_i;\mathbb{P}_0)   + \frac{1}{\sqrt{n}} \sum_{k=1}^{K} \sum_{i \in I_k} \frac{1 - R_i}{1 - e(X_i)}\varphi_{Z, X}^{(\mathrm{CS})}(Z_i, X_i;\mathbb{P}_0)  \notag \\
	& \quad +\frac{1}{\sqrt{n}} \sum_{k=1}^{K} \sum_{i \in I_k} \Bigl[M^{\left(\mathrm{CS}\right)}\left(x;\mathbb{\mathbb{P}}_0 \right) - \theta_{\mathrm{U}}^{(\mathrm{CS})}\Bigr] \notag \\
	& \quad + \sum_{k=1}^{K} \sqrt{n}(\frac{w_k}{n_k} - \frac{1}{n})  \sum_{i \in I_k}  \frac{R_i}{e(X_i)} \varphi_{Y, X}^{(\mathrm{CS})}(Y_i, X_i;\mathbb{P}_0) + \sum_{k=1}^{K} \sqrt{n}(\frac{w_k}{n_k} - \frac{1}{n_Z}) \sum_{i \in I_k} \frac{1 - R_i}{1 - e(X_i)} \varphi_{Z, X}^{(\mathrm{CS})}(Z_i, X_i;\mathbb{P}_0)  \notag \\
	& \quad +\sum_{k=1}^{K} \sqrt{n}(\frac{w_k}{n_k} - \frac{1}{n})\sum_{i \in I_k} \Bigl[M^{\left(\mathrm{CS}\right)}\left(x;\mathbb{\mathbb{P}}_0 \right) - \theta_{\mathrm{U}}^{(\mathrm{CS})}\Bigr] \notag \\
	& \overset{def}{=} I_Y + I_Z + I_X + \Tilde{I}_Y + \Tilde{I}_Z +\Tilde{I}_X
\end{align}

Notice that by unconfoundedness assumption,
\begin{align}
	&\mathbb{E}_{\mathbb{P}_0}\left[  \frac{R_i}{e(X_i)} \varphi_{Y, X}^{(\mathrm{CS})}(Y_i, X_i;\mathbb{P}_0) \Bigr] \right] \notag \\
	& =  \mathbb{E}_{\mathbb{P}_0}\left[\mathbb{E}_{\mathbb{P}_0}\left[  \frac{R_i}{e(X_i)} \varphi_{Y, X}^{(\mathrm{CS})}(Y_i, X_i;\mathbb{P}_0) \Big|X_i\right] \right] \notag \\
	&= \mathbb{E}_{\mathbb{P}_0}\left[\mathbb{E}_{\mathbb{P}_0}\left[  \frac{R_i}{e(X_i)} \Big|X_i\right] \mathbb{E}\left[ \varphi_{Y, X}^{(\mathrm{CS})}(Y_i, X_i;\mathbb{P}_0)  \right]\Big|X_i\right] \notag \\
	&= \mathbb{E}_{\mathbb{P}_0}\left[ \mathbb{E}_{\mathbb{P}_0}\left[ \varphi_{Y, X}^{(\mathrm{CS})}(Y_i, X_i;\mathbb{P}_0)  \Big|X_i\right] \right]\notag \\
	&= 0
\end{align}

Similarly, 
\begin{align}
	\mathbb{E}_{\mathbb{P}_0}\left[\frac{1 - R_i}{1 - e(X_i)} \varphi_{Z, X}^{(\mathrm{CS})}(Z_i, X_i;\mathbb{P}_0) \right] = 0
\end{align}

It is clear that
\begin{align}
	\mathbb{E}_{\mathbb{P}_0}\left[M^{\left(\mathrm{CS}\right)}\left(x;\mathbb{\mathbb{P}}_0 \right) - \theta_{\mathrm{U}}^{(\mathrm{CS})} \right] = 0
\end{align}

Therefore, by central limit theorem, $I_Y + I_Z + I_X$ \eqref{equ:T_0} converges in distribution to $N(0, \sigma^2)$, where
$$ \sigma^2 = \mathrm{var}_{\mathbb{P}_0}\left[\frac{R_i}{e(X_i)}\varphi_{Y, X}^{(\mathrm{CS})}(Y, X;\mathbb{P}_0) + \frac{1 - R_i}{1 - e(X_i)}\varphi_{Z, X}^{(\mathrm{CS})}(Z, X;\mathbb{P}_0) + M^{\left(\mathrm{CS}\right)}\left(x;\mathbb{\mathbb{P}}_0 \right)\right]$$

On the other hand, for $\Tilde{I}_Y$, we have

$$ \mathbb{E}_{\mathbb{P}_0}\Tilde{I}_Y =  \mathbb{E}\Bigl[ \sum_{k=1}^{K} \sqrt{n}(\frac{w_k}{n_k} - \frac{1}{n})  \sum_{i \in I_k}  \frac{R_i}{e(X_i)} \varphi_{Y, X}^{(\mathrm{CS})}(Y_i, X_i;\mathbb{P}_0) \Bigr] = 0 $$

\begin{align}
	\mathrm{var}_{\mathbb{P}_0}(\Tilde{I}_Y) &= \mathrm{var}_{\mathbb{P}_0}\Bigl[ \sum_{k=1}^{K} \sqrt{n}(\frac{w_k}{n_k} - \frac{1}{n})  \sum_{i \in I_k}  \frac{R_i}{e(X_i)} \varphi_{Y, X}^{(\mathrm{CS})}(Y_i, X_i;\mathbb{P}_0) \Bigr] \notag \\
	&= \sum_{k=1}^{K} n(\frac{w_k}{n_k} - \frac{1}{n})^2 \mathrm{var}_{\mathbb{P}_0}\Bigl[\frac{R_i}{e(X_i)} \varphi_{Y, X}^{(\mathrm{CS})}(Y_i, X_i;\mathbb{P}_0) \Bigr] \notag \\
	&\le \sum_{k=1}^{K} \frac{n}{n^2 n_k^2} \mathrm{var}_{\mathbb{P}_0}\Bigl[\frac{R_i}{e(X_i)} \varphi_{Y, X}^{(\mathrm{CS})}(Y_i, X_i;\mathbb{P}_0)\Bigr] \notag \\
	&\le   \sum_{k=1}^{K} \frac{1}{n} \mathrm{var}_{\mathbb{P}_0}\Bigl[\frac{R_i}{e(X_i)} \varphi_{Y, X}^{(\mathrm{CS})}(Y_i, X_i;\mathbb{P}_0)\Bigr] \notag \\
	& \to 0
\end{align}
Hence $\Tilde{I}_Y \overset{L_2}{\to} 0 $, which implies that  $\Tilde{I}_Y \overset{p}{\to} 0$. Similarly,  $\Tilde{I}_Z \overset{p}{\to} 0$, $\Tilde{I}_X \overset{p}{\to} 0$. 

In all, by Slutsky's theorem, we have 
$$ \sqrt{n} T_0 \overset{d}{\to} N(0, \sigma^2)$$

\subsection{Analysis of $T_{1k}$}
In this section, we prove that $\sqrt{n}T_{1k}\overset{p}{\to} 0$. To show this, it suffices to prove that $\sqrt{n}T_{1k}\overset{L_2}{\to} 0$. 
For any function $h_Y(y, x)$ of $(y, x)$, by definition, $\mathbb{E}_{Y, X} [(\mathbb{E}_k - \mathbb{E}_{Y, X}) h_Y] = 0$. As a result, 

\begin{align}
	\mathbb{E}_{Y, X} [\sqrt{n}(\mathbb{E}_k - \mathbb{E}_{Y, X}) h_Y]^2 &=  n\mathrm{var}_{Y, X} [(\mathbb{E}_k - \mathbb{E}_{Y, X}) h_Y] \notag \\
	&= n\sum_{i \in I_k} \mathrm{var}_{Y, X}[\frac{1}{n_k} var(h_Y)] \notag \\
	&= \frac{n}{n_k} var(h_Y) \notag \\
	&\le \frac{1}{c_0} \mathbb{E}[(h_Y)^2]
\end{align}

Hence, to show that $\sqrt{n}(\mathbb{E}_k - \mathbb{E}_{Y, X}) h_Y \overset{L_2}{\to} 0$, it suffices to prove that $\mathbb{E}(h_Y)^2 \to 0$ 

Therefore, to show that $\sqrt{n} T_{1k}\overset{p}{\to} 0$,  it suffices to have
\begin{equation} \label{equ:t1k:1}
	\mathbb{E}_{Y, X}\left[\frac{R_i}{\hat{e}^{(-k)}(X_i)}\varphi_{Y, X}^{(\mathrm{CS})}(Y_i, X_i;\hat{\mathbb{P}_0}^{(-k)}) - \frac{R_i}{e(X_i)}\varphi_{Y, X}^{(\mathrm{CS})}(Y_i, X_i;\mathbb{P}_0) \right]^2 \to 0
\end{equation}
\begin{equation} \label{equ:t1k:2}
	\mathbb{E}_{Z, X}\left[\frac{1 - R_i}{1 - \hat{e}^{(-k)}(X_i)}\varphi_{Z, X}^{(\mathrm{CS})}(Z_i, X_i;\hat{\mathbb{P}_0}^{(-k)}) - \frac{1 - R_i}{1 - e(X_i)} \varphi_{Z, X}^{(\mathrm{CS})}(Z_i, X_i;\mathbb{P}_0) \right]^2 \to 0
\end{equation}
\begin{equation} \label{equ:t1k:3}
	\mathbb{E}_{X}\left[M^{\left(\mathrm{CS}\right)}\left(x;\hat{\mathbb{P}}^{(-k)}\right)  - M^{\left(\mathrm{CS}\right)}\left(x;\mathbb{\mathbb{P}}_0 \right) \right]^2 \to 0
\end{equation}

The proof of \eqref{equ:t1k:1} and \eqref{equ:t1k:2} are exactly the same. We will focus on \eqref{equ:t1k:1}. 

By ignorability,
\begin{align}
	&\mathbb{E}_{Y, X}\left[\frac{R_i}{\hat{e}^{(-k)}(X_i)}\varphi_{Y, X}^{(\mathrm{CS})}(Y_i, X_i;\hat{\mathbb{P}_0}^{(-k)}) - \frac{R_i}{e(X_i)}\varphi_{Y, X}^{(\mathrm{CS})}(Y_i, X_i;\mathbb{P}_0) \right]^2 \notag \\
	&= \mathbb{E}_{Y, X}\left[R_i\left[\frac{1}{\hat{e}^{(-k)}(X_i)}\varphi_{Y, X}^{(\mathrm{CS})}(Y_i, X_i;\hat{\mathbb{P}_0}^{(-k)}) - \frac{1}{e(X_i)}\varphi_{Y, X}^{(\mathrm{CS})}(Y_i, X_i;\mathbb{P}_0) \right]^2\right] \notag \\
	&= \mathbb{E}_{Y, X}\left[\mathbb{E}_{Y, X}[R_i\mid X_i]\left[\frac{1}{\hat{e}^{(-k)}(X_i)}\varphi_{Y, X}^{(\mathrm{CS})}(Y_i, X_i;\hat{\mathbb{P}_0}^{(-k)}) - \frac{1}{e(X_i)}\varphi_{Y, X}^{(\mathrm{CS})}(Y_i, X_i;\mathbb{P}_0) \right]^2\right] \notag \\
	&= \mathbb{E}_{Y, X}\left[e(X_i)\left[\frac{1}{\hat{e}^{(-k)}(X_i)}\varphi_{Y, X}^{(\mathrm{CS})}(Y_i, X_i;\hat{\mathbb{P}_0}^{(-k)}) - \frac{1}{e(X_i)}\varphi_{Y, X}^{(\mathrm{CS})}(Y_i, X_i;\mathbb{P}_0) \right]^2\right] 
\end{align}

Define $\epsilon_{m, Y}(x) = \hat{m}_Y^{(-k)}(x) - m_Y(x)$, $\epsilon_{m, Z}(x) = \hat{m}_Z^{(-k)}(x) - m_Z(x)$, $\epsilon_{v, Y}(x) = \hat{v}_Y^{(-k)}(x) - v_Y(x)$, $\epsilon_{v, Z}(x) = \hat{v}_Z^{(-k)}(x) - v_Z(x)$, $\epsilon_{v, 0.5, Y}(x) = \sqrt{\hat{v}_Y^{(-k)}(x)} - \sqrt{v_Y(x)}$, $\epsilon_{v, 0.5,  Z}(x) = \sqrt{\hat{v}_Z^{(-k)}(x)} - \sqrt{v_Z(x)}$, $\epsilon_{v, -0.5, Y}(x) = 1 / \sqrt{\hat{v}_Y^{(-k)}(x)} -1 / \sqrt{v_Y(x)}$, $\epsilon_{v, -0.5,  Z}(x) = 1 / \sqrt{\hat{v}_Z^{(-k)}(x)}-1 / \sqrt{ v_Z(x)}$, $\epsilon_{e}(x) = 1 / \hat{e}^{(-k)}(x) - 1 / e(x)$. Then, we can write 

\begin{align}
	&\mathbb{E}_{Y, X}\left[\frac{R_i}{\hat{e}^{(-k)}(X_i)}\varphi_{Y, X}^{(\mathrm{CS})}(Y_i, X_i;\hat{\mathbb{P}_0}^{(-k)}) - \frac{R_i}{e(X_i)}\varphi_{Y, X}^{(\mathrm{CS})}(Y_i, X_i;\mathbb{P}_0) \right]^2 \notag \\
	&= \mathbb{E}_{Y, X}\Bigg[e(X_i) \Bigl(\Bigl(\frac{1}{e(X_i)} + \epsilon_{e}(X_i)\Bigr) \Bigl((f(Y_i) - m_Y(X_i) - \epsilon_{m, Y}(X_i))(m_Z(X_i) + \epsilon_{m, Z})  \notag \\ & \qquad +0.5 ((f(Y_i) - m_Y(X_i) - \epsilon_{m, Y}(X_i))^2 - v_Y(X_i) - \epsilon_{v, Y}(X_i)) (\sqrt{v_Z(X_i)} + \epsilon_{v, 0.5, Z}(X_i)) 
	\notag \\
	&\qquad \cdot (1 / \sqrt{v_Y(X_i)} + \epsilon_{v, -0.5, Y}(X_i)) \Bigr) - \varphi_{Y, X}^{(\mathrm{CS})}(\cdot;\mathbb{P}_0) \Bigr)^2 \Bigg] \notag \\
	&\le  100 \mathbb{E}_{Y, X}[e(X_i) \epsilon_e(X_i)^2 (f(Y_i) - m_Y(X_i))^2 m_z(X_i)^2] + 100 \mathbb{E}_{Y, X}[e(X_i) \epsilon_e(X_i)^2 (f(Y_i) - m_Y(X_i))^2 \epsilon_{m, z}(X_i)^2] \notag \\
	& \qquad + 100 \mathbb{E}_{Y, X}[e(X_i) \epsilon_e(X_i)^2 \epsilon_{m, Y}(X_i)^2 m_z(X_i)^2] + 100 \mathbb{E}_{Y, X}[e(X_i) \epsilon_e(X_i)^2 \epsilon_{m, Y}(X_i)^2 \epsilon_{m, z}(X_i)^2] \notag \\
	& \qquad + 100 \mathbb{E}_{Y, X}[\frac{1}{e(X_i)} \epsilon_{m, Y}(X_i)^2 m_z(X_i)^2] + 100 \mathbb{E}_{Y, X}[\frac{1}{e(X_i)} \epsilon_{m, Y}(X_i)^2 \epsilon_{m, z}(X_i)^2] \notag \\
	&\qquad + 100 \mathbb{E}_{Y, X}[ \frac{1}{e(X_i)}  (f(Y_i) - m_Y(X_i))^2 \epsilon_{m, z}(X_i)^2] \notag \\
	&\qquad + 50 \mathbb{E}_{Y, X}[e(X_i) \epsilon_e(X_i)^2 ((f(Y_i) - m_Y(X_i))^2 - v_Y(X_i))^2 v_Z(X_i) / v_Y(X_i)] \notag \\
	&\qquad+ 50 \mathbb{E}_{Y, X}[e(X_i) \epsilon_e(X_i)^2 ((f(Y_i) - m_Y(X_i))^2 - v_Y(X_i))^2 v_Z(X_i) \epsilon_{v, -0.5, Y}(X_i)^2] \notag \\
	&\qquad + 50 \mathbb{E}_{Y, X}[e(X_i) \epsilon_e(X_i)^2 ((f(Y_i) - m_Y(X_i))^2 - v_Y(X_i))^2 \epsilon_{v, 0.5, Z}(X_i)^2 / v_Y(X_i)] \notag \\
	&\qquad+ 50 \mathbb{E}_{Y, X}[e(X_i) \epsilon_e(X_i)^2 ((f(Y_i) - m_Y(X_i))^2 - v_Y(X_i))^2 \epsilon_{v, 0.5, Z}(X_i)^2 \epsilon_{v, -0.5, Y}(X_i)^2] \notag \\
	%
	&\qquad - 200 \mathbb{E}_{Y, X}[e(X_i) \epsilon_e(X_i)^2 (f(Y_i) - m_Y(X_i))^2 \epsilon_{m, Y}(X_i)^2 v_Z(X_i) / v_Y(X_i)] \notag \\
	&\qquad- 200 \mathbb{E}_{Y, X}[e(X_i) \epsilon_e(X_i)^2 (f(Y_i) - m_Y(X_i))^2 \epsilon_{m, Y}(X_i)^2 v_Z(X_i) \epsilon_{v, -0.5, Y}(X_i)^2] \notag \\
	&\qquad - 200 \mathbb{E}_{Y, X}[e(X_i) \epsilon_e(X_i)^2 (f(Y_i) - m_Y(X_i))^2 \epsilon_{m, Y}(X_i)^2 \epsilon_{v, 0.5, Z}(X_i)^2 / v_Y(X_i)] \notag \\
	&\qquad- 200 \mathbb{E}_{Y, X}[e(X_i) \epsilon_e(X_i)^2 (f(Y_i) - m_Y(X_i))^2 \epsilon_{m, Y}(X_i)^2 \epsilon_{v, 0.5, Z}(X_i)^2 \epsilon_{v, -0.5, Y}(X_i)^2] \notag \\
	%
	%
	&\qquad + 50 \mathbb{E}_{Y, X}[e(X_i) \epsilon_e(X_i)^2 \epsilon_{m, Y}(X_i)^4 v_Z(X_i) / v_Y(X_i)] \notag \\
	&\qquad+ 50 \mathbb{E}_{Y, X}[e(X_i) \epsilon_e(X_i)^2 \epsilon_{m, Y}(X_i)^4 v_Z(X_i) \epsilon_{v, -0.5, Y}(X_i)^2] \notag \\
	&\qquad + 50 \mathbb{E}_{Y, X}[e(X_i) \epsilon_e(X_i)^2 \epsilon_{m, Y}(X_i)^4 \epsilon_{v, 0.5, Z}(X_i)^2 / v_Y(X_i)] \notag \\
	&\qquad+ 50 \mathbb{E}_{Y, X}[e(X_i) \epsilon_e(X_i)^2\epsilon_{m, Y}(X_i)^4 \epsilon_{v, 0.5, Z}(X_i)^2 \epsilon_{v, -0.5, Y}(X_i)^2] \notag \\
	&\qquad - 50 \mathbb{E}_{Y, X}[e(X_i) \epsilon_e(X_i)^2  \epsilon_{v, Y}(X_i)^2 v_Z(X_i) / v_Y(X_i)] \notag \\
	&\qquad- 50 \mathbb{E}_{Y, X}[e(X_i) \epsilon_e(X_i)^2 \epsilon_{v, Y}(X_i)^2 v_Z(X_i) \epsilon_{v, -0.5, Y}(X_i)^2] \notag \\
	&\qquad - 50 \mathbb{E}_{Y, X}[e(X_i) \epsilon_e(X_i)^2 \epsilon_{v, Y}(X_i)^2 \epsilon_{v, 0.5, Z}(X_i)^2 / v_Y(X_i)] \notag \\
	&\qquad- 50 \mathbb{E}_{Y, X}[e(X_i) \epsilon_e(X_i)^2 \epsilon_{v, Y}(X_i)^2 \epsilon_{v, 0.5, Z}(X_i)^2 \epsilon_{v, -0.5, Y}(X_i)^2] \notag \\
	&\qquad+ 50 \mathbb{E}_{Y, X}[\frac{1}{e(X_i)} ((f(Y_i) - m_Y(X_i))^2 - v_Y(X_i))^2 v_Z(X_i) \epsilon_{v, -0.5, Y}(X_i)^2] \notag \\
	&\qquad + 50 \mathbb{E}_{Y, X}[\frac{1}{e(X_i)} ((f(Y_i) - m_Y(X_i))^2 - v_Y(X_i))^2 \epsilon_{v, 0.5, Z}(X_i)^2 / v_Y(X_i)] \notag \\
	&\qquad+ 50 \mathbb{E}_{Y, X}[\frac{1}{e(X_i)} ((f(Y_i) - m_Y(X_i))^2 - v_Y(X_i))^2 \epsilon_{v, 0.5, Z}(X_i)^2 \epsilon_{v, -0.5, Y}(X_i)^2] \notag \\
	%
	&\qquad - 200 \mathbb{E}_{Y, X}[\frac{1}{e(X_i)} (f(Y_i) - m_Y(X_i))^2 \epsilon_{m, Y}(X_i)^2 v_Z(X_i) / v_Y(X_i)] \notag \\
	&\qquad- 200 \mathbb{E}_{Y, X}[\frac{1}{e(X_i)} (f(Y_i) - m_Y(X_i))^2 \epsilon_{m, Y}(X_i)^2 v_Z(X_i) \epsilon_{v, -0.5, Y}(X_i)^2] \notag \\
	&\qquad - 200 \mathbb{E}_{Y, X}[\frac{1}{e(X_i)} (f(Y_i) - m_Y(X_i))^2 \epsilon_{m, Y}(X_i)^2 \epsilon_{v, 0.5, Z}(X_i)^2 / v_Y(X_i)] \notag \\
	&\qquad- 200 \mathbb{E}_{Y, X}[\frac{1}{e(X_i)} (f(Y_i) - m_Y(X_i))^2 \epsilon_{m, Y}(X_i)^2 \epsilon_{v, 0.5, Z}(X_i)^2 \epsilon_{v, -0.5, Y}(X_i)^2] \notag \\
	%
	%
	&\qquad + 50 \mathbb{E}_{Y, X}[\frac{1}{e(X_i)} \epsilon_{m, Y}(X_i)^4 v_Z(X_i) / v_Y(X_i)] \notag \\
	&\qquad+ 50 \mathbb{E}_{Y, X}\frac{1}{e(X_i)} \epsilon_{m, Y}(X_i)^4 v_Z(X_i) \epsilon_{v, -0.5, Y}(X_i)^2] \notag \\
	&\qquad + 50 \mathbb{E}_{Y, X}[\frac{1}{e(X_i)}\epsilon_{m, Y}(X_i)^4 \epsilon_{v, 0.5, Z}(X_i)^2 / v_Y(X_i)] \notag \\
	&\qquad+ 50 \mathbb{E}_{Y, X}[\frac{1}{e(X_i)} \epsilon_{m, Y}(X_i)^4 \epsilon_{v, 0.5, Z}(X_i)^2 \epsilon_{v, -0.5, Y}(X_i)^2] \notag \\
	&\qquad - 50 \mathbb{E}_{Y, X}[\frac{1}{e(X_i)}  \epsilon_{v, Y}(X_i)^2 v_Z(X_i) / v_Y(X_i)] \notag \\
	&\qquad- 50 \mathbb{E}_{Y, X}[\frac{1}{e(X_i)} \epsilon_{v, Y}(X_i)^2 v_Z(X_i) \epsilon_{v, -0.5, Y}(X_i)^2] \notag \\
	&\qquad - 50 \mathbb{E}_{Y, X}[\frac{1}{e(X_i)} \epsilon_{v, Y}(X_i)^2 \epsilon_{v, 0.5, Z}(X_i)^2 / v_Y(X_i)] \notag \\
	&\qquad- 50 \mathbb{E}_{Y, X}[\frac{1}{e(X_i)} \epsilon_{v, Y}(X_i)^2 \epsilon_{v, 0.5, Z}(X_i)^2 \epsilon_{v, -0.5, Y}(X_i)^2] \notag \\
	&\le  100 \mathbb{E}_{Y, X}[e(X_i) \epsilon_e(X_i)^2 v_Y(X_i) m_z(X_i)^2] + 100 \mathbb{E}_{Y, X}[e(X_i) \epsilon_e(X_i)^2 v_Y(X_i)\epsilon_{m, z}(X_i)^2] \notag \\
	& \qquad + 100 \mathbb{E}_{Y, X}[e(X_i) \epsilon_e(X_i)^2 \epsilon_{m, Y}(X_i)^2 m_z(X_i)^2] + 100 \mathbb{E}_{Y, X}[e(X_i) \epsilon_e(X_i)^2 \epsilon_{m, Y}(X_i)^2 \epsilon_{m, z}(X_i)^2] \notag \\
	& \qquad + 100 \mathbb{E}_{Y, X}[\frac{1}{e(X_i)} \epsilon_{m, Y}(X_i)^2 m_z(X_i)^2] + 100 \mathbb{E}_{Y, X}[\frac{1}{e(X_i)} \epsilon_{m, Y}(X_i)^2 \epsilon_{m, z}(X_i)^2] \notag \\
	&\qquad + 100 \mathbb{E}_{Y, X}[ \frac{1}{e(X_i)} v_Y(X_i) \epsilon_{m, z}(X_i)^2] \notag \\
	&\qquad + 50 \mathbb{E}_{Y, X}[e(X_i) \epsilon_e(X_i)^2 ((f(Y_i) - m_Y(X_i))^2 - v_Y(X_i))^2 v_Z(X_i) / v_Y(X_i)] \notag \\
	&\qquad+ 50 \mathbb{E}_{Y, X}[e(X_i) \epsilon_e(X_i)^2 ((f(Y_i) - m_Y(X_i))^2 - v_Y(X_i))^2 v_Z(X_i) \epsilon_{v, -0.5, Y}(X_i)^2] \notag \\
	&\qquad + 50 \mathbb{E}_{Y, X}[e(X_i) \epsilon_e(X_i)^2 ((f(Y_i) - m_Y(X_i))^2 - v_Y(X_i))^2 \epsilon_{v, 0.5, Z}(X_i)^2 / v_Y(X_i)] \notag \\
	&\qquad+ 50 \mathbb{E}_{Y, X}[e(X_i) \epsilon_e(X_i)^2 ((f(Y_i) - m_Y(X_i))^2 - v_Y(X_i))^2 \epsilon_{v, 0.5, Z}(X_i)^2 \epsilon_{v, -0.5, Y}(X_i)^2] \notag \\
	&\qquad + 50 \mathbb{E}_{Y, X}[e(X_i) \epsilon_e(X_i)^2 \epsilon_{m, Y}(X_i)^4 v_Z(X_i) / v_Y(X_i)] \notag \\
	&\qquad+ 50 \mathbb{E}_{Y, X}[e(X_i) \epsilon_e(X_i)^2 \epsilon_{m, Y}(X_i)^4 v_Z(X_i) \epsilon_{v, -0.5, Y}(X_i)^2] \notag \\
	&\qquad + 50 \mathbb{E}_{Y, X}[e(X_i) \epsilon_e(X_i)^2 \epsilon_{m, Y}(X_i)^4 \epsilon_{v, 0.5, Z}(X_i)^2 / v_Y(X_i)] \notag \\
	&\qquad+ 50 \mathbb{E}_{Y, X}[e(X_i) \epsilon_e(X_i)^2\epsilon_{m, Y}(X_i)^4 \epsilon_{v, 0.5, Z}(X_i)^2 \epsilon_{v, -0.5, Y}(X_i)^2] \notag \\
	%
	%
	&\qquad+ 50 \mathbb{E}_{Y, X}[\frac{1}{e(X_i)} ((f(Y_i) - m_Y(X_i))^2 - v_Y(X_i))^2 v_Z(X_i) \epsilon_{v, -0.5, Y}(X_i)^2] \notag \\
	&\qquad + 50 \mathbb{E}_{Y, X}[\frac{1}{e(X_i)} ((f(Y_i) - m_Y(X_i))^2 - v_Y(X_i))^2 \epsilon_{v, 0.5, Z}(X_i)^2 / v_Y(X_i)] \notag \\
	&\qquad+ 50 \mathbb{E}_{Y, X}[\frac{1}{e(X_i)} ((f(Y_i) - m_Y(X_i))^2 - v_Y(X_i))^2 \epsilon_{v, 0.5, Z}(X_i)^2 \epsilon_{v, -0.5, Y}(X_i)^2] \notag \\
	&\qquad + 50 \mathbb{E}_{Y, X}[\frac{1}{e(X_i)} \epsilon_{m, Y}(X_i)^4 v_Z(X_i) / v_Y(X_i)] \notag \\
	&\qquad+ 50 \mathbb{E}_{Y, X}[\frac{1}{e(X_i)} \epsilon_{m, Y}(X_i)^4 v_Z(X_i) \epsilon_{v, -0.5, Y}(X_i)^2] \notag \\
	&\qquad + 50 \mathbb{E}_{Y, X}[\frac{1}{e(X_i)}\epsilon_{m, Y}(X_i)^4 \epsilon_{v, 0.5, Z}(X_i)^2 / v_Y(X_i)] \notag \\
	&\qquad+ 50 \mathbb{E}_{Y, X}[\frac{1}{e(X_i)} \epsilon_{m, Y}(X_i)^4 \epsilon_{v, 0.5, Z}(X_i)^2 \epsilon_{v, -0.5, Y}(X_i)^2] \notag 
	%
\end{align}

By Cauchy-Schwarz Inequality, for random variables $W_1, W_2, W_3$,
\begin{align*}
	\mathbb{E} [W_1 W_2 W_3] \le  [\mathbb{E} [W_1^2]]^{0.5} [\mathbb{E} [W_2^2W_3^2]]^{0.5} \le \frac{1}{2}[\mathbb{E} [W_1^2]]^{0.5} [\mathbb{E} [W_2^4] + \mathbb{E} [W_3^4]]^{0.5} \le [\mathbb{E} [W_1^2]]^{0.5} [\mathbb{E} [W_2^4] + \mathbb{E} [W_3^4]]^{0.5}
\end{align*}

Similarly, for random variables $W_1, W_2, W_3, W_4$,
\begin{align*}
	\mathbb{E} [W_1 W_2 W_3 W_4] \le [\mathbb{E} [W_1^2]]^{0.5} [\mathbb{E} [W_2^8] + \mathbb{E} [W_3^8] +  \mathbb{E} [W_4^8]]^{0.5}
\end{align*}

Hence, notice that $e(x) \in [0, 1]$ for any $x \in \mathbb{R}$, we have that the above term
\begin{align}
	&\le   100 (\mathbb{E}_{Y, X}[ \epsilon_e(X_i)^4])^{0.5} (\mathbb{E}_{Y, X}[v_Y(X_i)^4] + \mathbb{E}_{Y, X}[m_z(X_i)^8] )^{0.5} \label{equ:st}\\
	&\qquad + 100 (\mathbb{E}_{Y, X}[ \epsilon_e(X_i)^4])^{0.5} (\mathbb{E}_{Y, X}[v_Y(X_i)^4] + \mathbb{E}_{Y, X}[\epsilon_{m, z}(X_i)^8] )^{0.5}\\
	& \qquad + 100 (\mathbb{E}_{Y, X}[\epsilon_e(X_i)^4])^{0.5} (\mathbb{E}_{Y, X}[\epsilon_{m, Y}(X_i)^8] + \mathbb{E}_{Y, X}[m_z(X_i)^8] )^{0.5} \\
	& \qquad + 100 (\mathbb{E}_{Y, X}[ \epsilon_e(X_i)^4])^{0.5} (\mathbb{E}_{Y, X}[\epsilon_{m, Y}(X_i)^8] +\epsilon_{m, z}(X_i)^8] )^{0.5} \\
	&\qquad + 100 (\mathbb{E}_{Y, X}[\epsilon_{m, Y}(X_i)^4])^{0.5} (\mathbb{E}_{Y, X}[\frac{1}{e(X_i)^4} + m_Z(X_i)^8] )^{0.5} \\
	&\qquad + 100 (\mathbb{E}_{Y, X}[\epsilon_{m, Y}(X_i)^4])^{0.5} (\mathbb{E}_{Y, X}[\frac{1}{e(X_i)^4} +\epsilon_{m, z}(X_i)^8] )^{0.5} \\
	&\qquad + 100 (\mathbb{E}_{Y, X}[\epsilon_{m, Z}(X_i)^4])^{0.5} (\mathbb{E}_{Y, X}[\frac{1}{e(X_i)^4} + m_Y(X_i)^8] )^{0.5} \\
	&\qquad + 50 (\mathbb{E}_{Y, X}[\epsilon_{e}(X_i)^4])^{0.5} (\mathbb{E}_{Y, X}[((f(Y_i) - m_Y(X_i))^2 - v_Y(X_i))^8 + v_Z(X_i)^8 + 1 / v_Y(X_i)^8])^{0.5} \\
	&\qquad + 50 (\mathbb{E}_{Y, X}[\epsilon_{e}(X_i)^4])^{0.5} (\mathbb{E}_{Y, X}[((f(Y_i) - m_Y(X_i))^2 - v_Y(X_i))^8 +  v_Z(X_i)^4 +  \epsilon_{v, -0.5, Y}(X_i)^{16}])^{0.5} \\
	&\qquad + 50 (\mathbb{E}_{Y, X}[\epsilon_{e}(X_i)^4])^{0.5} (\mathbb{E}_{Y, X}[((f(Y_i) - m_Y(X_i))^2 - v_Y(X_i))^8 +  \epsilon_{v, 0.5, Z}(X_i)^{16} + 1 / v_Y(X_i)^4])^{0.5} \\
	&\qquad + 50 (\mathbb{E}_{Y, X}[\epsilon_{e}(X_i)^4])^{0.5} (\mathbb{E}_{Y, X}[((f(Y_i) - m_Y(X_i))^2 - v_Y(X_i))^8 + \epsilon_{v, 0.5, Z}(X_i)^{16} + \epsilon_{v, -0.5, Y} (X_i)^8])^{0.5} \\
	&\qquad + 50 (\mathbb{E}_{Y, X}[\epsilon_{e}(X_i)^4])^{0.5} (\mathbb{E}_{Y, X}[\epsilon_{m, Y}(X_i)^{16} + v_Z(X_i)^8 + 1 / v_Y(X_i)^8])^{0.5} \\
	&\qquad + 50 (\mathbb{E}_{Y, X}[\epsilon_{e}(X_i)^4])^{0.5} (\mathbb{E}_{Y, X}[\epsilon_{m, Y}(X_i)^{16} +  v_Z(X_i)^8 + \epsilon_{v, -0.5, Y}(X_i)^{16}])^{0.5} \\
	&\qquad + 50 (\mathbb{E}_{Y, X}[\epsilon_{e}(X_i)^4])^{0.5} (\mathbb{E}_{Y, X}[\epsilon_{m, Y}(X_i)^{16} +  \epsilon_{v, 0.5, Z}(X_i)^{16} + 1 / v_Y(X_i)^8])^{0.5} \\
	&\qquad + 50 (\mathbb{E}_{Y, X}[\epsilon_{e}(X_i)^4])^{0.5} (\mathbb{E}_{Y, X}[\epsilon_{m, Y}(X_i)^{16} + \epsilon_{v, 0.5, Z}(X_i)^{16} + \epsilon_{v, -0.5, Y}(X_i)^{16}])^{0.5} \\
	&\qquad + 50 (\mathbb{E}_{Y, X}[\epsilon_{v, -0.5, Y}(X_i)^4])^{0.5} (\mathbb{E}_{Y, X}[((f(Y_i) - m_Y(X_i))^2 - v_Y(X_i))^8 + v_Z(X_i)^4 + 1 / e(X_i)^8 ])^{0.5} \\
	&\qquad + 50 (\mathbb{E}_{Y, X}[\epsilon_{v, 0.5, Z}(X_i)^4])^{0.5} (\mathbb{E}_{Y, X}[((f(Y_i) - m_Y(X_i))^2 - v_Y(X_i))^8 + 1  / v_Y(X_i)^4 +1 / e(X_i)^8])^{0.5} \\
	&\qquad + 50 (\mathbb{E}_{Y, X}[\epsilon_{v, 0.5, Z}(X_i)^4 ])^{0.5} (\mathbb{E}_{Y, X}[((f(Y_i) - m_Y(X_i))^2 - v_Y(X_i))^8 + \epsilon_{v, -0.5, Y} (X_i)^{8}])^{0.5} \\
	&\qquad + 50 (\mathbb{E}_{Y, X}[\epsilon_{m, Y}(X_i)^4])^{0.5} (\mathbb{E}_{Y, X}[\frac{1}{e(X_i)^4} + v_Z(X_i)^8 + 1 / v_Y(X_i)^8])^{0.5} \\
	&\qquad + 50 (\mathbb{E}_{Y, X}[\epsilon_{m, Y}(X_i)^4])^{0.5} (\mathbb{E}_{Y, X}[\frac{1}{e(X_i)^4} + v_Z(X_i)^8 +  \epsilon_{v, -0.5, Y}(X_i)^{16}])^{0.5} \\
	&\qquad + 50 (\mathbb{E}_{Y, X}[\epsilon_{m, Y}(X_i)^4])^{0.5} (\mathbb{E}_{Y, X}[\frac{1}{e(X_i)^4} + \epsilon_{v, 0.5, Z}(X_i)^{16} + 1 / v_Y(X_i)^8])^{0.5} \\
	&\qquad + 50 (\mathbb{E}_{Y, X}[\epsilon_{m, Y}(X_i)^4])^{0.5} (\mathbb{E}_{Y, X}[\frac{1}{e(X_i)^4} + \epsilon_{v, 0.5, Z}(X_i)^{16} + \epsilon_{v, -0.5, Y}(X_i)^{16}])^{0.5} \label{equ:end}
	%
\end{align}

Notice that
$$\mathbb{E}_{Y, X}[((f(Y_i) - m_Y(X_i))^2 - v_Y(X_i))^8] \le \mathbb{E}_{Y, X}[f(Y_i)^{16}]$$
$$ \mathbb{E}_{Y, X}[v_Y(X_i)^8] \le \mathbb{E}_{Y, X}[f(Y_i)^{16}]$$
$$ \mathbb{E}_{Y, X}[v_Z(X_i)^8] \le \mathbb{E}_{Y, X}[g(Z_i)^{16}]$$
Hence, all the terms \eqref{equ:st}-\eqref{equ:end} converge to 0. 
Therefore, $\sqrt{n}T_{1k}\overset{p}{\to} 0$.

\subsection{Analysis of $T_{2k}$}
By ignorability, one can rewrite $T_{2k}$ as follows:
\begin{align} \label{equ:T2k}
	T_{2k} &= - U_{21k} + \frac{1}{2} U_{22k} + \frac{1}{2} U_{23k} + \frac{1}{2} U_{24k}-  \mathbb{E}_X\left[(\hat{e}^{(-k)}(X) - e(X)) \Bigl[\frac{1}{\hat{e}^{(-k)}(X)} U_{25k} - \frac{1}{1 - \hat{e}^{(-k)}(X)} U_{26k} \Bigr]\right]
\end{align}

where 
\begin{align}
	U_{21k} &= \mathbb{E}_X[(\hat{m}^{(-k)}_Y(X) - m_Y(X)) (\hat{m}^{(-k)}_Z(X) - m_Z(X))] \\
	U_{22k} &= \mathbb{E}_X\Bigl[\sqrt{\frac{\hat{v}^{(-k)}_Z(X)}{\hat{v}^{(-k)}_Y(X)}} (\hat{m}^{(-k)}_Y(X) - m_Y(X))^2\Bigr] \\
	U_{23k} &= \mathbb{E}_X\Bigl[\sqrt{\frac{\hat{v}^{(-k)}_Y(X)}{\hat{v}^{(-k)}_Z(X)}} (\hat{m}^{(-k)}_Z(X) - m_Z(X))^2\Bigr] \\
	U_{24k} &= \mathbb{E}_X \left[\sqrt{\frac{\hat{v}^{(-k)}_Y(X)}{\hat{v}^{(-k)}_Z(X)}} v_Z(X) + \sqrt{\frac{\hat{v}^{(-k)}_Z(X)}{\hat{v}^{(-k)}_Y(X)}} v_Y(X) - 2 \sqrt{v_Y(X)} \sqrt{v_Z(X)}\right] \\
	U_{25k} &=  \hat{m}^{(-k)}_Z(X) (\hat{m}^{(-k)}_Y(X) - m_Y(X)) + \frac{1}{2} \sqrt{\frac{\hat{v}^{(-k)}_Y(X)}{\hat{v}^{(-k)}_Z(X)}} \Bigl((\hat{v}^{(-k)}_Y(X) - v_Y(X)) - (\hat{m}^{(-k)}_Y(X) - m_Y(X))^2\Bigr) \\
	U_{26k} &=  \hat{m}^{(-k)}_Y(X) (\hat{m}^{(-k)}_Z(X) - m_Z(X)) + \frac{1}{2} \sqrt{\frac{\hat{v}^{(-k)}_Z(X)}{\hat{v}^{(-k)}_Y(X)}} \Bigl((\hat{v}^{(-k)}_Z(X) - v_Z(X)) - (\hat{m}^{(-k)}_Z(X) - m_Z(X))^2\Bigr)
\end{align}
For $U_{21k}$, we have
\begin{align}
	U_{21k} =&\Bigl|-\mathbb{E}_X[(\hat{m}^{(-k)}_Y(X) - m_Y(X)) (\hat{m}^{(-k)}_Z(X) - m_Z(X))]\Bigr| \notag \\
	\le& \frac{1}{2}( \mathbb{E}_X[(\hat{m}^{(-k)}_Y(X) - m_Y(X))^2] + \mathbb{E}_X[(\hat{m}^{(-k)}_Z(X) - m_Z(X))^2]) \notag \\
	=& o(n^{-0.5}) \notag
\end{align}

For $U_{22k}$, by Cauchy-Schwarz inequality and Minkowski inequality, we have
\begin{align}
	U_{22k} & \le \mathbb{E}_X\Bigl[\frac{1}{\hat{v}^{(-k)}_Y(X)}\Bigr] \mathbb{E}_X\Bigl[\hat{v}^{(-k)}_Z(X) (\hat{m}^{(-k)}_Y(X) - m_Y(X))^4\Bigr] \notag \\
	&\le  \mathbb{E}_X\Bigl[\frac{1}{\hat{v}^{(-k)}_Y(X)}\Bigr] \mathbb{E}_X\Bigl[\hat{v}^{(-k)}_Z(X)\Bigr]^2 \mathbb{E}_X\Bigl[(\hat{m}^{(-k)}_Y(X) - m_Y(X))^8\Bigr] \notag \\
	&\le \left(\mathbb{E}_X\Bigl|\frac{1}{\hat{v}^{(-k)}_Y(X)} - \frac{1}{v_Y(X)}\Bigr| + \mathbb{E}_X\Bigl[\frac{1}{v_Y(X)}\Bigr] \right) (\|\hat{v}^{(-k)}_Y(X) - v_Y(X)\| + \|v_Y(X) \|)^2 \mathbb{E}_X\Bigl[(\hat{m}^{(-k)}_Y(X) - m_Y(X))^8\Bigr] \notag \\
	&= o(n^{-0.5}) \notag
\end{align}

Similarly, $U_{23k} = o(n^{-0.5})$.

For $U_{24k}$, 
\begin{align}
	U_{24k} = &\frac{1}{2} \mathbb{E}_X \Bigl[\sqrt{\frac{\hat{v}^{(-k)}_Y(X)}{\hat{v}^{(-k)}_Z(X)}} v_Z(X) + \sqrt{\frac{\hat{v}^{(-k)}_Z(X)}{\hat{v}^{(-k)}_Y(X)}} v_Y(X) - 2 \sqrt{v_Y(X)} 
	\sqrt{v_Z(X)}\Bigr] \notag \\
	=& \mathbb{E}\Bigl[\sqrt{\hat{v}^{(-k)}_Y(X)\hat{v}^{(-k)}_Z(X)}\Bigl(\sqrt{\frac{\hat{v}^{(-k)}_Y(X)}{v_Y(X)}} - \sqrt{\frac{\hat{v}^{(-k)}_Z(X)}{v_Z(X)}}\Bigr)^2\Bigr]  \notag \\
	=& o(n^{-0.5})
\end{align}

For $U_{25k}$ and $U_{26k}$, using exactly the same approach addressing $U_{22k}$, by Cauchy-Schwarz inequality and Minkowski inequality, we have

$$ \mathbb{E}_X\left[(\hat{e}^{(-k)}(X) - e(X)) \Bigl[\frac{1}{\hat{e}^{(-k)}(X)} U_{25k} - \frac{1}{1 - \hat{e}^{(-k)}(X)} U_{26k} \Bigr]\right] = o(n^{-0.5}) $$

\section{Proof of Theorem~\ref{thm:var}} \label{appendix:var}


We focus on the proof for the convergence of $\widehat{V}_\mathrm{U}$. The proof for $\widehat{V}_\mathrm{L}$ is exactly the same.

Notice that 
\begin{align} \label{equ:V_U:der}
	\widehat{V}_\mathrm{U} &= \frac{1}{n}\sum_{k=1}^{K}\sum_{i \in I_k}\left[ \varphi^{(\mathrm{CS})}_\mathrm{U}(Y_i, Z_i , X_i, R_i; \hat{\mathbb{P}}^{(-k)}) \right] \notag \\
	&=  \frac{1}{n}\sum_{k=1}^{K}\sum_{i \in I_k}\left[ \varphi^{(\mathrm{CS})}_\mathrm{U}(Y_i, Z_i , X_i, R_i; \hat{\mathbb{P}}^{(-k)})  - \varphi^{(\mathrm{CS})}_\mathrm{U}(Y_i, Z_i , X_i, R_i; \mathbb{P}) + \varphi^{(\mathrm{CS})}_\mathrm{U}(Y_i, Z_i , X_i, R_i; \mathbb{P})\right]^2 \notag \\
	&= \frac{1}{n}\sum_{k=1}^{K}\sum_{i \in I_k}\left[\varphi^{(\mathrm{CS})}_\mathrm{U}(Y_i, Z_i , X_i, R_i; \mathbb{P})\right]^2 + \frac{1}{n}\sum_{k=1}^{K}\sum_{i \in I_k}\left[\varphi^{(\mathrm{CS})}_\mathrm{U}(Y_i, Z_i , X_i, R_i; \hat{\mathbb{P}}^{(-k)})- \varphi^{(\mathrm{CS})}_\mathrm{U}(Y_i, Z_i , X_i, R_i; \mathbb{P})\right]^2 \notag \\
	&\qquad + \frac{2}{n}\sum_{k=1}^{K}\sum_{i \in I_k}\left[\varphi^{(\mathrm{CS})}_\mathrm{U}(Y_i, Z_i , X_i, R_i; \hat{\mathbb{P}}^{(-k)})\right]\left[\varphi^{(\mathrm{CS})}_\mathrm{U}(Y_i, Z_i , X_i, R_i; \hat{\mathbb{P}}^{(-k)})- \varphi^{(\mathrm{CS})}_\mathrm{U}(Y_i, Z_i , X_i, R_i; \mathbb{P})\right] \notag \\
	&\le T_5^2 + T_6^2 + 2 \sqrt{T_5} \sqrt{T_6}
\end{align}
where 
\begin{align*}
	T_5 &\overset{def}{=} \frac{1}{n}\sum_{k=1}^{K}\sum_{i \in I_k}\left[\varphi^{(\mathrm{CS})}_\mathrm{U}(Y_i, Z_i , X_i, R_i; \mathbb{P})\right]^2 \\
	T_6 &\overset{def}{=}\frac{1}{n}\sum_{k=1}^{K}\sum_{i \in I_k}\left[\varphi^{(\mathrm{CS})}_\mathrm{U}(Y_i, Z_i , X_i, R_i; \hat{\mathbb{P}}^{(-k)})- \varphi^{(\mathrm{CS})}_\mathrm{U}(Y_i, Z_i , X_i, R_i; \mathbb{P})\right]^2
\end{align*}

By strong law of large numbers, $T_5 \overset{a.s.}{\to} \mathbb{E}[\varphi^{(\mathrm{CS})}_\mathrm{U}(Y, Z , X, R; \mathbb{P})^2]$

Notice that $T_6 \ge 0$. Also,
$$ \mathbb{E}[T_6] = \mathbb{E}\left[\varphi^{(\mathrm{CS})}_\mathrm{U}(Y_i, Z_i , X_i, R_i; \hat{\mathbb{P}}^{(-k)})- \varphi^{(\mathrm{CS})}_\mathrm{U}(Y_i, Z_i , X_i, R_i; \mathbb{P})\right]^2 $$

Through \eqref{equ:t1k:1} \eqref{equ:t1k:2} \eqref{equ:t1k:3}, it is clear that $\mathbb{E}[T_6] \to 0$. Hence, $T_6 \overset{L_1}{\to} 0$.

Consequently, $T_5 \overset{p}{\to} \mathbb{E}[\varphi^{(\mathrm{CS})}_\mathrm{U}(Y, Z , X, R; \mathbb{P})^2]$, $T_6 \overset{p}{\to} 0$. Plugging into \eqref{equ:V_U:der}, we have
$$\widehat{V}_\mathrm{U} \overset{p}{\to} \mathbb{E}[\varphi^{(\mathrm{CS})}_\mathrm{U}(Y, Z , X, R; \mathbb{P})^2]$$

This concludes the proof.
\end{document}